\documentclass[prd,twocolumn,superscriptaddress,nofootinbib,amsmath,amssymb,preprintnumbers]{revtex4-2}

\usepackage[T1]{fontenc}
\usepackage{graphicx}
\usepackage{bm}
\usepackage{microtype}
\usepackage{enumitem}

\usepackage{natbib}
\usepackage[dvipsnames]{xcolor}
\usepackage[colorlinks=true,linkcolor=blue,citecolor=blue,urlcolor=blue]{hyperref}
\usepackage[capitalize]{cleveref}

\usepackage{etoolbox}
\usepackage{orcidlink}

\newcommand{\Iin}{\mathcal{I}}     
\newcommand{\Mact}{\mathcal{M}}    
\newcommand{\rhode}{\rho_{\rm de}}
\newcommand{\pde}{p_{\rm de}}
\newcommand{\wde}{w_{\rm de}}
\newcommand{\Iinde}{\mathcal{I}_{\rm de}}   
\newcommand{\Mactde}{\mathcal{M}_{\rm de}}

\newcommand{\Ls}{\Lambda_{\rm s}}
\newcommand{\rhoLs}{\rho_{\Ls}}
\newcommand{\pLs}{p_{\Ls}}

\newcommand{\zdag}{z_{\dagger}}
\newcommand{\zrep}{z_{\rm rep}}
\newcommand{\zacc}{z_{\rm acc}}
\newcommand{\rhoc}{\rho_{\rm cr0}}
\newcommand{\Omm}{\Omega_{\rm m0}}

\newcommand{\rhom}{\rho_{\rm m}}

\newcommand{\OmLs}{\Omega_{\Ls0}}
\newcommand{\Oml}{\Omega_{\ell0}}
\newcommand{\OmL}{\Omega_{\Lambda0}}

\newcommand{\tp}{_{\rm tp}}
\newcommand{\bw}{_{\rm bw}}

\def\Mpl{M_{\rm P}}

\newcommand{\zdagval}{2} 

\begin{document}


\title{Defocusing dark energy: Raychaudhuri diagnostics beyond \texorpdfstring{$w<-1/3$}{w < -1/3} and the phantom divide}

\author{\"{O}zg\"{u}r Akarsu\,\orcidlink{0000-0001-6917-6176}}
\email{akarsuo@itu.edu.tr}
\affiliation{Department of Physics, Istanbul Technical University, Maslak 34469 Istanbul, T\"{u}rkiye}

\author{Antonio De Felice\,\orcidlink{0000-0002-5556-4693}}
\email{antonio.defelice@yukawa.kyoto-u.ac.jp}
\affiliation{Center for Gravitational Physics and Quantum Information, Yukawa Institute for Theoretical Physics,
Kyoto University, 606-8502 Kyoto, Japan}

\author{N. Merve Uzun\,\orcidlink{0000-0001-5555-4500}}
\email{uzunmer@itu.edu.tr}
\affiliation{Department of Physics, Istanbul Technical University, Maslak 34469 Istanbul, T\"{u}rkiye}


\begin{abstract}
In general relativity, cosmic acceleration is timelike defocusing of the comoving congruence---a violation of the timelike convergence condition---and, through the field equations, requires a negative total active gravitational mass density, $\mathcal{M}_{\rm tot}=\rho_{\rm tot}+3p_{\rm tot}<0$. The familiar criterion $w\equiv p/\rho<-1/3$ therefore diagnoses a repulsive sector only for $\rho>0$: the inequality reverses for $\rho<0$, and ratio variables become ill-defined at $\rho=0$ even when the stress--energy tensor remains finite. For effective dark energy (DE) densities that change sign---as in $\Lambda_{\rm s}$CDM and related scenarios---we organize sector-level statements around the signed density $\rho_{\rm de}$ and two branch-independent combinations: the null energy condition (NEC) boundary $\mathcal{I}_{\rm de}=\rho_{\rm de}+p_{\rm de}=0$, which replaces the phantom divide line (PDL) $w_{\rm de}=-1$ as the regular separator between quintessence-like and phantom-like behavior, and $\mathcal{M}_{\rm de}=\rho_{\rm de}+3p_{\rm de}<0$, which governs sector-level Raychaudhuri repulsion. For a separately conserved effective DE sector with a smooth negative-to-positive density crossing of finite odd order $n$ at $z_\dagger$, we prove that $\mathcal{I}_{\rm de}$ and $\mathcal{M}_{\rm de}$ are negative in a punctured neighborhood of the crossing and non-positive at the zero, with strict inequalities for a simple zero. The ratio $w_{\rm de}$ develops a kinematic pole with universal residue $n(1+z_\dagger)/3$. If $\mathcal{M}_{\rm de}>0$ at some sufficiently high redshift, continuity requires at least one repulsion boundary $z_{\rm rep}>z_\dagger$, showing that the sector is already Raychaudhuri-repulsive while $\rho_{\rm de}<0$. We derive the exact range of $\rho_{\rm de}'(z_\dagger)$ for which the Universe accelerates at the crossing while the total NEC is satisfied there. Under the single-impulse and stationary-point assumptions specified in the text, the deceleration parameter has either one or three sign-changing zeros,
the latter defining an additional intermediate acceleration window. A smooth $\Lambda_{\rm s}$CDM profile, an exponential infrared teleparallel $f(T)$ model, and the minimal phantom brane illustrate these results, with $\Lambda$CDM and CPL as reference histories. At linear order, the perturbation equations can be formulated in unnormalized stress--energy variables without dividing by $\rho_{\rm de}$ or $\mathcal{I}_{\rm de}$, establishing kinematic regularity while leaving microphysical stability and closure relations dependent on the chosen completion. These results motivate organizing late-time inference around $(\rho_{\rm de},\mathcal{I}_{\rm de},\mathcal{M}_{\rm de})$ rather than around $w_{\rm de}$ alone.
\end{abstract}

\maketitle

\section{Introduction}
\label{sec:intro}

The late-time accelerated expansion of the Universe~\cite{SupernovaSearchTeam:1998fmf,SupernovaCosmologyProject:1998vns} is commonly discussed, within general relativity (GR), in terms of the equation of state (EoS) parameter $w_{\rm de}\equiv p_{\rm de}/\rho_{\rm de}$ of a dark energy (DE) component. On a positive-density branch, $\rhode>0$, the familiar condition $\wde<-1/3$ is simply the ratio form of $\rhode+3\pde<0$, namely the condition that the DE contribution to the Raychaudhuri source is repulsive. Likewise, for $\rhode\neq0$, the line $\wde=-1$ is the ratio representation of $\rhode+\pde=0$ and is conventionally identified as the phantom divide line (PDL) separating quintessence-like~\cite{Ratra:1987rm,Caldwell:1997ii} from phantom-like~\cite{Caldwell:1999ew} behavior~\cite{Weinberg:1988cp,Peebles:2002gy,Copeland:2006wr,Frieman:2008sn}. The interpretation of these ratio-based conditions is therefore conditional on the sign of the energy density. If $\rhode<0$, the inequalities inferred from $\wde$ reverse, whereas at $\rhode=0$ the ratio itself ceases to be a regular variable. Nevertheless, EoS-based descriptions of DE, including widely used phenomenological parametrizations such as Chevallier--Polarski--Linder (CPL)~\cite{Chevallier:2000qy,Linder:2002et,Linder:2008pp}, have become a standard language for cosmological inference. Growing interest in late-time scenarios with sign-changing \emph{effective} DE densities therefore exposes a limitation of this language. Indeed, a variety of independent analyses have explored late-time histories in which the effective DE density is negative at high redshift---or have reported hints of such behavior---with conclusions ranging from no evidence to fits competitive with or preferred over $\Lambda$ cold dark matter ($\Lambda$CDM)~\cite{Visinelli:2019qqu,Akarsu:2019hmw,
Calderon:2020hoc,Sen:2021wld,Malekjani:2023ple}. Poles in an effective EoS associated with a zero of the inferred DE density have long appeared in screened braneworld cosmologies~\cite{Alam:2005pb,Alam:2016wpf}, and the general connection between an isolated zero of a separately conserved homogeneous density and a pole in $w=p/\rho$ was established in Ref.~\cite{Ozulker:2022slu}. Our aim is not to regard such a pole as a physical singularity, but to identify the regular quantities that characterize the gravitational role of the sector across the density zero.

The underlying issue is that the EoS parameter is a \emph{ratio} variable, whereas the Friedmann-Lema\^{i}tre-Robertson-Walker (FLRW) equations of GR select particular linear combinations of $(\rho,p)$. The Hamiltonian constraint relates the total density to $H^2+K/a^2$, while the evolution equations single out the inertial mass density
$\Iin\equiv\rho+p$ and the active gravitational mass density
$\Mact\equiv\rho+3p$. For a separately conserved sector, $\Iin$ governs the evolution of its signed density through the continuity equation, whereas $\Mact$ determines whether that sector contributes attractively or repulsively to the timelike Raychaudhuri equation~\cite{Raychaudhuri:1953yv,Wald:1984rg,Hawking:1973uf}. By contrast, $w\equiv p/\rho$ is a derived parametrization whose interpretation depends on the density branch. The cosmological constant provides the simplest illustration: a Lorentz-invariant vacuum is fundamentally characterized by $T_{\mu\nu}^{(\Lambda)}=-\rho_\Lambda g_{\mu\nu}$, implying $\Iin_\Lambda=0$, while $w_\Lambda=-1$ is the corresponding ratio-based fluid description in an FLRW background. More generally, when an effective DE sector crosses $\rhode=0$, the variables $\wde=\pde/\rhode$ and $\delta_{\rm de}=\delta\rho_{\rm de}/\rhode$ can diverge even though the background stress--energy tensor remains finite; in a regular completion, the corresponding unnormalized perturbations can remain finite as well. In this setting, the usual $w$-based landmarks lose their branch-independent interpretive status, whereas the combinations selected directly by the field equations remain regular whenever the stress--energy tensor does. This perspective is reinforced by recent reconstruction studies. Model-agnostic reconstructions of $H(z)$ can remain regular while the corresponding GR-mapped effective DE density becomes very small or crosses zero; near such a crossing, large excursions of $\wde$ are algebraic ratio effects, whereas $\rhode$, $\pde$, $\Iinde$, and $q$ retain a direct physical interpretation, although the derivative-based quantities remain sensitive to the reconstruction prior~\cite{Akarsu:2026pom,Akarsu:2026anp}. A recent Raychaudhuri equation-informed spline reconstruction provides a concrete example by imposing $p_{\rm tot}<0$ over $0<z<2.33$ as a regularizing prior~\cite{GuptaChoudhury:2026gsl}. In a spatially flat FLRW spacetime within GR, this condition is equivalent to $q<1/2$. Within the standard late-time split into pressureless matter and DE, it does not exclude a local $\rhode<0$ phase with $\pde<0$, but it excludes any portion of such a branch with $\pde>0$, such as an asymptotically negative-vacuum plateau, for which $\pde=-\rhode>0$, as on a negative-$\Lambda$ branch. The prior therefore restricts, without generically forbidding, sign-changing DE density histories. In a complementary approach, an inverse scalar field reconstruction formulated directly in terms of $\rhode$ and $\rhode'$ remains well defined through a density zero and identifies $\Iinde$, rather than $\wde$, as the relevant kinetic and null energy condition (NEC) classifier~\cite{Akarsu:2026pom,Adil:2026kfn}.

To make sector-level statements precise, we adopt throughout the standard GR-like background split in which the matter (dust) and radiation sectors are separately conserved with their usual background evolution, and the remainder is identified as a possibly inferred, effective DE sector; for brevity, we refer to it below simply as the \emph{effective} DE sector, keeping in mind that it may be reconstructed from data or from a modified gravity rewriting rather than corresponding to a fundamental field. The partition into sectors is therefore description dependent---a manifestation of the dark degeneracy~\cite{Kunz:2007rk}---whereas the curvature contraction governing $\ddot a/a$ is geometric; within GR, it is sourced by the total combination $\sum_i(\rho_i+3p_i)$. Within this specified split, we call the DE sector \emph{Raychaudhuri-repulsive} whenever
$\Mactde=\rhode+3\pde<0$, meaning that its contribution raises $\ddot a/a$. We use the sign of $\Iinde=\rhode+\pde$ to characterize its sector-level NEC character and, as shorthand, refer to $\Iinde>0$ and $\Iinde<0$ as quintessence-like and phantom-like, respectively. The corresponding NEC boundary (NECB) is $\Iinde=0$. These ratio-free criteria remain meaningful on either density branch and through a smooth $\rhode=0$ crossing, provided the stress--energy remains finite. They also enforce a basic logical distinction: a sector can be Raychaudhuri-repulsive, $\Mactde<0$, without the Universe accelerating, because cosmic acceleration is controlled by the \emph{total} source $\sum_i\Mact_i$, not by any single component in isolation. Likewise, $\Iinde<0$ for an inferred sector does not by itself imply violation of the NEC by the total source~\cite{Caldwell:2025inn,Mishra:2026tzn}.

Beyond this conceptual issue, density-sign-aware diagnostics have become particularly relevant in the Dark Energy Spectroscopic Instrument (DESI) era. Recent combinations of baryon acoustic oscillations (BAO), cosmic microwave background (CMB), and Type Ia supernova data have revived interest in evolving DE descriptions, especially CPL-like parametrizations~\cite{DESI:2024mwx,DESI:2025zgx}, but the inferred dynamics can depend sensitively on dataset choices and on the functional prior used to represent the late-time expansion history~\cite{Xu:2026sbw}; the statistical evidence for the implied crossing of the PDL has itself been quantified and stress-tested in Refs.~\cite{Ozulker:2025ehg,Gokcen:2026pkq}. Smooth low-dimensional EoS parametrizations such as CPL~\cite{Chevallier:2000qy,Linder:2002et,Linder:2008pp} (see Refs.~\cite{Jassal:2004ej,Jassal:2005qc,Barboza:2008rh,Pan:2019brc,Najafi:2024qzm,Tripathi:2016slv} for other well-known examples) remain useful diagnostics, but in their standard implementation, once a positive present-day density, $\rhode(z=0)>0$, is fixed, they generate sign-preserving density histories by construction. If the actual or inferred expansion history contains localized intermediate-redshift structure, or if $\rhode$ becomes very small, such parametrizations can redistribute that structure into smoother evolution of $\wde(z)$ rather than representing it directly at the density level~\cite{Akarsu:2026pom}. Parametrizations formulated instead directly at the level of the density~\cite{DiValentino:2020naf,Adil:2023exv,Gokcen:2026pkq,Montefalcone:2026iga} or the pressure~\cite{Sen:2007gk,Cheng:2025lod}---or at the level of the linear combinations selected by the field equations, such as the inertial mass density~\cite{Akarsu:2019hmw,Acquaviva:2021jov,Escamilla:2026eks} and,
prospectively, the active gravitational mass density---need not enforce a fixed density sign by construction and would correspondingly enlarge the space of late-time histories available for probing the dynamics and nature of DE in the precision cosmology era. Recent discussions of CPL with sign-switching density have also emphasized that, once $\rhode$ is allowed to change sign, the PDL is not the appropriate separator; the regular boundary is the NECB, $\Iinde=0$~\cite{Gokcen:2026pkq}. Relatedly, model-agnostic reconstructions of the late-time expansion history allow the inferred DE density to approach zero and, for some dataset combinations, to change sign around $z\sim1.5$--$2.5$, together with localized intermediate-redshift structure in the reconstructed deceleration history~\cite{Escamilla:2023shf,Sabogal:2024qxs,Akarsu:2026pom,Akarsu:2026anp}.

Density-level sign-switching models make this possibility concrete and
suggest why a single late-time modification can bear on several
cosmological tensions at once~\cite{Verde:2019ivm,DiValentino:2020zio,Riess:2021jrx,eBOSS:2020yzd,DiValentino:2021izs,Perivolaropoulos:2021jda,Abdalla:2022yfr,Akarsu:2024qiq,CosmoVerse:2025wtd}. For otherwise fixed early-time physics and standard
matter and radiation, a negative effective contribution at intermediate
redshift lowers $H(z)$ relative to a matched $\Lambda$CDM reference and
increases the comoving distance accumulated across that interval.
Because the CMB constrains both the acoustic scale
$\theta_\ast=r_s/D_M(z_\ast)$ and the physical matter density
$\omega_{\rm m}=\Omega_{\rm m}h^2$~\cite{Planck:2018vyg}, this altered
distance budget can shift joint inference toward a larger $H_0$ and a
smaller $\Omega_{\rm m}$; the reduced matter fraction, together with
the modified post-transition growth, can in turn lower
$S_8$~\cite{Akarsu:2025ijk}. Thus, a localized change in the density
history propagates coherently into both the distance and growth
observables entering several cosmological tensions.
A particularly economical realization is provided by the
$\Lambda_{\rm s}$CDM
framework~\cite{Akarsu:2019hmw,Akarsu:2021fol,Akarsu:2022typ,Akarsu:2023mfb},
together with its VCDM embedding,
$\Lambda_{\rm s}$VCDM~\cite{Akarsu:2024qsi}, in which an AdS-to-dS-like
density sign switch occurs around $\zdag\sim2$. Multi-probe analyses report improved overall fits, with the $H_0$ (and
the associated $M_{\rm B}$) and $S_8$ tensions alleviated
together~\cite{Akarsu:2024eoo}. A dedicated growth-index analysis
further finds that the framework can alleviate the closely related
$\gamma$ tension while maintaining low $S_8$
values~\cite{Escamilla:2025imi}. Beyond linear growth, the transition
has also been studied in connection with bound cosmic structures and
the nonlinear matter power spectrum
~\cite{Paraskevas:2024ytz,Akarsu:2025nns}. When neutrino properties are
allowed to vary, the framework can retain its $H_0$--$S_8$ performance
while yielding $N_{\rm eff}$ consistent with the Standard Model
prediction and constraints on $\sum m_\nu$ compatible with
neutrino-oscillation experiments~\cite{Yadav:2024duq}. A DESI DR2 analysis further finds
that, for the dataset combinations considered, the framework can shift
the posterior for $\sum m_{\nu,\rm eff}$ from the unphysical negative
preference obtained under $\Lambda$CDM toward the physical
positive-mass region~\cite{Kibris:2026cqq}.
The quantitative conclusions depend on dataset and model choices, and
we do not perform parameter inference here. Their relevance is instead
conceptual: the density histories invoked to address these tensions
enter precisely the regime in which $w_{\rm de}$ can cease to be a
regular variable. The gravitational role of such histories is therefore
more faithfully captured by the signed density and the Raychaudhuri
combinations $(\Iinde,\Mactde)$, providing the phenomenological
motivation for the density-sign-aware description developed below.

In this work, we develop such a description. Rather than treating the
pole in $\wde$ as pathological, we organize sector-level statements
around the signed density $\rhode$ and the two combinations that enter
the field equations linearly: the inertial mass density
$\Iinde=\rhode+\pde$, whose zero is the regular NECB replacing the PDL, and the active gravitational mass density
$\Mactde=\rhode+3\pde$, whose sign governs the sector's contribution to
Raychaudhuri focusing. Our main results are as follows:
\begin{enumerate}
\item[(i)] For a separately conserved sector whose density undergoes a smooth
negative-to-positive crossing of finite odd order $n$ at $\zdag$, we prove that
$\Iinde$ and $\Mactde$ are negative on both sides sufficiently near the
crossing and non-positive at the zero (strictly negative for a simple
zero), while $\wde$ develops a kinematic pole with the universal
residue $n(1+\zdag)/3$, independent of all other profile details.
\item[(ii)] If $\Mactde>0$ at some sufficiently high redshift,
continuity implies at least one root $\zrep>\zdag$ of $\Mactde$. In
profiles with a single such root, it is the repulsion-onset redshift.
In all cases, the sector is already Raychaudhuri-repulsive while its
density is still negative, so the familiar proxy $\wde<-1/3$
misdiagnoses the onset.
\item[(iii)] We derive the exact range of the density slope
$\rho_{\rm de}'(\zdag)$ for which the Universe accelerates at the
crossing while the total NEC is satisfied there, both in the
matter-dominated approximation and including radiation.
\item[(iv)] Under the single-impulse and stationary-point assumptions specified below, we show that the
expansion history admits either one or three sign-changing zeros of the
deceleration parameter, the latter defining an additional intermediate
acceleration window. When the intermediate window contains the density zero and remains separated from the late-time acceleration phase, we establish the
ordering $\zrep>\zacc^{\rm int,begin}>\zdag>\zacc^{\rm int,end}>\zacc^{\rm late}$ [Eq.~\eqref{eq:order-z}].
\item[(v)] We illustrate these results with three backgrounds treated in the
same GR-like split: a smooth $\Ls$CDM density profile representing a
late-time AdS-to-dS-like transition~\cite{Akarsu:2022typ,Akarsu:2024qsi,Akarsu:2024eoo,Souza:2024qwd}, the exponential infrared
teleparallel $f(T)$ model~\cite{Awad:2017yod,Hashim:2020sez,Hashim:2021pkq,
Akarsu:2024nas}, and the minimal
phantom brane~\cite{Sahni:2002dx,Sahni:2000ubn,Bag:2021cqm}, with $\Lambda$CDM and CPL shown as
reference histories. The modified gravity examples demonstrate that the
separation between density sign change and repulsion onset is not
profile-specific; for the brane we further show analytically that the
$q=0$ crossing is unique.
\item[(vi)] Because zeros of $\rhode$ or $\Iinde$ make variables
normalized by these quantities ill-conditioned, we formulate a
ratio-safe linear perturbation description based on unnormalized
stress--energy variables. For a minimally modified gravity (VCDM)
embedding, the scalar-sector modification relevant to the Boltzmann
implementation can be encoded in a single regular, scale-dependent
factor fixed by the background history
~\cite{DeFelice:2020eju,DeFelice:2022uxv,Akarsu:2024qsi}.
\end{enumerate}
The value of this analysis lies in the exact, branch-independent criteria it yields: they remain well defined precisely where the standard EoS language breaks down and make clear that kinematic regularity does not by itself establish microphysical stability.

The paper is organized as follows. Section~\ref{sec:geometry} recalls
the Raychaudhuri-based, ratio-free description of focusing and
defocusing selected by the field equations.
Section~\ref{sec:diagnostics} introduces the sector-level diagnostics,
proves the local theorems near a density zero, and establishes
$\zrep>\zdag$. Section~\ref{sec:acceleration} connects the sector-level
statements to cosmic acceleration, deriving the exact
acceleration-with-total NEC window and the one-or-three counting of
$q=0$ crossings. Section~\ref{sec:examples} presents the worked
examples and a comparative summary, including the sensitivity of the
extracted redshifts to the adopted split.
Section~\ref{sec:perturbations} develops the ratio-safe perturbation
formulation, and Sec.~\ref{sec:conclusions} concludes.
Appendix~\ref{app:microphysics} surveys microphysical realizations and
their stability conditions, and App.~\ref{app:vcdm} details the VCDM
embedding and the corresponding minimal Boltzmann-code modification.
All parameter choices in the figures are illustrative and are used
solely to expose these logical distinctions.

\section{Geometry first: focusing, defocusing, and the variables selected by the field equations}
\label{sec:geometry}
Focusing and defocusing are geometrical notions before they are dynamical ones. For a timelike congruence with unit tangent $u^\mu$, expansion $\Theta=\nabla_\mu u^\mu$, four-acceleration $A^\mu\equiv\dot{u}^\mu\equiv u^\nu\nabla_\nu u^\mu$, shear $\sigma_{\mu\nu}$, and vorticity $\omega_{\mu\nu}$, the Raychaudhuri equation~\cite{Raychaudhuri:1953yv} (see also Refs.~\cite{Wald:1984rg,Hawking:1973uf}) reads
\begin{equation}
\frac{{\rm d}\Theta}{{\rm d}\tau}
=
-\frac{1}{3}\Theta^2
-\sigma_{\mu\nu}\sigma^{\mu\nu}
+\omega_{\mu\nu}\omega^{\mu\nu}
-R_{\mu\nu}u^\mu u^\nu
+\nabla_\mu A^\mu.
\end{equation}
No EoS parameter, and indeed no stress--energy tensor, appears in this identity. A gravitational field equation is needed only when one asks which source gives rise to the curvature term. Thus, accelerated expansion is not fundamentally a statement about an EoS ratio; it is a statement about the geometry of a cosmological congruence, which in GR is translated into a statement about particular stress--energy combinations.

In an FLRW spacetime written in cosmic time, the fundamental comoving congruence has unit tangent $u^\mu=\delta^\mu{}_0$, expansion scalar $\Theta=3H$, and vanishing four-acceleration, shear, and vorticity. The timelike Raychaudhuri equation therefore reduces to
\begin{equation}
\frac{\ddot a}{a}
=-\frac{1}{3}\,R_{\mu\nu}u^\mu u^\nu,
\label{eq:defocusing}
\end{equation}
so that $\ddot a>0$ is equivalent to $R_{\mu\nu}u^\mu u^\nu<0$ for the comoving congruence~\cite{Raychaudhuri:1953yv,Wald:1984rg,Hawking:1973uf}.
Throughout, by \emph{defocusing} we mean precisely this violation of the timelike convergence condition, $R_{\mu\nu}u^\mu u^\nu\geq0$, by the comoving congruence: equivalently, $\ddot a>0$, i.e., $\dot\Theta>-\Theta^2/3$ for the expansion scalar $\Theta=3H$. Defocusing in this sense does not require $\dot\Theta>0$. In the worked examples below, the total inertial mass density remains positive over the displayed interval $0\le z\le7$, so $\Theta$ continues to decrease there.

For the corresponding null contraction, one must keep track of the normalization of the null vector. For any future-directed null vector $k^\mu$, let $\mathcal{E}\equiv-u_\mu k^\mu>0$ denote the energy measured by comoving observers. FLRW symmetry gives the exact geometric identity
\begin{equation}
\dot H-\frac{K}{a^2}
=-\frac{1}{2\mathcal{E}^2}\,
R_{\mu\nu}k^\mu k^\nu.
\label{eq:Hrate}
\end{equation}
An affine rescaling of $k^\mu$ may be used to set $\mathcal{E}=1$ at a chosen event, but $\mathcal{E}$ is not constant along a cosmological null geodesic; for the comoving congruence it redshifts as $\mathcal{E}\propto1/a$. Equations~\eqref{eq:defocusing} and~\eqref{eq:Hrate} are purely geometric identities; they become statements about the cosmic sources only after a gravitational field equation is specified.

We now specialize to GR. Setting $c=1$ and moving a cosmological constant to the source side whenever it is represented as a fluid, Einstein's field equations read
\begin{equation}
R_{\mu\nu}-\frac{1}{2}R g_{\mu\nu}
=8\pi G\,T^{\rm tot}_{\mu\nu},
\quad
T^{\rm tot}_{\mu\nu}\equiv\sum_i T^{(i)}_{\mu\nu},
\label{eq:einstein}
\end{equation}
equivalently, in the trace-reversed form
\begin{equation} \label{eq:reversed}
R_{\mu\nu}
=8\pi G\left(T^{\rm tot}_{\mu\nu}
-\frac{1}{2}T^{\rm tot}g_{\mu\nu}\right).
\end{equation}
For a source written as a sum of comoving perfect fluids,
$T^\mu{}_{\nu(i)}=\mathrm{diag}[-\rho_i,p_i,p_i,p_i]$, we define
\begin{equation}
\Iin_i\equiv\rho_i+p_i,
\quad
\Mact_i\equiv\rho_i+3p_i.
\label{eq:I_M_defs}
\end{equation}
 Eq.~\eqref{eq:reversed} then gives
\begin{equation}
R_{\mu\nu}k^\mu k^\nu
=
8\pi G\,\mathcal{E}^2\sum_i\Iin_i,
\end{equation}
and
\begin{equation}
R_{\mu\nu}u^\mu u^\nu
=
4\pi G\sum_i\Mact_i.
\end{equation}
Thus, geometry selects the curvature contractions $R_{\mu\nu}k^\mu k^\nu$ and $R_{\mu\nu}u^\mu u^\nu$, while GR maps them to the total combinations $\sum_i(\rho_i+p_i)$ and $\sum_i(\rho_i+3p_i)$, respectively. Within a specified source decomposition, the linearity of Einstein's equations assigns a corresponding contribution to each sector.

In FLRW, the same field equations give the Hamiltonian constraint
\begin{equation}
H^2+\frac{K}{a^2}
=\frac{8\pi G}{3}\sum_i\rho_i,
\label{eq:friedmann}
\end{equation}
the Hubble-rate evolution equation
\begin{equation}
\dot H-\frac{K}{a^2}
=-4\pi G\sum_i(\rho_i+p_i)
\equiv-4\pi G\sum_i\Iin_i,
\label{eq:Hdot}
\end{equation}
and the acceleration equation
\begin{equation}
\frac{\ddot a}{a}
=-\frac{4\pi G}{3}\sum_i(\rho_i+3p_i)
\equiv-\frac{4\pi G}{3}\sum_i\Mact_i.
\label{eq:acc}
\end{equation}
Only two of Eqs.~\eqref{eq:friedmann}--\eqref{eq:acc} are independent once total stress--energy conservation is imposed. Here $K=-1,0,+1$ labels open, flat, and closed spatial sections, and the index $i$ runs over the chosen cosmic components, including $\Lambda$ if it is treated as a source with $p_\Lambda=-\rho_\Lambda$.

The quantity $\Iin_i$ is the enthalpy density, conventionally also called the inertial mass density, because it multiplies the four-acceleration $\dot u_\mu\equiv u^\nu\nabla_\nu u_\mu$ ($=A_\mu$)---the covariant proper-time derivative of the four-velocity along the flow---in the relativistic Euler equation,
\begin{equation}
(\rho+p)\,\dot u_\mu
=-h_\mu{}^\nu\nabla_\nu p,
\quad
h_{\mu\nu}=g_{\mu\nu}+u_\mu u_\nu.
\end{equation}
The quantity $\Mact_i$ is the active gravitational mass density entering the FLRW timelike-convergence source~\cite{Whittaker:1935lr,Tolman:1934bk}. Both are linear in $(\rho_i,p_i)$ and remain well defined for arbitrary finite real values of those variables. For a barotropic source, the functional relation $p_i=f_i(\rho_i)$ is the EoS; by contrast, $w_i\equiv p_i/\rho_i$ is a derived dimensionless ratio and is undefined at $\rho_i=0$.

Besides its contribution to the Hubble-rate evolution in Eq.~\eqref{eq:Hdot}, the inertial mass density controls the redshift derivative of any separately conserved component. If $\nabla_\mu T^{\mu\nu}_{(i)}=0$, then $\dot\rho_i+3H\Iin_i=0$, implying
\begin{equation}
\rho_i'(z)=\frac{3}{1+z}\,\Iin_i(z),
\label{eq:cont_I}
\end{equation}
where a prime denotes ${\rm d}/{\rm d}z$. Equivalently,
\begin{equation}
\frac{{\rm d}\rho_i}{{\rm d}\ln a}=-3\Iin_i.
\end{equation}
Here and throughout we assume an expanding universe, $H(z)>0$. On the physical domain $z>-1$, Eq.~\eqref{eq:cont_I} yields $\mathrm{sgn}(\rho_i')=\mathrm{sgn}(\Iin_i)$. Thus, at any redshift where $\Iin_i>0$, $\rho_i$ increases with redshift, i.e., decreases as the Universe expands, whereas at any redshift where $\Iin_i<0$, $\rho_i$ decreases with redshift, i.e., increases as the Universe expands. If $\Iin_i$ has a fixed sign over an interval, this gives the corresponding monotonic behavior throughout that interval. This statement concerns the signed density $\rho_i$ itself and requires no ratio variable such as the EoS parameter $w_i$. If sectors exchange energy, Eq.~\eqref{eq:cont_I} acquires the corresponding transfer term, while the definitions in Eq.~\eqref{eq:I_M_defs} and their roles in Eqs.~\eqref{eq:Hdot} and~\eqref{eq:acc} remain unchanged.

These same combinations make transparent the relation between matter energy conditions and curvature convergence. The NEC is the matter inequality $T_{\mu\nu}k^\mu k^\nu\ge0$ for every null vector $k^\mu$; for reviews of the energy conditions and their current status, see Refs.~\cite{Barcelo:2002bv,Curiel:2014zba,Kontou:2020bta}. In GR, because $g_{\mu\nu}k^\mu k^\nu=0$, it is equivalent to null convergence, $R_{\mu\nu}k^\mu k^\nu\ge0$. For the total FLRW perfect fluid source this reduces to
\begin{equation}
\Iin_{\rm tot}\equiv\sum_i\Iin_i\ge0.
\end{equation}
The strong energy condition (SEC),
$(T_{\mu\nu}-\tfrac12Tg_{\mu\nu})v^\mu v^\nu\ge0$ for every timelike vector $v^\mu$, is equivalent for a perfect fluid source to
\begin{equation}
\Iin_{\rm tot}\ge0,
\quad
\Mact_{\rm tot}\equiv\sum_i\Mact_i\ge0,
\end{equation}
and, through Einstein's equations, implies timelike convergence~\cite{Hawking:1973uf,Wald:1984rg}. Energy conditions may be applied to any specified stress--energy tensor, but in GR the curvature of spacetime is sourced by the total stress--energy tensor. Consequently, an inferred sector can have $\Iin_i<0$ or $\Mact_i<0$ while the corresponding total inequality remains satisfied; the interpretation of such a sector-level statement is necessarily tied to the adopted decomposition~\cite{Caldwell:2025inn}.

Equation~\eqref{eq:acc} shows that accelerated expansion requires $\Mact_{\rm tot}<0$ and therefore violates the timelike-convergence part of the SEC. Likewise, Eq.~\eqref{eq:Hdot} gives
\begin{equation}
\Iin_{\rm tot}\ge0
\quad\Longleftrightarrow\quad
\dot H-\frac{K}{a^2}\le0.
\end{equation}
For $K=0$, a phase with $\dot H>0$ is a super-accelerating phase and corresponds, within GR, to violation of the NEC by the total source.  In many concrete models such regimes are associated with issues such as ghost degrees of freedom, gradient instabilities, or future singularities~\cite{Rubakov:2014jja,Nojiri:2005sx}. These conclusions, however, cannot be drawn from the background inequality alone; they depend on the underlying degrees of freedom and their perturbation dynamics.\footnote{In a modified gravity theory recast in the standard GR-like split, the inferred expansion history can accommodate phases with $\dot H-K/a^2>0$ at the background level. This corresponds to a negative total effective enthalpy in that rewriting, but it need not originate from a fundamental NEC-violating matter field or, by itself, signal a fundamental instability. Whether such behavior is physically viable must be assessed from the perturbation-level covariant field equations, or equivalently, from the action.}

Finally, the Hamiltonian constraint restricts the total source rather than every term in a chosen decomposition. Within the FLRW source split adopted in Eq.~\eqref{eq:friedmann}, a real expansion rate requires
\begin{equation}
H^2
=\frac{8\pi G}{3}\rho_{\rm tot}-\frac{K}{a^2}
\ge0,
\quad
\rho_{\rm tot}\equiv\sum_i\rho_i.
\end{equation}
Thus, for $K=0$, the total source must satisfy $\rho_{\rm tot}=3H^2/(8\pi G)\ge0$, while for arbitrary curvature one has $\rho_{\rm tot}-3K/(8\pi G a^2)\ge0$.\footnote{In the $1{+}3$ covariant split with a timelike congruence $u^\mu$, if $\Lambda$ is kept on the geometric side, the expansion scalar $\Theta\equiv\nabla_\mu u^\mu$ obeys the Gauss--Hamiltonian constraint
$\frac{1}{3}\Theta^2=8\pi G\rho+\Lambda+\sigma^2-\omega^2-\frac{1}{2}\,{}^{(3)}\!R$, where $\rho\equiv T_{\mu\nu}u^\mu u^\nu$, ${}^{(3)}\!R$ is the projected spatial-curvature scalar (defined via the Gauss--Codazzi projection orthogonal to $u^\mu$; for $\omega\neq0$ the congruence is not hypersurface orthogonal and ${}^{(3)}\!R$ is not the intrinsic curvature of any foliation), $\sigma^2\equiv\frac12\sigma_{\mu\nu}\sigma^{\mu\nu}$, and $\omega^2\equiv\frac12\omega_{\mu\nu}\omega^{\mu\nu}$. For FLRW, $\sigma=\omega=0$, ${}^{(3)}\!R=6K/a^2$, and $\Theta=3H$, so this reduces to $H^2+\frac{K}{a^2}=\frac{8\pi G}{3}\rho+\frac{\Lambda}{3}$. If $\Lambda$ is instead absorbed into the source, one defines $\rho_{\rm tot}=\rho+\rho_\Lambda$ with $\rho_\Lambda=\Lambda/(8\pi G)$ and obtains Eq.~\eqref{eq:friedmann}. Thus, the Hamiltonian constraint fixes a combination of expansion, curvature, and the total source in the adopted bookkeeping convention; it does not impose positivity on each term in an effective decomposition. See Ref.~\cite{Ellis:1998ct}.}
No corresponding positivity condition follows for every individual sector. A negative or sign-changing inferred component is therefore not excluded by the homogeneous Friedmann constraint alone, provided the total source yields a real-valued Hubble rate, $H\in\mathbb R$. Its physical viability remains a separate, model-dependent question.

\section{Sector-level diagnostics beyond the equation of state}
\label{sec:diagnostics}

\subsection{Repulsive sector versus accelerating Universe: diagnostic versus proxy}
\label{subsec:proxy}
Equation~\eqref{eq:acc} makes a basic distinction immediate. Within a specified source decomposition, a perfect fluid sector $i$ contributes \emph{repulsively to the Raychaudhuri equation} whenever $\Mact_i=\rho_i+3p_i<0$, because its contribution to the right-hand side of the acceleration equation raises $\ddot a/a$. The Universe itself accelerates, however, only when the total source satisfies $\Mact_{\rm tot}=\sum_i\Mact_i<0$. Within this split, DE-sector repulsion is therefore necessary but not sufficient for cosmic acceleration: the repulsive contribution must overcome the attractive contributions of the other sectors. In the standard GR-like split adopted here, the matter (baryons plus CDM) and radiation (photons plus neutrinos) sectors are separately conserved with their usual background evolution, and both contribute attractively at the background level. We therefore adopt the following operational sector-level diagnostic: the effective DE sector is called \emph{Raychaudhuri-repulsive} at a given redshift whenever
\begin{equation}
\Mactde=\rhode+3\pde<0,
\label{eq:de_rep}
\end{equation}
and Raychaudhuri-repulsive over a chosen epoch of interest, e.g., the post-recombination era, if this condition holds throughout that epoch. The total Raychaudhuri source is tied to the geometry through Einstein's equations, whereas the partition into individual sectors is description-dependent; the criterion~\eqref{eq:de_rep} is branch independent, but its interpretation as a sector-level statement is tied to the adopted split. We state the scope of all sector-level claims once here rather than repeating it: (i) sector-level repulsion, $\Mactde<0$, does not imply cosmic acceleration, which is governed by the total source; (ii) sector-level phantom-like behavior, $\Iinde<0$, does not imply violation of the NEC by the total source, $\Iin_{\rm tot}\ge0$~\cite{Caldwell:2025inn,Mishra:2026tzn}; and (iii) neither property, being a statement about an effective sector within a specified split, identifies a fundamental phantom field or, by itself, signals a microphysical instability---that assessment requires a concrete completion (Sec.~\ref{sec:perturbations} and App.~\ref{app:microphysics}).
 
Away from $\rhode=0$, one may introduce the EoS parameter $\wde=\pde/\rhode$, so that $\Mactde=(1+3\wde)\rhode$. It follows that
\begin{equation}
\Mactde<0
\quad\Longleftrightarrow \quad\!
\begin{aligned}
&\wde<-\dfrac{1}{3},\quad  \rhode>0,\\[1mm]
&\wde>-\dfrac{1}{3},\quad  \rhode<0.
\end{aligned}
\label{eq:neg_rho_repulsive}
\end{equation}
Hence, the familiar condition $\wde<-1/3$ is not the general criterion for a repulsive DE sector; it is the positive-density-branch proxy for $\Mactde<0$, and the proxy reverses on the negative-density branch. At $\rhode=0$ the ratio $\wde$ is kinematically undefined, whereas the state of the sector remains regular when described by $(\rhode,\pde)$ or, equivalently, by the combinations $(\Iinde,\Mactde)$:
\begin{equation}
\Iinde=\pde,
\quad
\Mactde=3\pde
\quad \text{at} \quad \rhode=0,
\label{eq:rho_zero_IM}
\end{equation}
both of which remain finite whenever $\pde$ is finite. If, in addition, the sector is separately conserved and $\rhode(z)$ is differentiable, Eq.~\eqref{eq:cont_I} fixes the pressure at the crossing,
\begin{equation}
\pde(\zdag)=\frac{1+\zdag}{3}\,\rhode'(\zdag)
\quad \text{at} \quad \rhode(\zdag)=0,
\label{eq:p_at_crossing}
\end{equation}
a relation we exploit systematically below. What fails at the crossing is only the ratio coordinate $\wde=\pde/\rhode$, not the background stress--energy tensor or the Raychaudhuri diagnostics.
 
A parallel observation applies to the inertial mass density $\Iinde=\rhode+\pde$. Its sign determines the sector-level NEC character and, for a separately conserved sector, controls the redshift evolution of the signed density through Eq.~\eqref{eq:cont_I}: at any redshift where $\Iinde>0$, the signed density $\rhode$ increases with redshift and decreases as the Universe expands, whereas at any redshift where $\Iinde<0$, $\rhode$ decreases with redshift and increases as the Universe expands. We use the terms quintessence-like and phantom-like as shorthand for the two sector-level NEC characters $\Iinde>0$ and $\Iinde<0$, respectively. This terminology refers to the sign of $\Iinde$, not directly to the sign of $\rhode$ or to the position of $\wde$ relative to $-1$. Indeed, for $\rhode\neq0$ one has $\Iinde=(1+\wde)\rhode$, and therefore
\begin{equation}
\begin{array}{ll}
\rhode>0: \quad
\!
\begin{aligned}
& \Iinde>0\Longleftrightarrow\wde>-1,\\
           & \Iinde<0\Longleftrightarrow\wde<-1,
\end{aligned}      \\[0.6cm]     
\rhode<0: 
\quad
\!
\begin{aligned}& \Iinde>0\Longleftrightarrow\wde<-1,\\
           & \Iinde<0\Longleftrightarrow\wde>-1.
\end{aligned}
\end{array}
\label{eq:Ide_w_branches}
\end{equation}
The line $\wde=-1$ is therefore the ratio representation of $\Iinde=0$ only on a regular nonzero-density branch, and the conventional assignment of its two sides is valid only on the positive-density branch. Once $\rhode$ can be negative, the interpretation of the two sides of $\wde=-1$ reverses; at $\rhode=0$, the ratio itself ceases to exist. We therefore define the \emph{sector-level null energy condition boundary} by
\begin{equation}
\text{NECB:}\quad \Iinde=0,
\label{eq:NECB}
\end{equation}
and regard it as the regular, branch-independent separator of the sector-level NEC character within the adopted split. Accordingly, a sector-level NECB crossing occurs whenever the effective DE sector passes through $\Iinde=0$. The customary PDL, $\wde=-1$, remains useful only when $\rhode$ has fixed nonzero sign and $\wde$ is regular.
 
This distinction is especially transparent for scalar fields. Consider a minimally coupled real scalar field with ${\cal L}_\phi=\epsilon X-V(\phi)$, where $X\equiv-\frac{1}{2}g^{\mu\nu}\partial_\mu\phi\,\partial_\nu\phi=\dot\phi^{\,2}/2\ge0$ in FLRW. Its homogeneous density and pressure are $\rho_\phi=\epsilon X+V$ and $p_\phi=\epsilon X-V$, and hence
\begin{equation}
\Iin_\phi\equiv\rho_\phi+p_\phi=2\epsilon X
=\epsilon\dot\phi^{\,2}.
\label{eq:I_scalar}
\end{equation}
A canonical scalar, $\epsilon=+1$, has $\Iin_\phi\ge0$, whereas a phantom scalar, $\epsilon=-1$, has $\Iin_\phi\le0$, independently of the sign of $\rho_\phi$. Canonicality therefore fixes the kinetic character---the sign of $\Iin_\phi$---not the sign of the component energy density. For example, a canonical scalar with a sufficiently negative potential can have $\rho_\phi=X+V<0$ while $p_\phi=X-V>0$; in that case $w_\phi$ lies below $-1$ even though the kinetic structure is canonical.
 
The same point clarifies the usual no-go statement for crossing $w=-1$~\cite{Vikman:2004dc}. For $\rho_\phi\neq0$, $w_\phi=-1+\Iin_\phi/\rho_\phi$. In this minimally coupled one-field class, fixed $\epsilon$ prevents a sign change of $\Iin_\phi$; generically the NECB can only be touched at isolated turning points with $\dot\phi=0$. Thus, the invariant content of the single-field no-go statement is that such a field cannot cross the NECB without changing its kinetic structure or introducing additional degrees of freedom. If instead $\rho_\phi$ changes sign while $\Iin_\phi$ remains finite and nonzero, then $w_\phi$ passes through a pole rather than through the finite value $w_\phi=-1$. At a generic zero of $\rho_\phi$, one has
\begin{equation}
p_\phi=\Iin_\phi=2\epsilon X\neq0,
\end{equation}
so the divergence of $w_\phi=p_\phi/\rho_\phi$ is purely a failure
of the ratio parametrization. Explicit phantom-scalar realizations
with bounded step-like potentials exhibit precisely such smooth
density zeros, while $H$, $\rho_\phi$, $p_\phi$, and the field
evolution remain regular~\cite{Akarsu:2025gwi}.
Within the same class of solutions, the sector can already become
Raychaudhuri-repulsive while
$\rho_\phi<0$~\cite{Akarsu:2026lva}.
 
Finally, the repulsion condition has a useful algebraic consequence. For an inferred DE sector, $\Mactde<0$ implies $\pde<-\rhode/3$ and $\Iinde<2\rhode/3$. For $\rhode>0$, repulsion requires negative pressure, but $\Iinde$ may take either sign; a repulsive positive-density sector can therefore be either quintessence-like or phantom-like in the sector-level sense. For $\rhode<0$, the same inequalities become $\pde<|\rhode|/3$ and $\Iinde<-2|\rhode|/3$. Thus, on the negative-density branch, repulsion does not fix the sign of the pressure---$\pde$ may be negative, zero, or positive, provided it remains below $|\rhode|/3$---but it does fix the sign of $\Iinde$. Equivalently,
\begin{equation}
\rhode<0,\quad \Mactde<0
\quad\Longrightarrow\quad
\Iinde<0.
\label{eq:negative_rho_repulsion_nec}
\end{equation}
Every Raychaudhuri-repulsive negative-density sector is therefore necessarily phantom-like when viewed as an individual sector. At $\rhode=0$, Eq.~\eqref{eq:rho_zero_IM} makes the statement an equivalence: the sector is repulsive if and only if $\pde<0$, in which case $\Iinde=\pde<0$ as well (a sector-level statement, in the sense specified below Eq.~\eqref{eq:de_rep}). The two regular hypersurfaces $\Mactde=0$ and $\Iinde=0$ are therefore the meaningful separators in the $(\rhode,\pde)$ plane, while $\wde=-1/3$ and $\wde=-1$ are only their branch-dependent ratio representations.

\subsection{When does a separately conserved sector become repulsive?}
\label{subsec:repulsive}
We now restrict attention to an effective DE sector defined within the GR-like split adopted above and assumed to be separately conserved at the background level. Then, Eq.~\eqref{eq:cont_I} fixes the pressure once the signed density history $\rhode(z)$ is specified:
\begin{equation}
\pde(z)=\frac{1+z}{3}\,\rhode'(z)-\rhode(z).
\label{eq:pres}
\end{equation}
The corresponding EoS parameter is therefore
\begin{equation}
\wde(z)
=-1+\frac{1}{3}(1+z)\frac{\rhode'(z)}{\rhode(z)}
\quad \text{when} \quad \rhode\neq0.
\label{eq:eospar}
\end{equation}
Equation~\eqref{eq:eospar} is useful for comparison with the usual ratio language, but $\wde$ is not a regular variable at a density zero. The ratio-free quantities~\eqref{eq:I_M_defs} are instead
\begin{align}
\Iinde(z)
&=\frac{1+z}{3}\,\rhode'(z),
\label{eq:Iin_de}
\\
\Mactde(z)
&=(1+z)\rhode'(z)-2\rhode(z).
\label{eq:Mact_de}
\end{align}
By the Raychaudhuri equation~\eqref{eq:acc}, the sector is Raychaudhuri-repulsive at redshift $z$ if and only if
\begin{equation}
\Mactde(z)<0
\quad\Longleftrightarrow\quad
(1+z)\rhode'(z)<2\rhode(z).
\label{eq:repulsion_condition_general}
\end{equation}
On an interval ${\cal Z}$, we call the sector \emph{always repulsive} if $\Mactde(z)<0$ for all $z\in{\cal Z}$, and \emph{never repulsive} if $\Mactde(z)\ge0$ for all $z\in{\cal Z}$.

Equation~\eqref{eq:Iin_de} shows that the sector-level phantom-like character, $\Iinde<0$, is equivalent to $\rhode'(z)<0$. Thus, at such a redshift, the signed density $\rhode$ decreases with redshift and increases as the Universe expands. Repulsion is more restrictive: by Eq.~\eqref{eq:repulsion_condition_general}, it depends not only on $\rhode'$ but also on the sign and magnitude of $\rhode$ itself. On fixed-sign density branches this gives a useful classification.

\emph{Positive-density branch.} If $\rhode>0$, then Eq.~\eqref{eq:repulsion_condition_general} is equivalent to $\frac{(1+z)\rhode'}{\rhode}<2$. Every positive-density phantom-like segment, $\rhode'<0$, together with its NECB limit $\rhode'=0$, is automatically Raychaudhuri-repulsive, because $(1+z)\rhode'\le0<2\rhode$. If $\rhode'\ge0$, repulsion requires sufficiently mild growth toward the past: $0\le (1+z)\rhode'<2\rhode$. If instead $(1+z)\rhode'\ge2\rhode>0$, the sector is non-repulsive. Since $\wde$ is regular on this branch, the criterion reduces to the familiar positive-density proxy $\wde<-1/3$.

\emph{Negative-density branch.} If $\rhode<0$, the inequality reverses when written as a logarithmic slope: $\frac{(1+z)\rhode'}{\rhode}>2$. Thus, if $\rhode'\le0$, the sector is repulsive precisely when the magnitude grows toward the past fast enough, $(1+z)|\rhode'|>2|\rhode|$, and non-repulsive if instead $0\le(1+z)|\rhode'|\le2|\rhode|$. For $\rhode<0$ and $\rhode'\ge0$, Eq.~\eqref{eq:repulsion_condition_general} cannot be satisfied, since $(1+z)\rhode'\ge0>2\rhode$; the sector is then never repulsive on that interval. Never-repulsiveness on this branch forces strictly positive pressure, $\pde\ge|\rhode|/3>0$, by Eq.~\eqref{eq:pres}. Away from the crossing, the same statement can be written in ratio language as the reversed proxy $\wde>-1/3$ for repulsion.

The most important case for the present work is a smooth negative-to-positive density crossing. Suppose that
\begin{equation}
\rhode(\zdag)=0,
\quad
\rhode(z)<0\, (z>\zdag),
\quad
\rhode(z)>0\, (z<\zdag),
\label{eq:sign_switch_assumption}
\end{equation}
and assume that $\rhode(z)$ is differentiable in a neighborhood of $\zdag$. Evaluating Eqs.~\eqref{eq:Iin_de} and~\eqref{eq:Mact_de} at the zero gives
\begin{equation}
\Mactde(\zdag)=3\Iinde(\zdag)
=(1+\zdag)\rhode'(\zdag).
\label{eq:IM_crossing}
\end{equation}
Thus, at the crossing, repulsion is equivalent to sector-level phantom-like behavior:
\begin{equation}
\Mactde(\zdag)<0
\,\,\Longleftrightarrow\,\,
\Iinde(\zdag)<0
\,\,\Longleftrightarrow\,\,
\rhode'(\zdag)<0.
\end{equation}
No analogous statement can be formulated in terms of $\wde$, because the ratio $\wde=\pde/\rhode$ is undefined at $\rhode=0$.

The sign of $\rhode'(\zdag)$ requires a small qualification. The sign pattern in Eq.~\eqref{eq:sign_switch_assumption} renders both one-sided difference quotients of $\rhode$ at $\zdag$ strictly negative, so differentiability alone forces $\rhode'(\zdag)\le0$. For a simple zero, where $\rhode'(\zdag)\neq0$ by definition, this gives $\rhode'(\zdag)<0$. For a higher-order zero, however, the first derivative at the crossing vanishes. More generally, suppose that the first non-vanishing derivative of $\rhode$ at $\zdag$ occurs at finite odd order $n$ and that $\rhode$ is $C^{n+1}$ in a neighborhood of $\zdag$---even orders are excluded, since they would not change the sign of $\rhode$. Then, locally,
\begin{equation}
\rhode(z)
=
A(z-\zdag)^n
+\mathcal O\!\left((z-\zdag)^{n+1}\right),
\label{eq:rho_finite_order_crossing}
\end{equation}
where $n$ is odd and $A<0$. The condition $A<0$ is precisely the negative-to-positive orientation in Eq.~\eqref{eq:sign_switch_assumption}. Substituting Eq.~\eqref{eq:rho_finite_order_crossing} into Eqs.~\eqref{eq:pres}--\eqref{eq:Mact_de}, and using Eq.~\eqref{eq:eospar} for the ratio variable, gives
\begin{align}
\pde(z)
&=
\frac{n(1+\zdag)A}{3}(z-\zdag)^{n-1}
+\mathcal O\!\left((z-\zdag)^n\right),
\label{eq:p_finite_order_crossing}
\\
\Iinde(z)
&=
\frac{n(1+\zdag)A}{3}(z-\zdag)^{n-1}
+\mathcal O\!\left((z-\zdag)^n\right),
\label{eq:I_finite_order_crossing}
\\
\Mactde(z)
&=
n(1+\zdag)A(z-\zdag)^{n-1}
+\mathcal O\!\left((z-\zdag)^n\right),
\label{eq:M_finite_order_crossing}
\\
\wde(z)
&=
\frac{n(1+\zdag)}{3(z-\zdag)}
+\mathcal O(1).
\label{eq:w_finite_order_crossing}
\end{align}
The residue $n(1+\zdag)/3$ is universal: it depends only on the crossing redshift and the order of the zero, not on the amplitude $A$ or on other model details. Since $n-1$ is even and $A<0$, both $\Iinde$ and $\Mactde$ are negative in a punctured neighborhood of the crossing. For a simple zero, $n=1$, they are strictly negative at $z=\zdag$ itself. For higher odd order, $n>1$, they vanish at the exact crossing but remain negative immediately on both sides. Hence a smooth finite-order negative-to-positive sign switch is phantom-like and Raychaudhuri-repulsive in a punctured neighborhood of $\zdag$, with equality at the crossing only in non-generic higher-order cases:
\begin{equation}
\Iinde(\zdag)\le0,
\quad
\Mactde(\zdag)\le0,
\label{eq:IM_crossing_inequality}
\end{equation}
with strict inequalities for a simple zero.

Equation~\eqref{eq:w_finite_order_crossing} makes explicit that the divergence of $\wde$ at the crossing is purely kinematic. For a simple zero ($n=1$), the expansions reduce to
\begin{align}
\rhode(z)
&=\rhode'(\zdag)(z-\zdag)
+\mathcal O\!\left((z-\zdag)^2\right),
\label{eq:rho_simple_cross}
\\
\pde(z)
&=\frac{1+\zdag}{3}\rhode'(\zdag)
+\mathcal O(z-\zdag),
\label{eq:p_simple_cross}
\\
\wde(z)
&=\frac{1+\zdag}{3(z-\zdag)}
+\mathcal O(1).
\label{eq:w_pole_simple_cross}
\end{align}
The leading term of Eq.~\eqref{eq:p_simple_cross} is precisely the finite crossing pressure anticipated in Eq.~\eqref{eq:p_at_crossing}; for a simple zero it is strictly negative, $\pde(\zdag)=\frac{1+\zdag}{3}\,\rhode'(\zdag)<0$, so that $\Mactde(\zdag)=3\pde(\zdag)<0$ at the crossing. Because $1+\zdag>0$, one has
\begin{equation}
\wde\to+\infty\quad (z\to\zdag^+),
\quad
\wde\to-\infty\quad (z\to\zdag^-).
\label{eq:w_pole_sides}
\end{equation}
The apparent passage from $\wde>-1$ to $\wde<-1$ is therefore not a crossing of a regular divide at all: since $\Iinde<0$ throughout a punctured neighborhood of $\zdag$, the finite value $\wde=-1$ is never attained there, and the two sides of the PDL are connected only through the pole of the ratio coordinate.

This also explains why the opposite local sign assignment in $\wde$ language---$\wde<-1$ on the negative-density side ($z>\zdag$) and $\wde>-1$ on the positive-density side ($z<\zdag$), i.e., $\Iinde>0$ near the transition---does not describe a smooth zero. By Eq.~\eqref{eq:Iin_de}, such a neighborhood would require $\rhode'>0$ there. This is impossible for a continuous zero crossing with the sign pattern of Eq.~\eqref{eq:sign_switch_assumption}: $\rhode'>0$ immediately above $\zdag$, together with $\rhode(\zdag)=0$, would give $\rhode>0$ just above the crossing, contradicting $\rhode<0$ for $z>\zdag$; consistently, the point value obeys $\rhode'(\zdag)\le0$, as established above. Realizing this assignment therefore requires leaving the smooth finite stress-energy class in favor of a sign flip without a zero. A jump-like sign flip makes $\rhode'$, and hence $\Iinde$ and $\Mactde$, distributional; a pole-like sign flip makes $\rhode$, $\Iinde$, and $\Mactde$ singular and generically renders $H^2$ singular as well. A cusp-like zero with divergent derivative also lies outside the class of smooth, finite stress--energy tensors, because $\Iinde$ and $\Mactde$ diverge at the crossing; however, it still respects the sign pattern of Eq.~\eqref{eq:sign_switch_assumption} and remains phantom-like, $\Iinde\to-\infty$ as $z\to \zdag$, so it does not realize the opposite assignment either.

\emph{Remark (fractional-order crossings).} The universal-residue formula is not restricted to integer order. For a fractional-order (cusp-like) crossing $\rhode\propto{\rm sgn}(\zdag-z)\,|z-\zdag|^{\alpha}$ with $0<\alpha<1$, the pole structure of Eq.~\eqref{eq:w_finite_order_crossing} persists with $n\to\alpha$, namely,
\begin{equation}
\wde=\frac{\alpha\,(1+\zdag)}{3\,(z-\zdag)}+\mathcal O(1),
\label{eq:w_fractional_crossing}
\end{equation}
so the residue $\alpha(1+\zdag)/3$ again encodes only the order of the zero and its location, at the price of singular sector stress--energy at the crossing: $\Iinde$ and $\Mactde$ diverge as $|z-\zdag|^{\alpha-1}$. None of these non-smooth histories will be considered here.

Finally, suppose that the sector is attractive at some high-redshift $z_{\rm hi}>\zdag$, i.e., $\Mactde(z_{\rm hi})>0$, and that $\rhode$ is continuously differentiable on $(\zdag,z_{\rm hi}]$. For a smooth finite-order negative-to-positive crossing, $\Mactde<0$ immediately above $\zdag$. By continuity, there must then exist at least one redshift
\begin{equation}
\zrep\in(\zdag,z_{\rm hi}),
\quad
\Mactde(\zrep)=0.
\label{eq:zrep_existence}
\end{equation}
Thus, under this mild high-redshift assumption, there exists at least one redshift $\zrep>\zdag$ at which the effective DE sector passes from an attractive to a repulsive phase before the density sign switch itself. In profiles with a single such root, this root is the repulsion-onset redshift; in general, uniqueness of $\zrep$ requires additional assumptions about the profile of $\rhode(z)$. Moreover, on any interval below $\zrep$ where $\Mactde<0$ and $\rhode<0$, the sector is necessarily phantom-like as well, in accordance with Eq.~\eqref{eq:negative_rho_repulsion_nec}.

The conclusion is that, for a separately conserved sign-switching sector, the smooth background dynamics are controlled by $\rhode$, $\pde$, $\Iinde$, and $\Mactde$, not by the ratio $\wde$. In particular, a smooth finite-order negative-to-positive sign switch is phantom-like and Raychaudhuri-repulsive in a punctured neighborhood of the crossing, and at the crossing itself for a simple zero; higher odd-order zeros instead saturate $\Iinde(\zdag)=\Mactde(\zdag)=0$. The associated pole in $\wde$ carries the universal, amplitude-independent residue $n(1+\zdag)/3$, and, under a mild high-redshift condition, sector repulsion sets in before the switch itself, $\zrep>\zdag$. A concrete realization of such a regular branch can be provided by minimally coupled phantom scalar dynamics~\cite{Akarsu:2025gwi}, and recent background reconstructions formulated directly in terms of $\rhode(z)$ and $\rhode'(z)$ similarly identify the sign of $\Iinde$ as the relevant consistency criterion~\cite{Adil:2026kfn}. Ratio variables such as $\wde$ and $\delta_{\rm de}$ are therefore ill-conditioned at $\rhode=0$, whereas the Raychaudhuri diagnostics remain finite and directly interpretable. The same point underlies recent CPL-based sign-switching analyses, in which the meaningful boundary is the NECB, $\Iinde=0$, rather than the conventional PDL, $\wde=-1$~\cite{Gokcen:2026pkq}.

\subsection{Two repulsive branches and the epoch intuition}
\label{subsec:branches}
As a benchmark, consider a fixed-sign, separately conserved DE component with a linear EoS,
$\pde=w_\star\,\rhode$, $w_\star={\rm const.}$ The sign of $\rhode$ is then fixed by the integration constant, and stress--energy conservation~\eqref{eq:cont_I} gives $\rhode\propto(1+z)^{3(1+w_\star)}$. Since $\Mactde=(1+3w_\star)\rhode$, a repulsive DE contribution, $\Mactde<0$, admits two branches: $w_\star<-1/3$ for $\rhode>0$ and $w_\star>-1/3$ for $\rhode<0$. For comparison with the attractive standard components, $\Mact_{\rm m}=\rhom\propto(1+z)^3$ and $\Mact_{\rm r}=\rho_{\rm r}+3p_{\rm r}=2\rho_{\rm r}\propto(1+z)^4$, one has
\begin{equation}
\frac{|\Mactde|}{\Mact_{\rm m}}\propto (1+z)^{3w_\star},
\quad
\frac{|\Mactde|}{\Mact_{\rm r}}\propto (1+z)^{3w_\star-1}.
\label{eq:scaling_ratios}
\end{equation}
Equation~\eqref{eq:scaling_ratios} encodes the basic epoch intuition. On the standard positive-density repulsive branch, $w_\star<-1/3$, the repulsive contribution grows relative to both matter and radiation as $z$ decreases, making late-time acceleration natural. On the negative-density repulsive branch, $w_\star>-1/3$, the epoch of relevance is controlled by the value of $w_\star$: for $-1/3<w_\star<0$, the repulsive term still grows relative to matter toward the future, and is therefore also naturally late-time, although with a milder scaling than for more negative $w_\star$; for $w_\star=0$, it tracks matter in magnitude; for $0<w_\star<1/3$, it grows toward the past relative to matter while remaining suppressed relative to radiation as $z\to\infty$, and is therefore naturally most relevant at intermediate redshifts during matter domination. For $w_\star\ge1/3$, the contribution is no longer automatically suppressed relative to radiation in the far past and would be strongly constrained by early-Universe consistency if its amplitude were appreciable; in particular, a negative-density component with $w_\star=1/3$ acts as a negative effective contribution to $\Delta N_{\rm eff}$ and is bounded by BBN and CMB constraints on the early expansion rate~\cite{Fields:2019pfx,Planck:2018vyg}.
This constant-$w_\star$ classification is only a fixed-branch guide. A separately conserved constant-$w_\star$ component cannot describe a smooth sign change of $\rhode$; its sign is fixed by the integration constant. Sign-switching histories instead require nontrivial DE dynamics, either from multiple effective contributions, for example alongside $\Lambda$, or from a single sector whose $\rhode(z)$ departs from a constant-$w_\star$ power law. In particular, if $\rhode(z)$ changes sign, the expansion history can interpolate between the branchwise regimes above and exhibit localized repulsive phases that no single constant $w_\star$, nor any globally regular $\wde(z)$ across $\rhode=0$, can capture. Moreover, if $|\rhode|$ grows too large for a negative-density component, which enters the Friedmann equation with its sign, it can drive $H^2\to0$ and lead to a turnaround or recollapse. Such histories require a separate analysis, and Eq.~\eqref{eq:scaling_ratios} is used here only to build intuition.

\section{Cosmic acceleration near a density zero}
\label{sec:acceleration}

\subsection{Link to cosmic acceleration near a \texorpdfstring{$\rhode=0$}{rho\_de = 0} crossing (matter era)}
\label{subsec:crossing}
We now connect the sector-level diagnostic to the acceleration of the Universe. For clarity, we first consider a spatially flat matter plus DE setting and neglect radiation. The Raychaudhuri equation~\eqref{eq:acc} reads
\begin{equation}
\label{eq:acc_dust_de}
\frac{\ddot a}{a}
=
-\frac{4\pi G}{3}\Big[\rhom(z)+\Mactde(z)\Big],
\end{equation}
with
$\rhom(z)=\rho_{\rm m0}(1+z)^3$, where a subscript $0$ denotes the present-day value. Equivalently, the deceleration parameter $q\equiv-\ddot a/(aH^2)$ is, for $K=0$,
\begin{equation}
q(z)
=
\frac{\Mact_{\rm tot}(z)}{2\rho_{\rm cr}(z)}
=
\frac{1}{2}\,
\frac{\rhom(z)+\Mactde(z)}
{\rhom(z)+\rhode(z)},
\label{eq:q_general}
\end{equation}
where $\rho_{\rm cr}(z)\equiv3H^2(z)/(8\pi G)=\rhom(z)+\rhode(z)$ in the spatially flat case. Hence the Universe accelerates, $\ddot a>0$, if and only if
\begin{equation}
\Mact_{\rm tot}(z)
\equiv
\rhom(z)+\Mactde(z)<0.
\label{eq:acc_condition_general}
\end{equation}
Since the matter sector is attractive, this is stronger than sector-level repulsion: cosmic acceleration requires
$\Mactde<-\rhom<0$, not merely $\Mactde<0$.
 
The same flat Friedmann constraint~\eqref{eq:friedmann},
$H^2=\frac{8\pi G}{3}\,[\rhom(z)+\rhode(z)]$, imposes
\begin{equation}
\rhom(z)+\rhode(z)>0.
\end{equation}
This condition is nontrivial when $\rhode<0$, where it becomes $\rhode>-\rhom$. Combining $H^2>0$ with the acceleration condition gives the double inequality
\begin{equation}
\rhode+3\pde<-\rhom<\rhode.
\label{eq:H2_acc_interval}
\end{equation}
For a given matter density, Eq.~\eqref{eq:H2_acc_interval} is the exact condition for simultaneous $H^2>0$ and $\ddot a>0$ in the flat matter plus DE setting. The interval for $-\rhom$ is nonempty if and only if $\pde<0$; negative DE pressure is therefore necessary, though not sufficient, because $-\rhom(z)$ must actually lie inside the interval at the epoch in question. In ratio language this algebraic remark corresponds to $\wde<0$ on a positive-density branch and $\wde>0$ on a negative-density branch, but the ratio language is again secondary.
 
For a sign-switching history, $\Mactde=(1+z)\rhode'-2\rhode$ [Eq.~\eqref{eq:Mact_de}] contains a derivative term, so a sufficiently sharp transition in $\rhode(z)$ can produce a localized negative excursion in the Raychaudhuri source. At a differentiable density zero, $\rhode(\zdag)=0$, Eq.~\eqref{eq:IM_crossing} gives $\Mactde(\zdag)=(1+\zdag)\rhode'(\zdag)$, so the Universe accelerates at the crossing if and only if
\begin{equation}
\rhode'(\zdag)
<
-\rho_{\rm m0}(1+\zdag)^2.
\label{eq:acc_at_crossing}
\end{equation}
For a simple negative-to-positive crossing this is a condition on the negative slope of $\rhode$ at the zero. Since $H^2=(8\pi G/3)\rhom(\zdag)>0$ at the crossing and $\Mact_{\rm tot}(z)$ is continuous for a smooth $\rhode(z)$, the strict inequality~\eqref{eq:acc_at_crossing} implies an open interval around $\zdag$ in which $q(z)<0$. This guarantees an open acceleration interval associated with the density sign switch.
 
If Eq.~\eqref{eq:acc_at_crossing} fails, the DE sector can still be Raychaudhuri-repulsive near $\zdag$, but the repulsive contribution is not large enough at the crossing to overcome matter. Acceleration may then occur only at low redshift (the standard late-time acceleration onset), or in a window at a finite offset from $\zdag$ if the negative excursion of $\Mactde$ deepens away from the crossing. For a higher odd-order zero, $\rhode'(\zdag)=0$, so acceleration cannot occur exactly at the crossing in a matter-dominated background, although it may still occur at a finite offset from $\zdag$ if $\Mactde$ becomes sufficiently negative.
 
It is also useful to separate acceleration at the crossing from violation of the total NEC. At $\rhode(\zdag)=0$,
\begin{equation}
\Iin_{\rm tot}(\zdag)
=
\rho_{\rm m0}(1+\zdag)^3
+
\frac{1+\zdag}{3}\rhode'(\zdag).
\end{equation}
Thus the total NEC is preserved at the crossing if and only if
\begin{equation}
\rhode'(\zdag)
\ge
-3\rho_{\rm m0}(1+\zdag)^2.
\label{eq:total_NEC_crossing}
\end{equation}
Combining Eqs.~\eqref{eq:acc_at_crossing} and~\eqref{eq:total_NEC_crossing}, transient acceleration at the crossing without total NEC violation occurs precisely in the window
\begin{equation}
-3\rho_{\rm m0}(1+\zdag)^2
\le
\rhode'(\zdag)
<
-\rho_{\rm m0}(1+\zdag)^2.
\label{eq:acc_without_total_NEC_window}
\end{equation}
If the slope is more negative than the lower bound, the crossing lies in a total-NEC-violating, super-accelerating regime in flat GR. If it is greater than or equal to the upper bound, the DE sector may be Raychaudhuri-repulsive but cannot drive acceleration exactly at the crossing.
 
Including radiation is straightforward. With $x_\dagger\equiv1+\zdag$, the acceleration condition at the crossing becomes $\rhode'(\zdag)
<
-\rho_{\rm m0}x_\dagger^2
-2\rho_{\rm r0}x_\dagger^3$,
whereas total NEC preservation requires $\rhode'(\zdag)
\ge
-3\rho_{\rm m0}x_\dagger^2
-4\rho_{\rm r0}x_\dagger^3$.
Thus, with radiation included, acceleration at the crossing without total NEC violation occurs in the exact window
\begin{equation}
-3\rho_{\rm m0}x_\dagger^2
-4\rho_{\rm r0}x_\dagger^3
\le
\rhode'(\zdag)
<
-\rho_{\rm m0}x_\dagger^2
-2\rho_{\rm r0}x_\dagger^3.
\end{equation}
Radiation is negligible for the matter-era examples below, but these expressions make clear that the distinction between sector-level repulsion, cosmic acceleration, and total NEC violation is exact and not tied to the dust-only approximation.

\subsection{How many \texorpdfstring{$q=0$}{q = 0} crossings can occur?}
\label{subsec:counting}

Assuming spatial flatness, $H^2>0$, and neglecting radiation as above, the boundaries between accelerated and decelerated expansion are determined by
\begin{equation}
q(z)=0
\quad\Longleftrightarrow\quad
\Mact_{\rm tot}(z)\equiv\rhom(z)+\Mactde(z)=0,
\label{eq:q_zero_condition}
\end{equation}
as follows from Eq.~\eqref{eq:q_general}. Thus, the number of transitions between deceleration and acceleration is controlled by the number of \emph{sign-changing} zeros of $q(z)$, equivalently of $\Mact_{\rm tot}(z)$. Each such zero is an instantaneous coasting point, $\ddot a=0$. Deep in the post-recombination matter-dominated regime, the histories considered here approach $q=1/2>0$, whereas the sign-switching histories considered in this work all have $q(0)<0$. A tangential zero, at which $q$ vanishes without changing sign, is also an instantaneous coasting point but not an acceleration transition and is therefore not included in this count. Therefore, assuming a finite number of sign-changing zeros, the number of $q=0$ transitions in the post-recombination era must be odd for these histories. We do not, however, elevate $q(0)<0$ to a viability requirement.
For instance, for the CPL history at the joint DESI DR2 BAO+CMB
central values used below~\cite{DESI:2025zgx}, one finds
$q(0)=\tfrac12+\tfrac32 w_0(1-\Omm)\simeq+0.06$, with two
sign-changing zeros at $z\simeq0.04$ and $z\simeq1.01$; hence
$q<0$ only for $0.04<z<1.01$. This is a central-value illustration,
not a posterior-level determination of the sign of $q(0)$; the CPL
history also lies outside the sign-switching class analyzed here.

To make the counting more explicit, suppose that $\rhode(z)$ has $N_{\rm sw}$ smooth sign switches and no additional oscillatory structure beyond the localized features induced by those switches. We assume that each sign switch generates at most one localized negative excursion, or ``impulse'', in $\Mactde(z)$. Let $N_{\rm imp}$ denote the number of such impulses that are sufficiently deep to overcome the attractive matter contribution and open a connected intermediate interval with $q(z)<0$. Then $N_{\rm imp}\le N_{\rm sw}$. Provided these intermediate acceleration windows are mutually disjoint and remain separate from the late-time acceleration phase, each contributes exactly two sign-changing zeros of $q(z)$, one marking the beginning and one the end of the transient acceleration phase. The standard late-time onset contributes one further sign-changing zero. Hence, under these assumptions,
\begin{equation}
N_{q=0}=2N_{\rm imp}+1 \le 2N_{\rm sw}+1.
\label{eq:q_crossing_count}
\end{equation}
If two acceleration windows merge, or if an intermediate window merges with the late-time acceleration phase, the number of distinct sign-changing zeros is smaller. Equation~\eqref{eq:q_crossing_count} should therefore be read as the count for separated windows, and as an upper bound for the corresponding non-oscillatory sign-switching class.

In this work, we consider smooth DE profiles exhibiting a \emph{single impulse}: $\rhode(z)$ crosses zero once at $\zdag$, and the derivative term $(1+z)\rhode'(z)$ generates at most one localized negative excursion in $\Mactde(z)$ around the sign switch, with no additional oscillatory structure. By an impulse we mean a connected component of the non-positive region
$\Mactde(z)\le0$ that contains a nonempty open subinterval on which
$\Mactde<0$. In the single-impulse class this component is a single
interval containing the sign switch. For a simple zero,
$\Mactde(\zdag)<0$, whereas for a higher-order zero
$\Mactde(\zdag)=0$ but $\Mactde<0$ immediately on both sides. Correspondingly, we assume that $\Mact_{\rm tot}(z)$ is continuously differentiable---as it is for the smooth histories considered here---with at most two stationary points in the post-recombination interval. Then $\Mact_{\rm tot}(z)$ can have at most three distinct zeros. Indeed, if it had four distinct zeros $z_1<z_2<z_3<z_4$, Rolle's theorem would imply the existence of three distinct points $c_1\in(z_1,z_2)$, $c_2\in(z_2,z_3)$, and $c_3\in(z_3,z_4)$ such that $\Mact_{\rm tot}'(c_i)=0$, contradicting the assumption of at most two stationary points. Therefore, within this single-impulse class, subject to the stated stationary-point bound and counting only sign-changing zeros, the acceleration history admits only two possibilities: either one $q=0$ transition, corresponding to the standard late-time onset of acceleration, or three $q=0$ transitions, corresponding to one transient intermediate acceleration window together with the standard late-time onset. This one-or-three pattern is thus tied not to any particular analytic ansatz but to the existence of a single sufficiently deep impulse together with the stated bound on stationary points; more than three transitions require additional structure, such as multiple sign switches, oscillatory behavior in $\rhode(z)$, or additional stationary points of $\Mact_{\rm tot}(z)$.

It is useful to distinguish three conceptually different redshifts:
\textbf{(i)} the \emph{density sign switch} redshift $\zdag$, defined by $\rhode(\zdag)=0$;
\textbf{(ii)} the \emph{repulsion-onset} redshift $\zrep>\zdag$ of the effective DE sector on its negative-density branch, defined when it exists by $\Mactde(\zrep)=0$;
and \textbf{(iii)} the redshift(s) $\zacc$ at which the \emph{Universe} transitions between deceleration and acceleration, defined by $q(\zacc)=0$, i.e., by Eq.~\eqref{eq:q_zero_condition} evaluated at $\zacc$---the moment when the repulsive Raychaudhuri contribution of the DE sector balances the attractive matter contribution.

We denote by $\zacc^{\rm late}$ the low-redshift solution corresponding to the standard late-time onset of acceleration:
\begin{equation}
q(\zacc^{\rm late})=0, \quad
q'(\zacc^{\rm late})>0, \quad
q(z)<0\quad \text{for}\quad z<\zacc^{\rm late}.
\label{eq:zacc_late_def}
\end{equation}
If the localized excursion in $\Mactde(z)$ around $z\simeq\zdag$ is sufficiently deep, then $q(z)$ has two additional sign-changing zeros, indicating a transient intermediate acceleration window. We denote these by $\zacc^{\rm int,begin}$ and $\zacc^{\rm int,end}$, with $\zacc^{\rm int,begin}>\zacc^{\rm int,end}$, and define them by
\begin{align}
q(\zacc^{\rm int,begin})&=0, &
q'(\zacc^{\rm int,begin})&>0,
\nonumber\\
q(\zacc^{\rm int,end})&=0, &
q'(\zacc^{\rm int,end})&<0,
\label{eq:zacc_int_defs}
\end{align}
so that
\begin{equation}
q(z)<0 \quad \text{for}\quad \zacc^{\rm int,end}<z<\zacc^{\rm int,begin}.
\label{eq:int_acc_interval}
\end{equation}
Whenever such an intermediate acceleration window exists, its onset can occur only after the DE sector itself has become Raychaudhuri-repulsive. In the single-impulse class, for $z>\zrep$ one has $\Mactde>0$, so the DE sector is still attractive and cannot drive acceleration. Hence
\begin{equation}
\zacc^{\rm int,begin}<\zrep.
\end{equation}
Moreover, whenever the intermediate acceleration condition~\eqref{eq:acc_at_crossing} holds, one has $q(\zdag)<0$, so the intermediate window contains $\zdag$ and straddles the density sign switch. For a distinct intermediate window that straddles the sign switch and remains separated from the late-time acceleration phase, the relevant redshifts obey
\begin{equation}
\zrep>
\zacc^{\rm int,begin}>
\zdag>
\zacc^{\rm int,end}>
\zacc^{\rm late}.
\label{eq:order-z}
\end{equation}
This ordering makes explicit the logical distinction between density sign-switching, sector-level repulsion, and cosmic acceleration. In what follows, we illustrate these general results with a compact analytic example and then show that the same pattern appears in effective sectors inferred from modified gravity backgrounds.

\section{Worked examples}
\label{sec:examples}

We now illustrate the general results of Secs.~\ref{sec:diagnostics} and~\ref{sec:acceleration} with three explicit sign-switching histories: a smooth $\Ls$CDM-type density profile (Sec.~\ref{subsec:lscdm}), an exponential infrared teleparallel $f(T)$ model (Sec.~\ref{subsec:ft}), and the minimal phantom brane (Sec.~\ref{subsec:brane}), using $\Lambda$CDM and CPL as reference histories. For the modified gravity examples, the mapping into an effective DE sector is not unique in general; throughout, we adopt the effective fluid interpretation in which the standard matter and radiation sectors follow their usual background evolution, while the remaining terms are packaged into an effective DE sector $(\rhode,\pde)$. Within this specified GR-like split, the distinction between sector-level repulsion and cosmic acceleration appears in the same ratio-free language emphasized above: the relevant inferred quantities are $\rhode(z)$, $\Iinde(z)$, and $\Mactde(z)$, whereas $\wde(z)$ is a secondary ratio variable and may become singular at a density zero. The key diagnostics of all the examples are collected in Table~\ref{tab:summary}, and the dependence of the extracted redshifts on the adopted split is quantified in Sec.~\ref{subsec:split}. The figures display the interval $0\le z\le7$; statements explicitly identified below as analytic or full-branch results refer to the full physical domain $z>-1$.

\subsection{A sign-switching counterexample: repulsion begins while \texorpdfstring{$\rhode<0$}{rho\_de < 0}}
\label{subsec:lscdm}

We first construct an explicit counterexample to the positive-density proxy: a separately conserved effective DE sector that is already Raychaudhuri-repulsive, $\Mactde<0$, while its density is still negative, a regime that the familiar condition $\wde<-1/3$ misdiagnoses as non-repulsive. Consider the smooth sign-switching density profile~\cite{Akarsu:2022typ,Akarsu:2024qsi,Akarsu:2024eoo,Souza:2024qwd}
\begin{equation}
\rhoLs(z) = \rho_{\Ls0}\,
\frac{\tanh\!\left[\eta(\zdag-z)\right]}{\tanh(\eta\zdag)},
\label{eq:rho_tanh}
\end{equation}
where $\rho_{\Ls0}>0$, the sign switch redshift satisfies $\zdag>0$, and $\eta>0$ controls the rapidity of the transition. By construction, $\rho_{\Ls}(\zdag)=0$, with $\rho_{\Ls}>0$ for $z<\zdag$ and $\rho_{\Ls}<0$ for $z>\zdag$. We refer to the resulting spatially flat matter plus DE cosmology as smooth-$\Ls$CDM in the figures. Closely related smooth sign-switching realizations have recently been explored at the background and perturbation levels and confronted with observational data; see, e.g., Refs.~\cite{Bouhmadi-Lopez:2025ggl,Bouhmadi-Lopez:2025spo,Ibarra-Uriondo:2026zbp,Bouhmadi-Lopez:2026vyc,Akarsu:2025gwi,Akarsu:2025dmj,Akarsu:2025nns,Akarsu:2026lva}. Operationally, we require the profile to be close to its saturated plateaus outside the transition region. For definiteness, we delimit the transition by
$\zdag-2.5\,\eta^{-1}\lesssim z\lesssim\zdag+2.5\,\eta^{-1}$,
so that the full width of the sign flip is
$\Delta z_{\rm tr}\simeq5\eta^{-1}$, corresponding to an order-percent tolerance to saturation,
$|\tanh[\eta(\zdag-z)]|\gtrsim0.987$~\cite{Akarsu:2024qsi}.
When $\eta\zdag\gtrsim3$, the two asymptotic branches approach approximately constant, mirror-symmetric plateaus,
$\rho_{\Ls}(z\gg\zdag)\simeq-\rho_{\Ls0}$ and
$\rho_{\Ls}(z\to-1^+)\simeq+\rho_{\Ls0}$, within the physical domain $z>-1$.\footnote{The exact high-redshift limit of
Eq.~\eqref{eq:rho_tanh} is
$\rho_{\Ls}(z\to\infty)=-\rho_{\Ls0}\coth(\eta\zdag)$; its magnitude
differs from $\rho_{\Ls0}$ by less than $0.5\%$ when
$\eta\zdag\gtrsim3$. The present-day value is
$\rho_{\Ls}(0)=\rho_{\Ls0}$ by construction, whereas the future limit is
$\rho_{\Ls}(z\to-1^+)=
\rho_{\Ls0}\coth(\eta\zdag)\tanh[\eta(1+\zdag)]$
and is within the same tolerance of $\rho_{\Ls0}$.}
Thus, for $\zdag\sim2$ as suggested in
Refs.~\cite{Akarsu:2019hmw,Akarsu:2021fol,Akarsu:2022typ,Akarsu:2023mfb},
values $\eta\gtrsim1.5$ already ensure nearly mirror-symmetric plateau amplitudes. This condition should be distinguished from the rapidity of the transition itself, which is controlled by
$\Delta z_{\rm tr}\simeq5/\eta$: progressively larger values of $\eta$ describe increasingly sharp switches, with the effectively abrupt limit recovered only as $\eta\to\infty$.

For comparison with the ratio language, away from $\rho_{\Ls}=0$ one may write
\begin{equation}
w_{\Ls}(z) = -1-\frac{2}{3}\eta(1+z)\,
\operatorname{csch}\!\left[2\eta(\zdag-z)\right].
\end{equation}
This expression makes the pole at $z=\zdag$ explicit, but the quantities selected by the field equations are regular. Indeed,
\begin{equation}
\rho_{\Ls}'(z) = -\rho_{\Ls0}\eta\,\coth(\eta\zdag)\,
\operatorname{sech}^{2}\!\left[\eta(\zdag-z)\right],
\end{equation}
and therefore
\begin{equation}
    \begin{aligned}
        \Iin_{\Ls}(z) &= -\frac{\rho_{\Ls0}\eta(1+z)}{3}\,
\coth(\eta\zdag)\,
\operatorname{sech}^{2}\!\left[\eta(\zdag-z)\right],\\
\Mact_{\Ls}(z) &= -\rho_{\Ls0}\coth(\eta\zdag)\,
\operatorname{sech}^{2}\!\left[\eta(\zdag-z)\right]\\
&\quad\quad\quad\times\left\{ \eta(1+z)+\sinh\!\left[2\eta(\zdag-z)\right] \right\}.
\label{eq:IM_Ls_tanh}
    \end{aligned}
\end{equation}
Since $\rho_{\Ls}'(z)<0$ and $1+z>0$ throughout $z>-1$,
separate conservation gives $\Iin_{\Ls}<0$ on this entire domain: the sector is phantom-like in the sector-level sense of Sec.~\ref{subsec:proxy}, so the profile cannot be realized by a canonical $X-V$ quintessence field~\cite{Caldwell:2025inn}. It can be reconstructed by a minimally coupled phantom scalar, $-X-V$, although such a restricted scalar realization raises the usual microphysical questions. On the other hand, a closely related quiescent AdS$\to$dS transition has been realized in the fully predictive $\Ls$VCDM framework, based on a type-II minimally modified gravity theory that is free from ghost and other perturbative instabilities~\cite{Akarsu:2024qsi}. We return to the distinction between kinematic regularity, scalar reconstructions, and microphysical stability in the perturbation discussion and in the appendices. A companion multi-probe analysis combining CMB, BAO, Type Ia supernovae, and cosmic-shear data shows that $\Ls$VCDM improves the overall fit and can alleviate both the $H_0$ and $S_8$ tensions~\cite{Akarsu:2024eoo}. The modified gravity curves shown in the figures below and defined in the following sections are, on the other hand, included here only to display the common diagnostic structure in the adopted GR-like split.

\begin{figure*}[t]
\centering
\includegraphics[width=0.92\columnwidth]{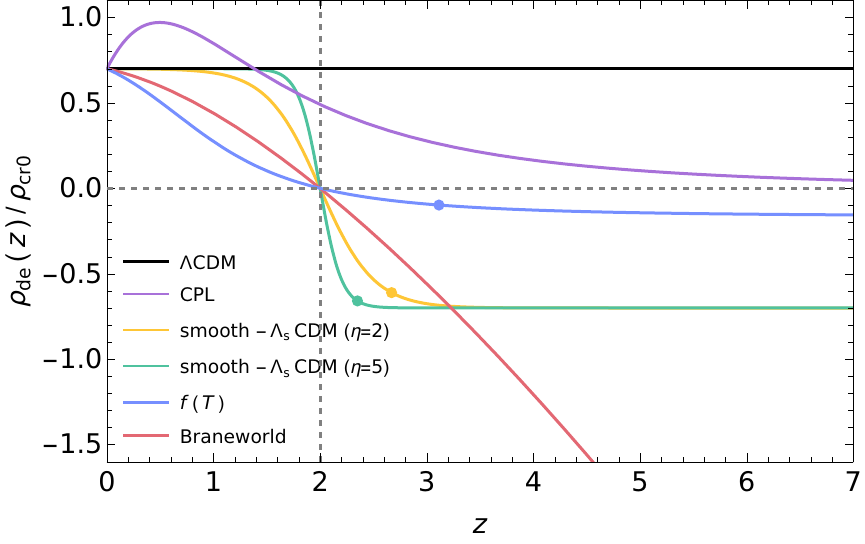}
\includegraphics[width=0.9\columnwidth]{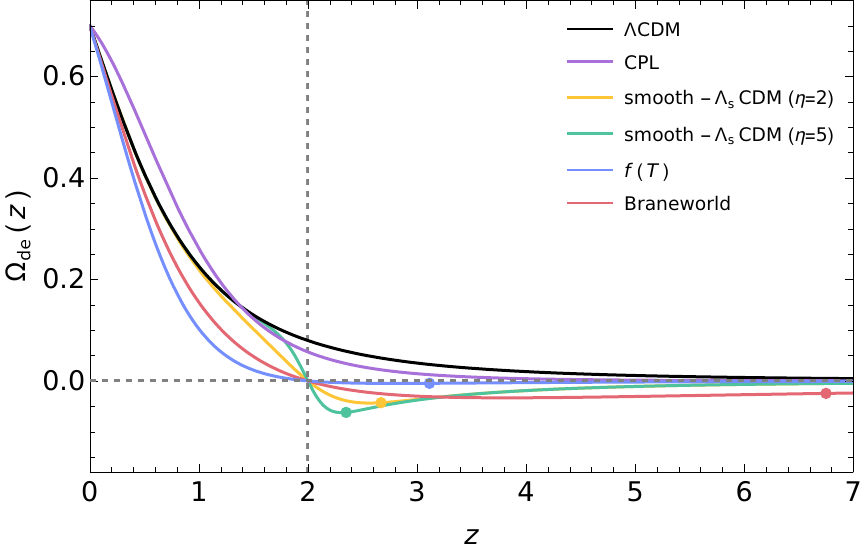}
\includegraphics[width=0.9\columnwidth]{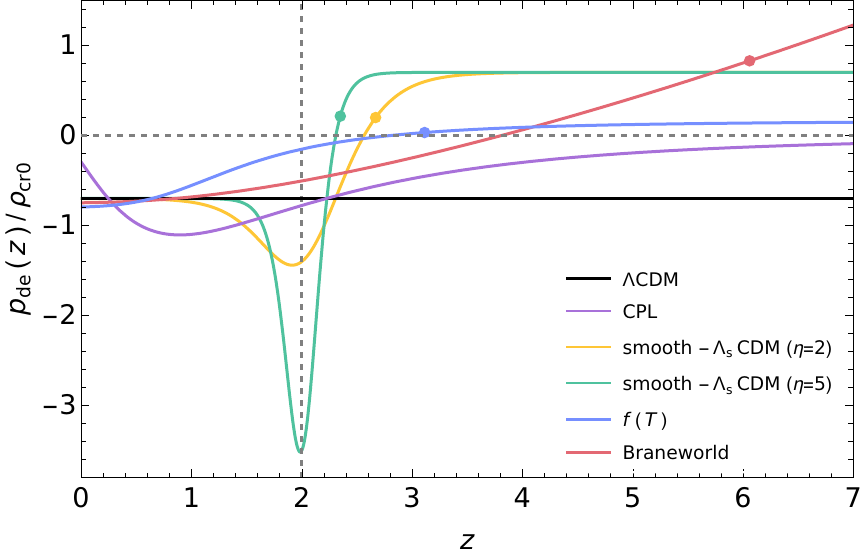}
\includegraphics[width=0.9\columnwidth]{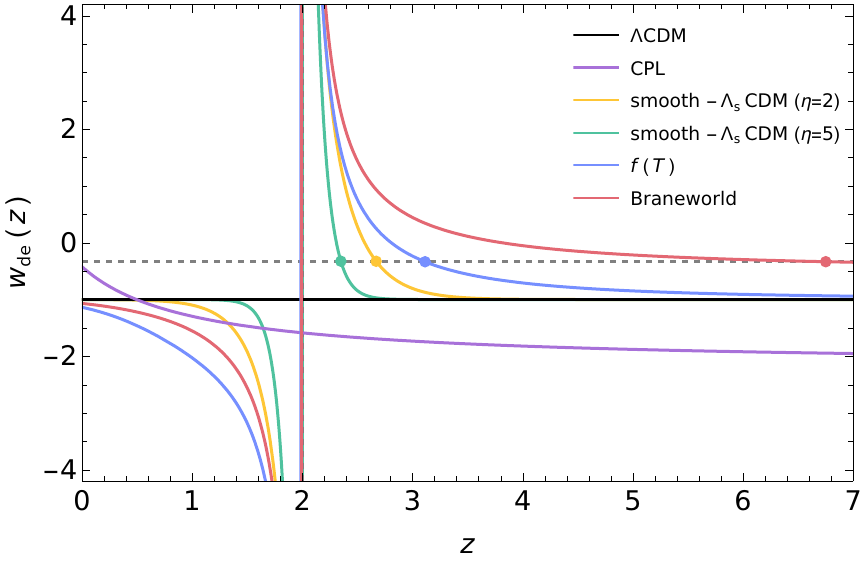}
\caption{\label{fig:rhopw}
Evolution of separately conserved sign-switching DE sectors on a spatially flat matter plus DE background, with radiation neglected. We set $\Omm=0.3$ for all models. Model-specific parameters are $\eta=2,5$ for smooth-$\Ls$CDM, $\beta\simeq-0.977$ (fixed by the condition $\zdag=\zdagval$) for the exponential infrared $f(T)$ model, and $\Oml=0.036$ for the minimal phantom brane. We also show $\Lambda$CDM and CPL for comparison; for CPL, we use the central values $(w_0,w_a)=(-0.42,-1.75)$ reported by the joint DESI DR2 BAO$+$CMB analysis~\cite{DESI:2025zgx}.
The vertical dashed lines mark the illustrative sign-switch redshift $\zdag=\zdagval$, where $\rhode(\zdag)=0$, with $\rhode>0$ for $z<\zdag$ and $\rhode<0$ for $z>\zdag$. Filled circles mark the repulsion-onset redshift $\zrep>\zdag$, defined on the negative-density branch by $\Mactde(\zrep)=0$. \textbf{Top left}: $\rhode(z)/\rhoc$;
\textbf{Top right}: $\Omega_{\rm de}(z)=\rhode(z)/\rho_{\rm cr}(z)$, where $\rho_{\rm cr}(z)=3H^2(z)/(8\pi G)$;
\textbf{Bottom left}: $\pde(z)/\rhoc$ (finite across $\zdag$, exhibiting a localized excursion set by $\eta$ for smooth-$\Ls$CDM);
\textbf{Bottom right}: $\wde(z)=\pde(z)/\rhode(z)$, which is kinematically ill-defined at $\zdag$ even though $T^\mu{}_{\nu\,({\rm de})}$ is finite. Away from $\rhode=0$, the condition $\wde=-1/3$ (horizontal dashed line) is equivalent to $\Mactde=0$, identifying $\zrep$ on the negative-density branch.}
\end{figure*}

\begin{figure*}[t]
\centering
\includegraphics[width=0.95\columnwidth]{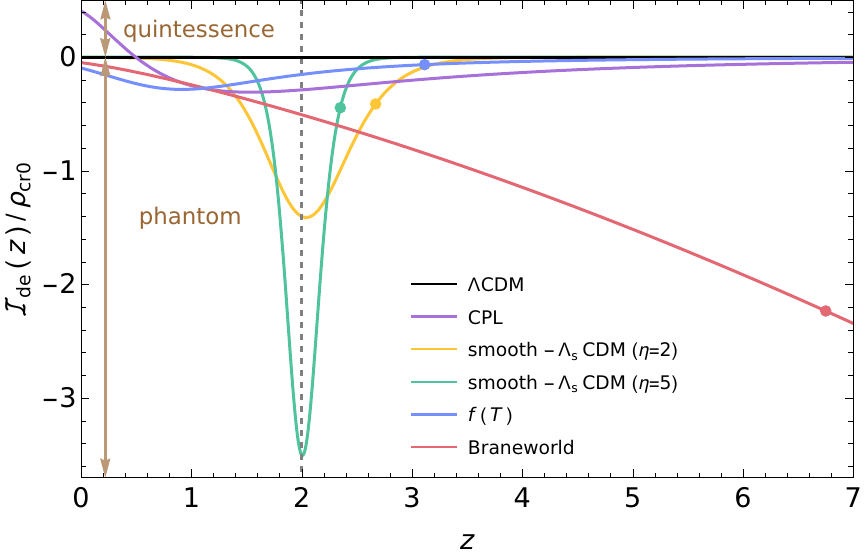}
\includegraphics[width=0.97\columnwidth]{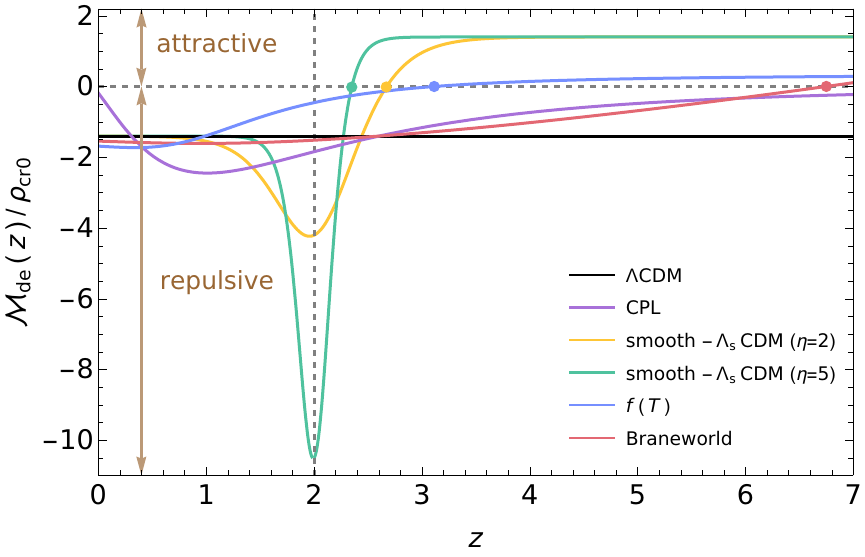}
\includegraphics[width=0.95\columnwidth]{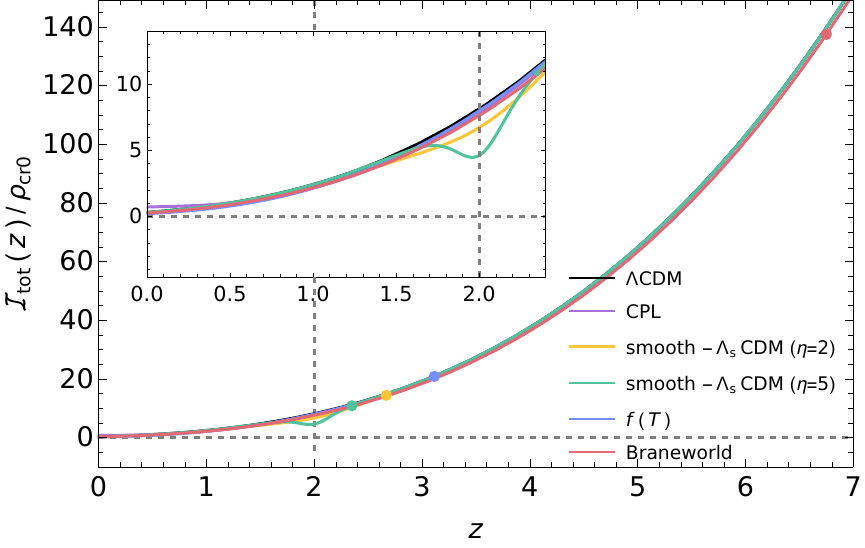}
\includegraphics[width=0.95\columnwidth]{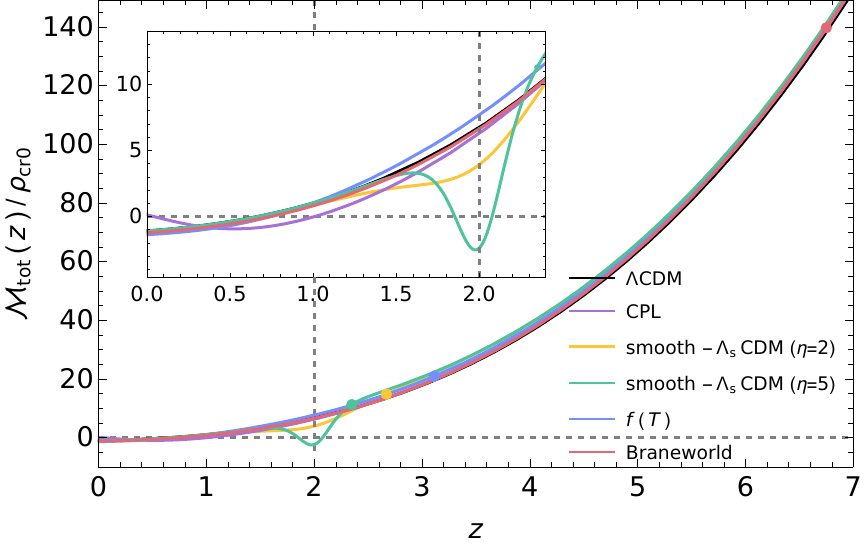}
\caption{\label{fig:qmu}
Inertial and active gravitational mass densities, defined in Eq.~\eqref{eq:I_M_defs} and scaled by $\rhoc$, for the DE sectors and total sources of the same models as in Fig.~\ref{fig:rhopw}. \textbf{Top left}: $\Iinde(z)/\rhoc$, with $\Iinde=0$ corresponding to the NECB; \textbf{Top right}: $\Mactde(z)/\rhoc$; \textbf{Bottom left}: total Hubble-rate source $\Iin_{\rm tot}(z)/\rhoc$; \textbf{Bottom right}: total Raychaudhuri source $\Mact_{\rm tot}(z)/\rhoc$. The vertical dashed lines mark $\zdag$, and filled circles mark $\zrep$ defined by $\Mactde(\zrep)=0$ on the negative-density branch. Over the displayed interval $0\le z\le7$, the total source of each curve satisfies $\Iin_{\rm tot}>0$, implying $\dot H<0$ in the spatially flat setting and avoiding super-acceleration over this interval. The DE sector is Raychaudhuri-repulsive if and only if $\Mactde<0$, while acceleration of the Universe requires $\Mact_{\rm tot}<0$. The top right panel makes explicit a finite interval with $\rhode<0$ but $\Mactde<0$, and shows how increasing $\eta$ in smooth-$\Ls$CDM deepens the localized negative excursion that can trigger intermediate acceleration.}
\end{figure*}

Implementing the (component-wise) repulsion-onset condition $\Mact_{\Ls}(\zrep)=0$ in Eq.~\eqref{eq:IM_Ls_tanh} yields
\begin{equation}
\sinh\!\left(2\eta[\zrep-\zdag]\right)
=
\eta(1+\zrep).
\label{eq:zrep_relation}
\end{equation}
Define $F(z)\equiv\sinh\!\left(2\eta[z-\zdag]\right)-\eta(1+z)$. Then $F(\zdag)=-\eta(1+\zdag)<0$, while $F'(z)=2\eta\cosh\!\left(2\eta[z-\zdag]\right)-\eta
\ge \eta>0$.
Since $F(z)\to+\infty$ as $z\to\infty$, Eq.~\eqref{eq:zrep_relation} has exactly one solution, and that solution satisfies $\zrep>\zdag$. Hence there is a finite interval
\begin{equation}
\rho_{\Ls}(z)<0
\,\,\text{but}\,\,
\Mact_{\Ls}(z)<0
\,\,
\text{for}
\,\,
\zdag<z<\zrep.
\end{equation}
The width of this interval follows implicitly from Eq.~\eqref{eq:zrep_relation}:
\begin{equation}
\begin{aligned}
  \delta z\equiv\zrep-\zdag
&=
\frac{1}{2\eta}
\operatorname{arcsinh}\!\left[\eta(1+\zrep)\right]\\
&\simeq
\frac{1}{2\eta}\ln\!\left[2\eta(1+\zdag)\right],
\end{aligned}
\end{equation}
where the last expression is the rapid-transition estimate. Thus $\delta z=\mathcal O(\eta^{-1}\ln\eta)$; for $\zdag=\zdagval$, one finds $\delta z\simeq0.67$ for $\eta=2$ and $\delta z\simeq0.35$ for $\eta=5$. This is the advertised counterexample: the sector becomes Raychaudhuri-repulsive while its density is still negative. The same possibility is consistent with recent model-agnostic reconstructions, which allow effective DE densities to descend toward zero and, for some dataset combinations, toward a sign change around $z\sim1.5$--$2.5$, together with localized intermediate-redshift structure in the reconstructed deceleration history~\cite{Akarsu:2026pom,Akarsu:2026anp}.

We illustrate these statements using representative spatially flat backgrounds with $\Omm=0.3$, neglecting radiation, and fixing $\zdag=\zdagval$ for the sign-switching models; present-day density parameters are defined by
$\Omega_{i0}\equiv\rho_{i0}/\rhoc$, while
$\Omega_i(z)\equiv\rho_i(z)/\rho_{\rm cr}(z)$ denotes the corresponding
time-dependent fraction. For the smooth-$\Ls$CDM profile we take $\OmLs=0.7$ and show $\eta=2$ and $\eta=5$. The bottom right panel of Fig.~\ref{fig:rhopw} displays the kinematic pole in $w_{\Ls}$ at $\zdag$, while the bottom left panel shows that $\pLs$ remains finite. Thus the singularity is not in the stress--energy tensor, but only in the ratio variable. The top panels of Fig.~\ref{fig:qmu} show the corresponding regular behavior of $\Iin_{\Ls}$ and $\Mact_{\Ls}$, and the top right panel marks the unique $\zrep$ on the negative-density branch. Numerically, one finds $\zrep=2.67$ for $\eta=2$ and $\zrep=2.35$ for $\eta=5$, both satisfying $\zrep>\zdag=\zdagval$. The two sides of the pole realize the assignment of Eq.~\eqref{eq:w_pole_sides}: $w_{\Ls}>-1$ on the negative-density side ($z>\zdag$) and $w_{\Ls}<-1$ on the positive-density side ($z<\zdag$), while the analytic result above gives $\Iin_{\Ls}<0$, i.e., phantom-like behavior, for every finite $z>-1$; the top left panel of Fig.~\ref{fig:qmu} illustrates this behavior over $0\le z\le7$. The $(\wde,z)$ plane thus suggests a quintessence-to-phantom transition that never occurs: once $\rhode$ changes sign, the regular branch-independent separator of the sector-level NEC character, within the adopted split, is the NECB, $\Iinde=0$, not the PDL, $\wde=-1$~\cite{Gokcen:2026pkq}. A distinct composite-sector mechanism is provided by combinations of standard and negative-density quintessence, for which the effective
$\wde$ can cross the PDL even though the constituent EoS parameters do
not~\cite{Gomez-Valent:2025mfl}. The CPL reference history makes the contrast explicit: for the DESI DR2 central values its density remains positive for every finite $z>-1$, and its unique NECB crossing at $z\simeq0.50$---where $\Iinde$ changes sign smoothly on the positive-density branch (top left panel of Fig.~\ref{fig:qmu})---is a bona fide PDL crossing, with $\wde$ passing regularly through $-1$ at the same redshift. Thus $\Iin_{\rm CPL}>0$ below the NECB crossing redshift and
$\Iin_{\rm CPL}<0$ above it, in agreement with
Table~\ref{tab:summary}. The sign-switching histories instead never attain $\wde=-1$ near $\zdag$: there, the two sides of the PDL are connected only through the kinematic pole, while the NEC character does not change at all.

\begin{figure*}[t]
\centering
\includegraphics[width=0.95\columnwidth]{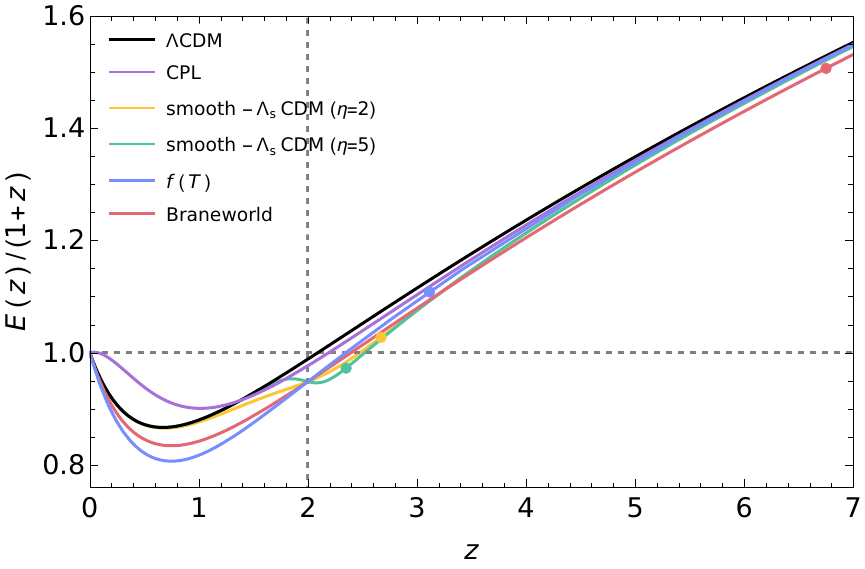}
\includegraphics[width=0.96\columnwidth]{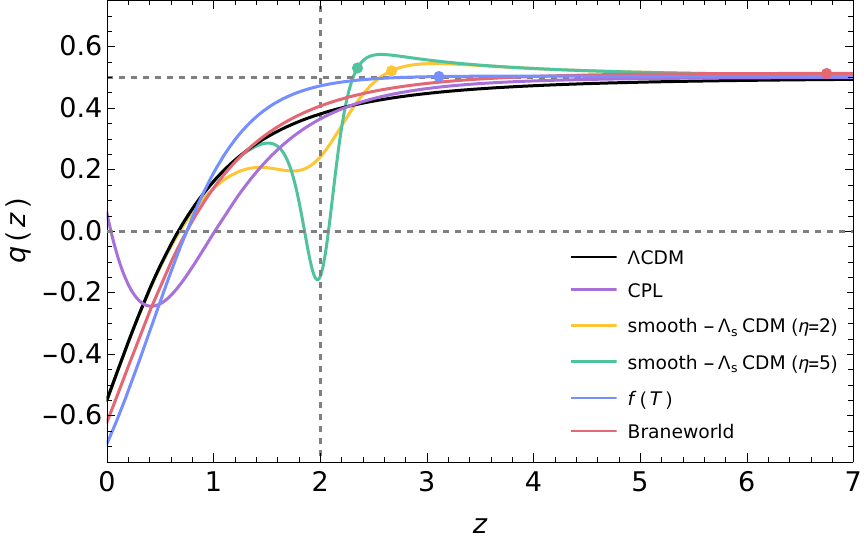}
\caption{\label{fig:comv-H-q}
\textbf{Left}: normalized comoving Hubble parameter $E(z)/(1+z)$, with $E(z)=H(z)/H_0$; \textbf{Right}: deceleration parameter $q(z)=-\ddot a/(aH^2)$. Acceleration corresponds to $q<0$, equivalently $\Mact_{\rm tot}<0$ in the bottom right panel of Fig.~\ref{fig:qmu}. For the parameters considered, the intermediate acceleration condition~\eqref{eq:acc_at_crossing} is satisfied in smooth-$\Ls$CDM for $\eta=5$, but not for $\eta=2$, producing a transient acceleration phase around $z\simeq\zdag$ only in the sharper transition.}
\end{figure*}

The bottom left panel of Fig.~\ref{fig:qmu} shows that the total NEC is
satisfied over the displayed interval $0\le z\le7$,
$\Iin_{\rm tot}>0$. This is important because the smooth-$\Ls$CDM
sector is sector-level phantom-like, $\Iin_{\Ls}<0$, throughout the
full physical domain $z>-1$, whereas the total source remains NEC-respecting over the displayed interval. This displayed-range statement does not extend to the asymptotic
future: for both illustrative tanh histories,
$\Iin_{\rm tot}\to0^{-}$ as $z\to-1^{+}$, corresponding to an
extremely late super-accelerating approach to de Sitter. For $\eta=5$,
a transient dip in $\Iin_{\rm tot}$ is nonetheless visible near
$\zdag$, signaling that still sharper transitions would drive
$\Iin_{\rm tot}<0$ near the sign switch; at the crossing this is
quantified by Eq.~\eqref{eq:eta_window_Ls} below, and a fuller
parameter-space analysis lies beyond the scope of the present work.
For the tanh profile, the intermediate acceleration condition~\eqref{eq:acc_at_crossing} becomes
\begin{equation}
\OmLs\,\frac{\eta}{\tanh(\eta\zdag)}
>
\Omm(1+\zdag)^2.
\label{eq:criterion_transient_Ls}
\end{equation}
Combining Eq.~\eqref{eq:criterion_transient_Ls} with the total-NEC bound~\eqref{eq:total_NEC_crossing}, the matter-era window~\eqref{eq:acc_without_total_NEC_window} becomes
\begin{equation}
\Omm(1+\zdag)^2
<
\OmLs\,\frac{\eta}{\tanh(\eta\zdag)}
\le
3\Omm(1+\zdag)^2.
\label{eq:eta_window_Ls}
\end{equation}
For the adopted parameters, using $\tanh(\eta\zdag)\simeq1$, this gives approximately $3.9\lesssim\eta\lesssim11.6$. Thus $\eta=5$ lies inside the window: the transition is sharp enough to generate acceleration at the crossing, but not so sharp as to violate the total NEC there. By contrast, $\eta=2$ fails the acceleration condition. 

For the tanh profile, a numerical scan of the full branch $z>-1$
gives the low-redshift roots $\zacc^{\rm late}=0.68$ for $\eta=2$ and $\zacc^{\rm late}=0.67$ for $\eta=5$, with no additional roots in the future interval $-1<z<0$. Since $q\to-1$ as $z\to-1^{+}$, the corresponding late-time accelerating intervals are $-1<z<0.68$ and $-1<z<0.67$, respectively. For $\eta=5$, Eq.~\eqref{eq:criterion_transient_Ls} is satisfied and $q(z)$ develops two additional roots,
\[
\zacc^{\rm int,begin}=2.08,
\quad
\zacc^{\rm int,end}=1.85,
\]
straddling $\zdag=\zdagval$, with the transient dip centered near $z=1.99$ (right panel of Fig.~\ref{fig:comv-H-q} and bottom right panel of Fig.~\ref{fig:qmu}). The ordering of Eq.~\eqref{eq:order-z} is therefore realized explicitly for the $\eta=5$ profile:
\[
2.35>2.08>\zdagval>1.85>0.67.
\]
This model cleanly separates density sign-switching, sector-level repulsion, and cosmic acceleration.


\subsection{Exponential infrared teleparallel model}
\label{subsec:ft}

We next consider the exponential infrared teleparallel model $f(T)=T\,e^{\beta T_0/T}$~\cite{Awad:2017yod,Hashim:2020sez,Hashim:2021pkq,Akarsu:2024nas}, with $T=6H^2$ and $T_0=6H_0^2$ in flat FLRW, focusing on the negative-$\beta$ branch relevant for the sign-changing effective density histories considered here. We include a cosmological constant through $f(T)\to f(T)+2\Lambda$, so that the $\beta=0$ limit recovers $\Lambda$CDM within the same flat-background parametrization~\cite{Akarsu:2024nas}.\footnote{The negative-$\beta$ branch is the branch naturally associated with the sign-changing effective DE histories emphasized in Ref.~\cite{Akarsu:2024nas}. With $\Lambda=0$, the density sign switch occurs only on this branch, while after the extension $f(T)\to f(T)+2\Lambda$ the $\beta>0$ branch can also realize sign-changing effective densities, although the negative branch does so more readily (see Secs.~IV and~V of Ref.~\cite{Akarsu:2024nas}). The same work also discusses the ghost-avoidance condition on the negative branch and argues that local gravity constraints can be evaded through an effective chameleon-like suppression in high-torsion environments. These issues concern the underlying modified gravity theory and are logically distinct from the GR-like effective fluid diagnostics used here.}
Within the standard GR-like split adopted throughout, a convenient parametric representation of the inferred effective DE density in terms of $E\equiv H/H_0$ is
\begin{equation}
\rho\tp(E) = \rhoc\Big[
E^2-\left(E^2-2\beta\right)e^{\beta/E^2}
+\OmL
\Big],
\label{eq:rhoT_LL}
\end{equation}
with
\begin{equation}
1+z(E) = \left[\frac{\left(E^2-2\beta\right)e^{\beta/E^2}-\OmL}{\Omm}\right]^{1/3},
\end{equation}
where $\Omm=(1-2\beta)\,e^{\beta}-\OmL$ follows from $z(E{=}1)=0$. In
this parametrization, the chain rule gives the closed form
\begin{equation}
\rho\tp'(z)
=
3\,\Omm(1+z)^{2}\,\rhoc\,
\frac{E^{4}-\mathcal{A}(E)}{\mathcal{A}(E)},
\label{eq:rhotp_prime}
\end{equation}
where $\mathcal{A}(E)\equiv\left(E^{4}-\beta E^{2}+2\beta^{2}\right)e^{\beta/E^{2}}$.
On the negative-$\beta$ branch, $\mathcal{A}(E)>0$, and
${\rm d}z/{\rm d}E\propto\mathcal{A}(E)/[(1+z)^{2}E^{3}]>0$, so $E$ and
$z$ are in monotonic correspondence; the sign of $\rho\tp'(z)$ is
therefore that of $E^{4}-\mathcal{A}(E)$. The ratio-free diagnostics
$\Iin\tp$ and $\Mact\tp$ then follow from
Eqs.~\eqref{eq:Iin_de} and~\eqref{eq:Mact_de}. For the illustrative history used below, $\beta\simeq-0.977$, the sign of $\rho\tp'$ can be established throughout the full physical domain. Writing $x\equiv-\beta/E^{2}>0$, one has $\frac{\mathcal{A}(E)}{E^{4}}=(1+x+2x^{2})e^{-x}$. Along this history the branch approaches $E_{\rm dS}\simeq0.894$ as $z\to-1^{+}$, where $(E_{\rm dS}^{2}-2\beta)e^{\beta/E_{\rm dS}^{2}}=\OmL$. Defining $x_{\rm dS}\equiv-\beta/E_{\rm dS}^{2}\simeq1.222$, monotonicity of $E(z)$ gives $0<x<x_{\rm dS}<3/2$ for every finite $z>-1$. Since $\frac{{\rm d}}{{\rm d}x}\left[(1+x+2x^{2})e^{-x}\right]=x(3-2x)e^{-x}>0$ on this interval and the function tends to unity as $x\to0$, it follows that $E^{4}<\mathcal{A}(E)$. Hence $\rho\tp'(z)<0$ and $\Iin\tp<0$ for all $z>-1$. In the adopted GR-like split, the inferred effective DE sector is therefore sector-level phantom-like throughout the physical domain; as always (Sec.~\ref{subsec:proxy}), this reflects the modified gravitational sector recast as an effective source, not a fundamental phantom field. At a density zero, Eq.~\eqref{eq:IM_crossing} gives
\begin{equation}
\Mact\tp(\zdag)=(1+\zdag)\rho\tp'(\zdag)<0,
\end{equation}
so the inferred sector is already Raychaudhuri-repulsive at the sign switch, as in the smooth-$\Ls$CDM profile~\eqref{eq:rho_tanh}. This realizes, within modified gravity, the same distinction between a density zero and the NECB that has been emphasized for general sign-changing effective histories~\cite{Gokcen:2026pkq,Adil:2026kfn}.
 
In the high-redshift limit, $E\to\infty$, Eq.~\eqref{eq:rhoT_LL} gives
\begin{equation}
\rho\tp\simeq \rhoc(\beta+\OmL),
\quad
\Mact\tp\simeq -2\rhoc(\beta+\OmL).
\end{equation}
High-redshift negativity of the inferred effective DE density requires $\beta+\OmL<0$, i.e., $\OmL<-\beta$, while spatial flatness fixes the present value to $\rho\tp(0)=\rhoc(1-\Omm)>0$; by continuity, the two together guarantee at least one density sign switch. The condition $\OmL<-\beta$ also makes the inferred effective DE sector asymptotically attractive in the matter-dominated regime. Since $\Mact\tp<0$ at the density zero but $\Mact\tp>0$ at sufficiently high redshift, continuity implies at least one repulsion-onset redshift $\zrep>\zdag$, defined by $\Mact\tp(\zrep)=0$. In this modified gravity example as well, the effective sector therefore becomes Raychaudhuri-repulsive before the density sign switch, while $\rho\tp<0$, exactly as anticipated by the general argument above.
 
For the illustrative choice $\Omm=0.3$, the condition $\rho\tp<0$ in the high-redshift past gives approximately $\beta\lesssim-0.85$ on the negative-$\beta$ branch. In the figures we fix the density sign switch at $\zdag=\zdagval$, which for $\Omm=0.3$ corresponds to $\beta\simeq-0.977$ and $\OmL\simeq0.81$. For these parameters, a numerical scan in $x\in(0,x_{\rm dS})$, which parametrizes the full physical branch $z>-1$, finds exactly one zero of $\Mact\tp$, at $\zrep=3.11$, and exactly one zero of $q$, at $\zacc^{\rm late}=0.75$. Since $q\to-1$ as $z\to-1^{+}$, the complete accelerating interval is $-1<z<0.75$. The corresponding portions of these histories are shown in Figs.~\ref{fig:qmu} and~\ref{fig:comv-H-q}. Hence, unlike the sharp $\eta=5$ realization of the smooth-$\Ls$CDM profile, this model does not open an intermediate acceleration window. In particular, the intermediate acceleration condition~\eqref{eq:acc_at_crossing} is not satisfied: for the parameters under consideration, $\rho\tp(z)$ does not exhibit a rapid transition, as seen in the top left panel of Fig.~\ref{fig:rhopw}, so the negative excursion in $\Mact\tp$ never becomes deep enough to overcome the matter term.
 
We note that a very recent analysis including linear perturbations~\cite{Hashim:2026yoy} decisively disfavors the sign-changing branch of the \emph{minimal} ($\Lambda=0$) exponential infrared model, in which $\beta$ is fixed by $\Omm$ and the background is therefore maximally rigid. The $\Lambda$-extended branch adopted here~\cite{Akarsu:2024nas} carries an additional free parameter and is not automatically subject to the same minimal-model overconstraint; in any case, our parameter choices are illustrative, and the example serves to exhibit the background-level diagnostics rather than to advocate a particular realization. The result nonetheless underlines the distinction, emphasized below, between kinematic regularity and full observational viability.

\subsection{Braneworld cosmology: Minimal phantom brane}
\label{subsec:brane}

As a second modified gravity example, we consider the spatially flat
minimal phantom brane, defined as the normal branch of the
induced-gravity (DGP) braneworld with vanishing bulk cosmological
constant and background dark radiation term, so that the background
bulk spacetime is flat~\cite{Bag:2021cqm}. For a review of the
associated phenomenology, see Ref.~\cite{Sahni:2005pf}. Neglecting
ordinary radiation on the brane, its expansion history can be written
as~\cite{Sahni:2002dx,Sahni:2000ubn,Bag:2018jle}
\begin{equation}
\begin{aligned}
    E(z)&=\sqrt{\Sigma(z)}-\sqrt{\Oml},\\
\Sigma(z)&\equiv\Omm(1+z)^3+\OmL+\Oml,
\end{aligned}
\label{eq:mpb_E}
\end{equation}
with $E(z)=H(z)/H_0$ as before. Here $\OmL$ denotes the
cosmological constant contribution on the brane, equivalently the
brane tension contribution, and not a bulk cosmological constant. The normalization $E(0)=1$ gives the usual minimal phantom brane constraint
\begin{equation}
\OmL=1-\Omm+2\sqrt{\Oml},
\label{eq:mpb_constraint}
\end{equation}
so that, once $\Omm$ and $\Oml$ are specified, the background has no additional free normalization parameter.
 
Within the standard GR-like split used throughout, the inferred effective DE density is obtained from
$E^2=\Omm(1+z)^3+\rho\bw/\rhoc$, namely,
\begin{equation}
\rho\bw(z)
=
\rhoc\left[
\OmL+2\Oml
-2\sqrt{\Oml}\sqrt{\Sigma(z)}
\right].
\label{eq:rho_bw}
\end{equation}
Since $\Sigma'(z)=3\Omm(1+z)^2>0$, one finds
\begin{equation}
\rho\bw'(z)
=
-\rhoc\sqrt{\Oml}\,
\frac{\Sigma'(z)}{\sqrt{\Sigma(z)}}<0.
\label{eq:rho_bw_prime}
\end{equation}
Therefore, within the adopted GR-like split, the inferred effective DE sector is sector-level phantom-like throughout $z>-1$, since $\Iin\bw=(1+z)\rho\bw'/3<0$ on this domain.
 
At a density zero, Eq.~\eqref{eq:IM_crossing} gives
\begin{equation}
\Mact\bw(\zdag)=(1+\zdag)\rho\bw'(\zdag)<0,
\end{equation}
so the inferred brane sector is already Raychaudhuri-repulsive at the density sign switch. At high redshift,
\begin{equation}
\begin{aligned}
  \rho\bw(z)\simeq
-2\rhoc\sqrt{\Oml\Omm}\,(1+z)^{3/2},\\
\Mact\bw(z)\simeq
\rhoc\sqrt{\Oml\Omm}\,(1+z)^{3/2}>0.  
\end{aligned}
\end{equation}
Thus, the sector is asymptotically attractive in the matter-dominated regime, while it is repulsive at the density zero, so by continuity a repulsion-onset redshift $\zrep>\zdag$ with $\Mact\bw(\zrep)=0$ exists. Here it can be located exactly: using $(1+z)\Sigma'(z)=3\left[\Sigma(z)-\OmL-\Oml\right]$, the condition $\Mact\bw=0$ reduces to a quadratic in $s\equiv\sqrt{\Sigma}$,
\begin{equation}
\sqrt{\Oml}\,s^{2}
-2\left(\OmL+2\Oml\right)s
+3\sqrt{\Oml}\left(\OmL+\Oml\right)=0.
\label{eq:mpb_zrep_quadratic}
\end{equation}
For $\Omega_{\Lambda0}>0$ and $\Omega_{\ell0}>0$, the left-hand side is negative at $s=\sqrt{\OmL+\Oml}$, i.e., at $z=-1$ (its value there is $-2\sqrt{\OmL+\Oml}\,[\sqrt{\OmL+\Oml}-\sqrt{\Oml}]^{2}<0$), so exactly one root lies in the physical domain: $\zrep$ is unique and follows from the larger root through $\Sigma(\zrep)=s_{+}^{2}$. Hence the minimal phantom brane realizes the same qualitative pattern identified above, with a finite interval on the negative-density branch in which $\rho\bw<0$ and $\Mact\bw<0$. At the phenomenological level, this is analogous to reconstructed histories containing a negative effective density phase and localized intermediate-redshift dynamics~\cite{Akarsu:2026anp}; the braneworld example makes the corresponding negative-density repulsive interval explicit in the Raychaudhuri language used here. The phantom brane has also been confronted directly with observations in the context of the Hubble tension, where its effective phantom-like behavior accommodates higher values of $H_0$ while respecting CMB constraints~\cite{Bag:2021cqm}.
 
For the illustrative curves (red in the figures) we use $\Omm=0.3$ and choose $\Oml=0.036$, which gives $\OmL=1.08$ through Eq.~\eqref{eq:mpb_constraint} and places the density sign switch at $\zdag=\zdagval$. Equivalently, the density-zero condition following from Eq.~\eqref{eq:rho_bw} is
\begin{equation}
\Omm(1+\zdag)^3
=
\frac{\OmL^2}{4\Oml}.
\label{eq:mpb_zdag_condition}
\end{equation}
For these parameters, the larger root of Eq.~\eqref{eq:mpb_zrep_quadratic} gives $\zrep=6.75$, so the interval between repulsion onset and the density sign switch is $\delta z=\zrep-\zdag\simeq4.75$, considerably broader than in the smooth-$\Ls$CDM and teleparallel examples shown in the figures.
 
The model nevertheless produces only the standard late-time acceleration. The bottom right panel of Fig.~\ref{fig:qmu} and the right panel of Fig.~\ref{fig:comv-H-q} show that $q(z)$, equivalently $\Mact_{\rm tot}(z)$, crosses zero only once, at $\zacc^{\rm late}=0.75$. More strongly, this one-crossing behavior can be established analytically and is not merely a property of the illustrative parameter choice. Define
\begin{equation}
y\equiv\Omm(1+z)^3,
\quad
C\equiv\OmL+\Oml,
\quad
r\equiv\sqrt{\Oml},
\end{equation}
so that $\Sigma=y+C$ and Eq.~\eqref{eq:mpb_E} becomes
$E=\sqrt{y+C}-r$. Since
$q=-1+(1+z)E'/E$ and
$(1+z)y'=3y$, one obtains
\begin{equation}
q(y)
=
-1+
\frac{3y}
{2\sqrt{y+C}\left(\sqrt{y+C}-r\right)}.
\label{eq:mpb_q_y}
\end{equation}
The condition $q=0$ is therefore equivalent to
\begin{equation}
F(y)\equiv
y+2r\sqrt{y+C}-2C=0.
\label{eq:mpb_q_root}
\end{equation}
For $\OmL>0$ and $\Oml>0$, one has $C>r^2$ and hence
\begin{equation}
F(0)=2\sqrt{C}\left(r-\sqrt{C}\right)<0,
\quad
F'(y)=1+\frac{r}{\sqrt{y+C}}>0.
\end{equation}
Moreover, $F(y)\to+\infty$ as $y\to\infty$. Thus $F(y)$ has exactly one root on the physical domain $y>0$, and consequently the minimal phantom brane background admits exactly one $q=0$ crossing. Since $q\to1/2$ as $z\to\infty$ and $q\to-1$ as $z\to-1^+$, this unique root separates the matter-era decelerating phase from the single late-time accelerating phase. For the parameters used in the figures, it occurs at $\zacc^{\rm late}=0.75$; hence the complete accelerating interval is $-1<z<0.75$.

Although this uniqueness result already excludes a distinct intermediate acceleration window, it is instructive to evaluate the acceleration condition specifically at the density sign switch. At $\rhode=\rho\bw=0$, Eq.~\eqref{eq:rho_bw_prime} gives
\begin{equation}
\rho\bw'(\zdag)
=
-\rhoc\,
\frac{6\Omm\Oml(1+\zdag)^2}{\OmL+2\Oml}.
\end{equation}
Equation~\eqref{eq:acc_at_crossing} then shows that acceleration at the sign switch would require
\begin{equation}
\frac{6\Oml}{\OmL+2\Oml}>1,
\label{eq:mpb_acc_condition}
\end{equation}
independently of $\zdag$ and $\Omm$, or equivalently $\OmL<4\Oml$. Together with the normalization constraint~\eqref{eq:mpb_constraint} and $\Omm\simeq0.3$, this would require $\Oml\gtrsim0.54$, far outside the phenomenologically relevant regime considered in observational analyses~\cite{Bag:2021cqm}. For $\Oml=0.036$ and $\OmL=1.08$, the left-hand side of Eq.~\eqref{eq:mpb_acc_condition} is approximately $0.19$, so the Universe remains decelerating at the density sign switch. Even if Eq.~\eqref{eq:mpb_acc_condition} were satisfied elsewhere in parameter space, the uniqueness result above shows that the sign switch would then lie within the model's single connected acceleration epoch rather than generate an additional transient episode. The minimal phantom brane consequently provides a clean example in which the inferred sector becomes Raychaudhuri-repulsive while $\rho\bw<0$, but cannot produce a separate intermediate acceleration window.

\subsection{Comparative summary and split dependence}
\label{subsec:split}

\begin{table*}[ht!]
\caption{\label{tab:summary}%
Characteristic redshifts, sector-level NEC character, and accelerating
intervals of the six background histories shown in
Figs.~\ref{fig:rhopw}--\ref{fig:comv-H-q}. All histories are spatially
flat with $\Omm=0.3$ and radiation neglected; the sign-switching
histories fix $\zdag=\zdagval$, with $\beta\simeq-0.977$ for the
$f(T)$ history. Although the figures display
$0\le z\le7$, the last two columns describe the full physical domain
$z>-1$ for the adopted parameter choices; numerical boundaries are
rounded to two decimal places. For CPL we use the joint DESI DR2
BAO+CMB central values $(w_0,w_a)=(-0.42,-1.75)$~\cite{DESI:2025zgx}.
The CPL accelerating interval refers to these central values and is not
a posterior-level statement about the sign of $q(0)$. For the reference
histories, dashes in the $\zdag$ and $\zrep$ columns indicate that these
sign-switching diagnostics are not applicable.}
\begingroup
\small
\setlength{\tabcolsep}{3.5pt}
\renewcommand{\arraystretch}{1.12}
\begin{ruledtabular}
\begin{tabular}{lcccc}
Model
& $\zdag$
& $\zrep$
& \shortstack[c]{Sector-level NEC character}
& \shortstack[c]{Accelerating interval(s) ($q<0$)} \\
\hline
$\Lambda$CDM
& ---
& ---
& $\Iin_\Lambda\equiv0$ for all $z>-1$
& $-1<z<0.67$ \\

CPL (DESI DR2)
& ---
& ---
& \begin{tabular}[c]{@{}c@{}}
regular NECB crossing at $z\simeq0.50$\\
$\Iin_{\rm CPL}>0$ below; $\Iin_{\rm CPL}<0$ above
\end{tabular}
& $0.04<z<1.01$ \\

Smooth-$\Ls$CDM ($\eta=2$)
& $2$
& $2.67$
& $\Iin_{\Ls}<0$ for all $z>-1$
& $-1<z<0.68$ \\

Smooth-$\Ls$CDM ($\eta=5$)
& $2$
& $2.35$
& $\Iin_{\Ls}<0$ for all $z>-1$
& $-1<z<0.67$;\quad $1.85<z<2.08$ \\

Exponential infrared $f(T)$
& $2$
& $3.11$
& $\Iin\tp<0$ for all $z>-1$
& $-1<z<0.75$ \\

Minimal phantom brane
& $2$
& $6.75$
& $\Iin\bw<0$ for all $z>-1$
& $-1<z<0.75$ \\
\end{tabular}
\end{ruledtabular}
\endgroup
\end{table*}

Table~\ref{tab:summary} collects the characteristic redshifts,
sector-level NEC character, and accelerating intervals of the six
background histories. Although
Figs.~\ref{fig:rhopw}--\ref{fig:comv-H-q} display $0\le z\le7$, the
last two columns of the table refer to the full physical domain $z>-1$
for the adopted parameter choices. The illustrative sign-switching
histories share the qualitative pattern established in
Secs.~\ref{sec:diagnostics} and~\ref{sec:acceleration}---a density zero
at $\zdag$, phantom-like sector-level NEC character throughout
$z>-1$, and a repulsion onset at $\zrep>\zdag$---while realizing it
with widely different separations,
$\zrep-\zdag\simeq0.35$--$4.75$. For every illustrative sign-switching background history, $\Iinde\to0^{-}$ as $z\to-1^{+}$. Only the sufficiently sharp $\eta=5$ profile opens an additional intermediate acceleration interval.

The two reference histories provide complementary controls:
$\Lambda$CDM lies identically on the NECB,
$\Iin_\Lambda\equiv0$, and accelerates for $-1<z<0.67$. CPL with the
joint DESI DR2 BAO+CMB central values~\cite{DESI:2025zgx} crosses the
NECB regularly at $z\simeq0.50$, with $\Iin_{\rm CPL}>0$ below and
$\Iin_{\rm CPL}<0$ above the crossing redshift, while its total background
accelerates only for $0.04<z<1.01$. Because neither reference history
has a density sign switch, the negative-branch repulsion-onset
diagnostic $\zrep$ is not assigned to them in
Table~\ref{tab:summary}.

Two caveats govern the use of such numbers in inference. First, all sector-level quantities inherit the split dependence emphasized in Sec.~\ref{subsec:proxy} (see also Ref.~\cite{Kunz:2007rk}), and this dependence is quantitatively significant. At fixed expansion history, a shift $\Delta\Omm$ in the assumed present-day matter density shifts the inferred DE density by $\Delta\rhode(z)=-\Delta\Omm\,\rhoc\,(1+z)^3$, and hence moves a simple crossing by
\begin{equation}
\Delta\zdag\simeq\frac{\rhoc\,(1+\zdag)^{3}}{\rhode'(\zdag)}\,\Delta\Omm,
\label{eq:split_sensitivity}
\end{equation}
i.e., toward lower redshift for $\Delta\Omm>0$, since $\rhode'(\zdag)<0$. For the tanh profile~\eqref{eq:rho_tanh} this gives $|\Delta\zdag|\simeq(1+\zdag)^3\tanh(\eta\zdag)\,\Delta\Omm/(\OmLs\,\eta)$, i.e., $|\Delta\zdag|\simeq0.19$ for $\eta=2$ and $\simeq0.08$ for $\eta=5$ per $\Delta\Omm=0.01$; the repulsion-onset redshift $\zrep$ is likewise split dependent. Reported values of $\zdag$ and $\zrep$ should therefore be quoted jointly with the matter-density posterior of the analysis that produced them. Second, by contrast, the total-source combinations are kinematic within the stated framework: for a spatially flat GR background,
\begin{equation}
\Mact_{\rm tot}=2\,q\,\rho_{\rm cr},
\quad
\Iin_{\rm tot}=\frac{(1+z)\,H H'}{4\pi G},
\label{eq:total_kinematic}
\end{equation}
with $\rho_{\rm cr}\equiv3H^2/(8\pi G)$, are functionals of the expansion history alone, so reconstructions of $H(z)$ determine them without reference to the split. The sector-level diagnostics $(\rhode,\Iinde,\Mactde)$ then follow by subtracting the assumed matter and radiation contributions, which is where the dependence on $\Omm$ (and on early-time physics) enters. When model independence is at a premium, $(\Iin_{\rm tot},\Mact_{\rm tot})$---equivalently $q(z)$ and $H'(z)$---are the primary quantities, with the sector-level statements conditional on the adopted split.

\section{Perturbations: kinematic regularity versus microphysical stability}
\label{sec:perturbations}

The discussion so far has been entirely at the background level. At the perturbative level, it is useful to distinguish \emph{kinematic regularity} from \emph{microphysical stability}. Conditions such as absence of ghosts, gradient stability, closure relations for pressure perturbations, and the presence or absence of anisotropic stress are properties of a specific realization of the effective DE sector: a thermodynamic fluid, a canonical or phantom scalar, a scalar--tensor theory, an effective source arising from modified gravity, or some other completion. Our concern here is more modest. If an effective DE sector crosses $\rhode=0$, or if its inertial mass density $\Iinde=\rhode+\pde$ crosses zero, then ratio variables such as
\[
\wde=\frac{\pde}{\rhode},
\quad
\delta_{\rm de}\equiv\frac{\delta\rhode}{\rhode},
\quad
\theta_{\rm de}=\frac{Q_{\rm de}}{\Iinde}
\]
can become ill-defined even in a completion for which the background stress--energy tensor and the corresponding unnormalized perturbations remain finite. Here $Q_{\rm de}$ denotes the momentum-density variable introduced below.

This distinction matters because many Boltzmann implementations are formulated in terms of fractional perturbations, such as $\delta=\delta\rho/\rho$, and use $w=p/\rho$ together with a closure prescription for $\delta p$ based on a sound speed and an entropy perturbation. Such formulations implicitly assume $\rho_i\neq0$, and often also $\Iin_i\equiv\rho_i+p_i\neq0$, throughout the fluid evolution. For an isolated zero of $\Iin_i$ alone on a positive-density branch---the conventional PDL crossing---the standard remedy in Boltzmann implementations is the parametrized post-Friedmann (PPF) prescription~\cite{Hu:2004kh,Fang:2008sn}, which replaces the fluid closure by a regular effective description in a neighborhood of the crossing; being formulated in terms of $w$ on a positive-density branch, however, it does not by itself cover a zero of $\rho_i$. A ratio-safe formulation instead evolves unnormalized perturbations. Let $Q_i$ denote the scalar momentum-density divergence of component $i$, which on a regular branch with $\Iin_i\neq0$ is related to the usual Fourier-space velocity-divergence potential by
\begin{equation}
Q_i\equiv\Iin_i\,\theta_i.
\label{eq:Q_def}
\end{equation}
Likewise, let $\Pi_i$ denote the unnormalized scalar anisotropic-stress contribution, so that on a nonzero-$\Iin_i$ branch one may write $\Pi_i=\Iin_i\sigma_i$, with $\sigma_i$ the conventional anisotropic-stress potential of Ref.~\cite{Ma:1995ey}. In Newtonian gauge,
\begin{equation}
{\rm d}s^2 = a^2(\tau)\left[-(1+2\Psi){\rm d}\tau^2
+(1-2\Phi){\rm d}{\bf x}^2\right],
\end{equation}
where $\tau$ is conformal time (used in this section and App.~\ref{app:vcdm}; distinct from the proper time along the congruence in the Raychaudhuri equation), the linearized conservation equations $\nabla_\mu T^{\mu\nu}_{(i)}=0$ for a separately conserved component can be written without dividing by $\rho_i$ or $\Iin_i$ as~\cite{Ma:1995ey,Mukhanov:2005sc,Malik:2008im}
\begin{align}
\frac{{\rm d}}{{\rm d}\tau}\delta\rho_i &=
-3\mathcal H(\delta\rho_i+\delta p_i)
+3\Iin_i\frac{{\rm d}\Phi}{{\rm d}\tau}
-Q_i,
\label{eq:drho_safe}
\\
\frac{{\rm d}}{{\rm d}\tau}Q_i &=
-4\mathcal H Q_i
+k^2\left(\delta p_i+\Iin_i\Psi-\Pi_i\right),
\label{eq:Q_safe}
\end{align}
where $\mathcal H\equiv a^{-1}{\rm d}a/{\rm d}\tau$ and $k$ is the comoving wavenumber. Equations~\eqref{eq:drho_safe}--\eqref{eq:Q_safe} contain no division by $\rho_i$ or $\Iin_i$. Thus, a zero of $\rho_i$ is not by itself a singular point of the perturbation kinematics, provided the unnormalized variables $\delta\rho_i$, $\delta p_i$, $Q_i$, and $\Pi_i$ remain finite. If $\Iin_i$ crosses zero, the normalized velocity variable $\theta_i=Q_i/\Iin_i$ becomes ill-defined unless $Q_i$ vanishes sufficiently fast; the unnormalized momentum density $Q_i$, read from the time--space component of the perturbed stress--energy tensor, is the regular variable. In simple scalar-field completions, for example, $\delta T^0{}_{j}\propto\dot\phi\,\partial_j\delta\phi$, so $Q_i$ vanishes at a turning point $\dot\phi=0$ if $\delta\phi$ remains finite. This last statement is model dependent, whereas the ratio-safe form of Eqs.~\eqref{eq:drho_safe}--\eqref{eq:Q_safe} is purely kinematic.

Microphysical stability, on the other hand, depends on the underlying model. A Boltzmann-solver implementation requires a closure prescription for $\delta p_i$ and $\Pi_i$, and this is where microphysics enters. For instance, a strictly barotropic perfect fluid with a linear EoS, $p=w\rho$ with constant $w$, has adiabatic sound speed $c_a^2={\rm d}p/{\rm d}\rho=w$. On a positive-density repulsive branch, such a fluid has negative pressure and hence $w<0$, so the adiabatic mode is gradient-unstable unless non-adiabatic stresses or more general microphysics are introduced. Canonical single-field quintessence provides a different example: it has rest-frame sound speed $c_s^2=1$ at linear order~\cite{Mukhanov:2005sc}, but its kinetic structure enforces $\Iin_\phi=\dot\phi^{\,2}\ge0$. Moreover,
\[
\Mact_\phi=\rho_\phi+3p_\phi=2(\dot\phi^{\,2}-V),
\]
so a repulsive canonical scalar requires $V>\dot\phi^{\,2}$ and therefore $\rho_\phi=\dot\phi^{\,2}/2+V>0$. A sign-switching sector of the kind discussed in this work therefore points beyond thermodynamic perfect fluids and canonical single-field quintessence, toward phantom or non-canonical scalar descriptions, multi-field or interacting dark-sector models, minimally modified gravity, or other effective sources. The viability of any particular realization must be assessed case by case. Perturbation-level data can indeed be decisive: a model with an acceptable background history may be excluded once linear perturbations are included~\cite{Hashim:2026yoy}.

A useful controlled embedding paradigm is provided by VCDM-type minimally modified gravity theories~\cite{DeFelice:2020eju,DeFelice:2022uxv,Akarsu:2024qsi}. Such theories can implement a chosen background history $H(z)$, equivalently an inferred $\rhode(z)$, without introducing a new propagating scalar degree of freedom; in this setting the departure from the standard Boltzmann implementation can be minimal. We do not attempt such a model-building or observational viability analysis here. Our aim is only to emphasize that divergences of $w_i$, $\delta_i$, or velocity variables normalized by $\Iin_i$ are kinematic artifacts of ratio variables, and that linear perturbations can be written in a ratio-safe form. Additional microphysical context is collected in App.~\ref{app:microphysics}, and the perturbation-level ingredient relevant for VCDM embeddings is summarized in App.~\ref{app:vcdm}.

\section{Conclusions}
\label{sec:conclusions}
The central message of this work is that the usual EoS language for DE is less fundamental than the Raychaudhuri language once effective sectors are allowed to change sign. The familiar condition $\wde<-1/3$ is a positive-density-branch proxy for a negative Raychaudhuri contribution, not a branch-independent definition of repulsion, and the PDL, $\wde=-1$, is only the ratio representation of $\Iinde=0$ on a regular nonzero-density branch. This limitation is no longer academic: multi-probe analyses addressing cosmological tensions have found improved fits for sign-switching effective density histories~\cite{Akarsu:2024eoo}, and model-agnostic reconstructions find, for some dataset combinations, an inferred density that crosses zero at intermediate redshift~\cite{Akarsu:2026pom,Akarsu:2026anp}. Complementary inverse scalar-field reconstructions formulated directly in terms of $\rhode$ and $\rhode'$ remain well defined through such a density zero~\cite{Adil:2026kfn}. When $\rhode$ approaches or crosses zero, the ratio variables become ill-conditioned, while the quantities selected directly by the field equations, $\Iinde=\rhode+\pde$ and $\Mactde=\rhode+3\pde$, remain the appropriate variables whenever the stress--energy tensor is regular.

Within the standard GR-like split used throughout, this leads to a simple but consequential reorganization of the DE diagnostics. The condition $\Mactde<0$ identifies a Raychaudhuri-repulsive effective sector, and the NECB, $\Iinde=0$, separates the two signs of the sector-level NEC character; these statements are sector-level and split-dependent, but branch-independent within the specified split. They make clear that density sign change, sector-level repulsion, cosmic acceleration, and total NEC violation are four distinct notions. For a separately conserved sector with a smooth finite-order
negative-to-positive density crossing, the pole in $\wde$ is kinematic
rather than a singularity of the background stress--energy
tensor---its residue, $n(1+\zdag)/3$, is fixed by the crossing redshift
and the order of the zero alone. If $\Mactde>0$ at some sufficiently
high redshift, continuity implies at least one repulsion boundary
$\zrep>\zdag$; when this boundary is unique, it is the repulsion-onset
redshift. A finite interval with $\rhode<0$ but $\Mactde<0$ is therefore not exceptional; it is the natural Raychaudhuri interpretation of a smooth sign-changing effective sector. Sector-level repulsion is nevertheless not cosmic acceleration, which is governed by the total source~\cite{Caldwell:2025inn,Mishra:2026tzn}. A sufficiently sharp sign switch can open a transient intermediate acceleration episode around the density crossing. We derive the exact matter-era range of crossing slopes for which the Universe accelerates at the crossing while the total NEC remains satisfied there. Under the single-impulse and stationary-point assumptions of Sec.~\ref{subsec:counting}, the acceleration history exhibits either the usual single late-time onset or a three-crossing pattern with one intermediate window.

The examples make these logical distinctions explicit. The smooth sign-switching profile provides a minimal counterexample to the positive-density intuition, with repulsion beginning while the inferred density is still negative and, for sufficiently rapid transitions, an intermediate acceleration window; the exponential infrared teleparallel model and the minimal phantom brane show that the same separation between density sign switch and repulsion onset appears when modified gravity backgrounds are rewritten in the same GR-like effective fluid split. Together with the general argument of Sec.~\ref{subsec:repulsive}, these three constructions show that this separation is not tied to any particular ansatz; the existence of a finite repulsion-onset redshift remains subject to the stated high-redshift condition. The examples are illustrative rather than parameter-inference models; their role is to expose the diagnostic structure.

At the perturbative level, the same lesson persists, together with its
limits. Divergences of $\wde$, $\delta_{\rm de}$, or velocity variables
normalized by $\Iinde$ are not by themselves singularities of the
stress--energy tensor: the conservation equations can be formulated in
the unnormalized variables
$(\delta\rho_i,\delta p_i,Q_i,\Pi_i)$ without dividing by $\rho_i$ or
$\Iin_i$. Once a regular microphysical closure for $\delta p_i$ and
$\Pi_i$ is supplied, these variables can be evolved through the
crossing. This establishes kinematic regularity of the formulation, not
the microphysical viability of an arbitrary completion; stability,
closure relations, and observational viability remain properties of
the completion, as discussed in App.~\ref{app:microphysics}. VCDM-type minimally modified gravity provides one controlled embedding paradigm in which the relevant scalar-sector modification of the Boltzmann system can be captured by a single scale-dependent factor in the momentum constraint, as summarized in App.~\ref{app:vcdm}.

The practical implication for future DE studies is correspondingly
concrete. Where the inferred density is comfortably positive, $\wde$
is a convenient coordinate, and nothing in our analysis calls that use
into question. Where $\rhode$ is instead allowed to become small or to
change sign---precisely the regime highlighted by current
reconstructions%
~\cite{Sabogal:2024qxs,Akarsu:2026pom,Akarsu:2026anp}---studies should
track $\rhode$, $\pde$, $\Iinde$, and $\Mactde$ directly and employ
parametrizations capable of admitting both density branches rather than
enforcing a fixed density sign by construction. Such parametrizations
may be formulated, for example, at the level of the
density~\cite{DiValentino:2020naf,Adil:2023exv,Montefalcone:2026iga},
the pressure~\cite{Sen:2007gk,Cheng:2025lod}, or the inertial mass
density~\cite{Akarsu:2019hmw,Acquaviva:2021jov,Escamilla:2026eks},
with the active gravitational mass density as a natural next step. At
a regular density crossing, any resulting pole in $\wde$ should be
read as a kinematic signature of the crossing rather than as a
singularity of the stress--energy tensor, and the NECB rather than the
PDL should be used as the regular sector-level
boundary~\cite{Akarsu:2026anp,Gokcen:2026pkq}. Studies should likewise
distinguish sector-level phantom behavior from total NEC violation and
separate the onset of sector-level repulsion from the onset of cosmic
acceleration. The field equations already single out these quantities;
we propose only that theoretical model building and observational
inference give them priority whenever both density branches are to be
explored.

\begin{acknowledgments}
The authors thank Varun Sahni and Yuri Shtanov for valuable discussions
and helpful comments. \"{O}.A. acknowledges support from the
Turkish Academy of Sciences through the Outstanding Young Scientist
Award programme (T\"{U}BA-GEB\.{I}P). N.M.U. is supported by the
Scientific and Technological Research Council of T{\"u}rkiye
(T\"{U}B\.{I}TAK) through the 2218 National Postdoctoral Research
Fellowship Programme, Project No.~124C450. This article is based upon
work from COST Action CA21136, ``Addressing observational tensions in
cosmology with systematics and fundamental physics'' (CosmoVerse),
supported by COST (European Cooperation in Science and Technology).
\end{acknowledgments}

\bigskip

\appendix

\section{Microphysical realizations and stability}
\label{app:microphysics}

This appendix elaborates on the distinction emphasized in the main text between
(i) \emph{kinematic regularity} of variables across $\rho=0$ and $\Iin=0$ crossings, and
(ii) \emph{microphysical stability}, which depends on the underlying completion of the effective DE sector. We summarize representative realizations and their associated stability conditions, emphasizing that requirements natural within one class of models, such as thermodynamic perfect fluids, should not be promoted to universal conditions on generic effective sectors.

\emph{Thermodynamic perfect fluids.}
Many Boltzmann-solver implementations model cosmic species using a fluid description. A perfect fluid understood as a thermodynamic system is specified by two independent state variables. The choice is conventional, although some choices are more convenient than others. For example, taking the number density $n$ and temperature $\mathcal T$ as independent variables, one may write
\begin{equation}
\rho=\rho(n,\mathcal T),
\quad
p=p(n,\mathcal T),
\quad
s=s(n,\mathcal T),
\end{equation}
where $s$ is the entropy per particle. The first law then takes the form
\begin{equation}
{\rm d}\rho
=
h_{\rm fl}\,{\rm d}n
+n\mathcal T\,{\rm d}s,
\quad
h_{\rm fl}\equiv\frac{\rho+p}{n}
=\frac{\Iin}{n},
\end{equation}
where $h_{\rm fl}$ is the enthalpy per particle. Equivalently, one may take a single fundamental relation $\rho=\rho(n,s)$, from which
$h_{\rm fl}=(\partial\rho/\partial n)_s$ and
$\mathcal T=n^{-1}(\partial\rho/\partial s)_n$ follow. If instead $h_{\rm fl}(n,s)$ and $\mathcal T(n,s)$ are specified independently, they must satisfy the integrability condition
\[
\left(\frac{\partial h_{\rm fl}}{\partial s}\right)_n
=
\left[\frac{\partial(n\mathcal T)}{\partial n}\right]_s,
\]
which guarantees that ${\rm d}\rho$ is an exact differential.

\emph{Fluid stability and the limitation of constant-$w$ repulsion.}
Within this restricted thermodynamic perfect fluid class, the adiabatic sound speed is
$c_a^2\equiv(\partial p/\partial\rho)_s$. For a propagating adiabatic mode, gradient stability requires $c_a^2\ge0$. The limiting case $c_a^2=0$ is non-propagating at this order, and whether it represents a regular limit or signals strong coupling depends on the underlying completion and its regime of validity. In an action-based perfect fluid description with $c_a^2>0$, the sign of the scalar kinetic weight is determined by the enthalpy density; absence of a ghost-like scalar mode therefore requires~\cite{DeFelice:2009bx}
\begin{equation}
\Iin=\rho+p>0.
\end{equation}
This is a stability condition for this particular thermodynamic-fluid realization, not a universal requirement on every effective DE sector.

The perfect fluid framework successfully describes standard components such as pressureless matter, radiation, and a stiff fluid. Suppose, however, that one requires the same perfect fluid component to be Raychaudhuri-repulsive,
$\Mact=\rho+3p<0$, while also satisfying the kinetic-weight condition $\Iin>0$. These two requirements imply
\begin{equation}
\Iin>0
\;\;\text{and}\;\;
\Mact<0
\;\;\Longleftrightarrow\;\;
\rho>0
\;\;\text{and}\;\;
-\rho<p<-\frac{1}{3}\rho.
\end{equation}
Thus, in this minimal perfect fluid realization, a Raychaudhuri-repulsive component must have positive density and negative pressure; pressureless matter and radiation cannot play this role. Moreover, for the commonly used strictly barotropic form
$p=w\rho$ with constant $w$, one has $c_a^2=w$. Within the
nondegenerate Raychaudhuri-repulsive class just identified,
$-1<w<-1/3$, and hence $c_a^2<0$; its propagating adiabatic mode is
therefore gradient-unstable unless non-adiabatic stresses or more
general microphysics are introduced. The minimal thermodynamic
perfect fluid description is consequently too restrictive to
represent generic repulsive effective sectors.

\emph{Canonical and phantom scalar contrasts.}
Consider next a minimally coupled canonical scalar field $\phi(t)$ with potential $V(\phi)$,
\begin{equation}
\rho_\phi
=
\frac{1}{2}\dot\phi^2+V(\phi),
\quad
p_\phi
=
\frac{1}{2}\dot\phi^2-V(\phi).
\end{equation}
Its kinetic structure gives
$\Iin_\phi=\rho_\phi+p_\phi=\dot\phi^2\ge0$, while the high-frequency propagation speed in the scalar rest frame is luminal, $c_s^2=1$. For the scalar to contribute Raychaudhuri-repulsively, one requires
\begin{equation}
\Mact_\phi
=
\rho_\phi+3p_\phi
=
2\dot\phi^2-2V(\phi)<0.
\end{equation}
Thus, the scalar is Raychaudhuri-repulsive if and only if
$V(\phi)>\dot\phi^2$, which in turn implies
$\rho_\phi=\dot\phi^2/2+V>0$ on the repulsive branch. There the ratio variable is regular and satisfies
\begin{equation}
w_\phi
=
\frac{p_\phi}{\rho_\phi}\ge-1,
\quad
-\rho_\phi
\le p_\phi
<-\frac{1}{3}\rho_\phi.
\end{equation}
Canonical single-field quintessence can therefore furnish a Raychaudhuri-repulsive contribution capable of driving cosmic acceleration, but only on a positive-density branch with $\Iin_\phi\ge0$. It cannot realize the smooth negative-to-positive sign switch considered in the main text, which requires $\Iinde<0$ in a punctured neighborhood of $\zdag$ and $\Iinde(\zdag)\le0$ at the crossing [see Eq.~\eqref{eq:IM_crossing_inequality}], nor can it realize a negative-density Raychaudhuri-repulsive phase, which necessarily has $\Iinde<0$ by Eq.~\eqref{eq:negative_rho_repulsion_nec}.

A minimally coupled phantom scalar with Lagrangian
$-X-V(\phi)$ [$\epsilon=-1$ in Eq.~\eqref{eq:I_scalar}] instead has
$\Iin_\phi=-\dot\phi^2\le0$, while its high-frequency squared propagation speed remains $c_s^2=1$. Such a field can reconstruct sector-level phantom-like sign-switching backgrounds at the level of homogeneous dynamics, and explicit scalar-field realizations of smooth sign-changing DE histories have been discussed in Ref.~\cite{Akarsu:2025gwi}. In this minimal realization, however, the negative kinetic term is a genuine microphysical ghost: its Hamiltonian is unbounded below, leading, once interactions are admitted, to the familiar classical runaway and quantum vacuum-instability problems~\cite{Carroll:2003st,Cline:2003gs}. The existence of a scalar reconstruction is therefore not, by itself, evidence of microphysical viability. This is precisely why the main text separates the kinematic regularity of $(\rho,p,\Iin,\Mact)$ and of unnormalized perturbations from the stability of any particular completion.

\emph{Beyond minimal fluids and single canonical fields.}
Viable late-time acceleration is not limited to the two minimal classes above. In particular, a condition such as $\Iin=\rho+p>0$ should not be promoted to a universal ``stability requirement'' on a generic effective DE sector: doing so would restrict attention to a particular subset of completions and can lead to incorrect general conclusions
~\cite{Caldwell:2025inn,Mishra:2026tzn}. When the NEC is instead
imposed as a model prior for a noninteracting DE fluid, it yields strong acoustic-scale restrictions~\cite{Lewis:2024cqj}. A wide range of models enlarges the space of possible accelerating background histories, including Horndeski-type scalar--tensor theories~\cite{Horndeski:1974wa,Kobayashi:2019hrl}, multi-field constructions such as quintom models~\cite{Feng:2004ad,Cai:2009zp}, interacting dark-sector scenarios~\cite{Wang:2016lxa}, non-canonical scalar theories, and modified gravity effective sources. Under appropriate conditions, some of these can realize stable sector-level phantom behavior, stable crossing of the sector-level NECB---corresponding to PDL crossing on a regular positive-density branch---and even stable violation of the total NEC~\cite{Creminelli:2006xe,Deffayet:2010qz}. This list is not exhaustive. The essential point is that different completions contain different degrees of freedom, closure relations, and stability conditions, thereby providing a powerful discriminator among background histories that may otherwise be degenerate.

This model dependence becomes most pressing for the inverse problem: given a desired background history $H(z)$, or equivalently an inferred $\rhode(z)$ within the adopted split, does there exist a stable completion that realizes it? A concrete instance is whether a CPL-like background history admits a stable microphysical embedding. In effective field theory (EFT) language~\cite{Gubitosi:2012hu,Gleyzes:2013ooa}, embedding a chosen background into a scalar---or vector~\cite{Heisenberg:2014rta,DeFelice:2016yws}---theory requires no-ghost conditions, avoidance of infinitely strong coupling, and non-negative squared propagation speeds throughout the relevant epoch. If the theory modifies the tensor sector, tensor-mode no-ghost and gradient-stability conditions must also be satisfied, together with the stringent observational constraint on the late-time propagation speed of gravitational waves~\cite{Creminelli:2017sry,Ezquiaga:2017ekz}. Even after these theoretical requirements are met, viability must still be tested against observables such as the matter power spectrum and gravitational lensing, since many theories modify the effective Newtonian coupling and/or the lensing potential. A recent EFT-based assessment of single scalar field DE provides a complementary illustration: background-expansion data leave substantial microphysical underdetermination, while growth and integrated Sachs--Wolfe (ISW) measurements, fifth-force bounds, and the viability of screening mechanisms supply complementary tests of extended scalar field completions~\cite{Garcia-Garcia:2026nzy}.

Minimally modified gravity theories of the VCDM type~\cite{DeFelice:2020eju,DeFelice:2022uxv,Akarsu:2024qsi} provide a useful controlled paradigm for this inverse problem. They can implement a broad class of sufficiently regular background histories without introducing an additional propagating gravitational scalar degree of freedom. Consequently, the no-ghost and sound-speed conditions associated with such an extra scalar mode do not arise, because the mode itself is absent. The standard matter and radiation sectors retain their usual conservation equations and microphysical properties, while the modified scalar-sector constraint structure determines the perturbative completion of the chosen background. The corresponding minimal modification of the Boltzmann system is summarized in App.~\ref{app:vcdm}.

\section{VCDM embedding and a minimal Boltzmann-code modification (CLASS conventions)}
\label{app:vcdm}

A useful embedding paradigm is to realize a prescribed background history within a type-II minimally modified gravity theory of the VCDM type~\cite{DeFelice:2020eju,DeFelice:2022uxv}, such that the chosen background dynamics selects a specific member of the VCDM family~\cite{Akarsu:2024qsi}.\footnote{VCDM propagates only the two tensorial gravitational degrees of freedom in the gravity sector. The scalar-sector modification is encoded in a non-propagating constraint structure. In non-cosmological settings, this structure can manifest itself through a ``shadowy'' mode governed by elliptic equations and boundary conditions; this is distinct from an additional propagating scalar degree of freedom~\cite{DeFelice:2020eju,DeFelice:2022uxv}.}
This paradigm is useful for several reasons:
(i) a broad class of sufficiently regular background histories $H(z)>0$ can be implemented, provided the corresponding auxiliary-field reconstruction remains well defined throughout the relevant domain;
(ii) by construction, no additional propagating scalar degree of freedom is introduced in the gravity sector, so the usual no-ghost and sound-speed conditions associated with such a mode do not arise;
(iii) gravitational waves propagate at the speed of light~\cite{DeFelice:2020eju}, in accordance with the
stringent multimessenger constraint~\cite{Creminelli:2017sry,Ezquiaga:2017ekz};
and (iv) in the short-wavelength regime, matter clustering approaches
the GR force law, $G_{\rm eff}/G_{\rm N}\to1$, although the
nonstandard background expansion continues to affect the growth
history.

Accordingly, the perturbative departure from GR can remain minimal even when the spatially flat background contains a nontrivial effective DE contribution to the Friedmann constraint~\eqref{eq:friedmann}, which we write as
\begin{equation}
3\Mpl^{2}H^{2}
=
\sum_{I}\rho_{I}
+\rhode.
\label{eq:vcdm_friedmann}
\end{equation}
Here $\Mpl\equiv(8\pi G)^{-1/2}$ is the reduced Planck mass, and $I$ labels only the standard matter and radiation components---photons, neutrinos, baryons, and CDM---each satisfying its usual conservation equation. All modified background dynamics is packaged into the inferred effective DE density $\rhode(z)$ defined by the same GR-like split used throughout. Once the standard components are separately conserved, the effective remainder is separately conserved by construction, so its pressure and inertial mass density follow from Eqs.~\eqref{eq:pres} and~\eqref{eq:Iin_de}.

Since VCDM introduces no additional propagating scalar degree of
freedom, the scalar perturbation system used in Boltzmann solvers
differs from its GR counterpart only through the modified gravitational
constraint structure. In the Newtonian (longitudinal) gauge of the main
text, with Bardeen potentials $(\Phi,\Psi)$, the VCDM modification
relevant to the evolution of these potentials is encoded in the scalar
constraint written below. For definiteness, we adopt CLASS
conventions~\cite{Blas:2011rf}: for any background component, we write
\[
\rho=3\Mpl^{2}\varrho,
\quad
p=3\Mpl^{2}\mathcal P,
\]
where $p$ is the physical pressure and $(\varrho,\mathcal P)$ are the corresponding CLASS-rescaled variables. Thus,
\[
\Iin_I
=
\rho_I+p_I
=
3\Mpl^{2}(\varrho_I+\mathcal P_I).
\]
We denote by $\theta_I$ the Fourier-space velocity-divergence potential of component $I$, related to the ratio-safe momentum-density variable of Eq.~\eqref{eq:Q_def} by
\[
Q_I
=
3\Mpl^{2}(\varrho_I+\mathcal P_I)\theta_I.
\]
Conformal time $\tau$ and the conformal Hubble rate
$\mathcal H\equiv a^{-1}{\rm d}a/{\rm d}\tau$ are defined as in the main text.

For scalar Fourier modes with nonzero comoving wavenumber $k$, the momentum constraint in the VCDM implementation reads~\cite{DeFelice:2020eju,Akarsu:2024qsi}
\begin{equation}
\frac{{\rm d}\Phi}{{\rm d}\tau}
+\mathcal H\Psi
=
\frac{
3\Bigl[
k^{2}
-3\bigl(
\frac{{\rm d}\mathcal H}{{\rm d}\tau}
-\mathcal H^{2}
\bigr)
\Bigr]
\sum_{I}(\varrho_I+\mathcal P_I)\theta_I
}{
\frac{2k^{2}}{a^{2}}
\Bigl[
k^{2}
+\frac{9}{2}a^{2}
\sum_{J}(\varrho_J+\mathcal P_J)
\Bigr]
}.
\label{eq:vcdm_momentum}
\end{equation}
Both sums run over the standard matter and radiation components appearing in Eq.~\eqref{eq:vcdm_friedmann}. The inferred effective DE sector does not carry an independent velocity perturbation associated with a new propagating scalar mode.

For implementation purposes, the linear scalar system may be organized
as follows. The background is Eq.~\eqref{eq:vcdm_friedmann} with the
prescribed $\rhode(z)$, whose pressure and enthalpy follow from separate
conservation as in the main text. Each standard species obeys its usual
unmodified Boltzmann or fluid hierarchy, including the standard
collision terms. For a separately conserved fluid component, the
lowest two moments may be written in the ratio-safe form
\eqref{eq:drho_safe}--\eqref{eq:Q_safe}; species-specific closure
relations and higher-multipole equations are supplied in the usual
way~\cite{Ma:1995ey,Blas:2011rf}. The traceless spatial equation
relating the two metric potentials retains its GR form, sourced by the
anisotropic stress of the standard species, so that $\Phi=\Psi$ when
that shear is negligible. In the formulation adopted here, the VCDM
modification entering the scalar evolution is encoded in
Eq.~\eqref{eq:vcdm_momentum}, equivalently in
Eqs.~\eqref{eq:vcdm_momentum_factor} and~\eqref{eq:vcdm_factor}
~\cite{DeFelice:2020eju,Akarsu:2024qsi,Akarsu:2024eoo}. Together with
the standard species hierarchies and the GR-form anisotropy relation,
this closes the linear scalar evolution. Any additional constraint
used to construct initial data or monitor the numerical integration
must be imposed in its corresponding VCDM form. Deep in the radiation
era, the effective enthalpy of the histories considered here is
negligible relative to the standard total,
$|\varrho_{\rm de}+\mathcal P_{\rm de}|
\ll\sum_J(\varrho_J+\mathcal P_J)$, so
$\mathcal F_{\rm VCDM}\to1$ and the standard adiabatic initial
conditions are recovered.

The structure of Eq.~\eqref{eq:vcdm_momentum} becomes transparent upon using the background identity
\begin{equation}
\frac{{\rm d}\mathcal H}{{\rm d}\tau}
-\mathcal H^{2}
=
-\frac{3}{2}a^{2}
\left[
\sum_{J}(\varrho_J+\mathcal P_J)
+
(\varrho_{\rm de}+\mathcal P_{\rm de})
\right],
\label{eq:vcdm_bg_identity}
\end{equation}
where the total source includes the effective DE enthalpy,
\[
\varrho_{\rm de}+\mathcal P_{\rm de}
=
\frac{\Iinde}{3\Mpl^{2}},
\]
itself fixed at the background level by Eq.~\eqref{eq:Iin_de}. Substitution of Eq.~\eqref{eq:vcdm_bg_identity} into Eq.~\eqref{eq:vcdm_momentum} shows that the numerator and denominator brackets differ precisely by
$\frac{9}{2}a^{2}(\varrho_{\rm de}+\mathcal P_{\rm de})$. The VCDM constraint can therefore be written as the standard GR momentum-constraint source constructed from the ordinary matter and radiation components, multiplied by a scale-dependent factor:
\begin{equation}
\frac{{\rm d}\Phi}{{\rm d}\tau}
+\mathcal H\Psi
=
\frac{
3\sum_{I}(\varrho_I+\mathcal P_I)\theta_I
}{
2k^{2}/a^{2}
}
\,
\mathcal F_{\rm VCDM}(k,\tau),
\label{eq:vcdm_momentum_factor}
\end{equation}
where
\begin{equation}
\mathcal F_{\rm VCDM}(k,\tau)
=
\frac{
k^{2}
+\frac{9}{2}a^{2}
\left[
\sum_{J}(\varrho_J+\mathcal P_J)
+
(\varrho_{\rm de}+\mathcal P_{\rm de})
\right]
}{
k^{2}
+\frac{9}{2}a^{2}
\sum_{J}(\varrho_J+\mathcal P_J)
}.
\label{eq:vcdm_factor}
\end{equation}

In the $\Lambda$CDM limit,
$\mathcal P_{\rm de}=-\varrho_{\rm de}$, the factor
$\mathcal F_{\rm VCDM}$ equals unity for every nonzero Fourier mode, and Eq.~\eqref{eq:vcdm_momentum_factor} reduces to the standard GR momentum constraint. The same GR limit is approached on subhorizon scales, $k\gg\mathcal H$, because the $k^2$ terms dominate both the numerator and denominator of Eq.~\eqref{eq:vcdm_factor}. For the standard matter and radiation components,
$\sum_J(\varrho_J+\mathcal P_J)>0$, so the denominator is positive. When the inferred effective DE sector is sector-level phantom-like,
$\varrho_{\rm de}+\mathcal P_{\rm de}<0$, one has
$\mathcal F_{\rm VCDM}<1$, with the departure from unity concentrated on near- and super-horizon scales. Over the displayed interval $0\le z\le7$, the histories considered
here satisfy $\Iin_{\rm tot}>0$ (bottom left panel of
Fig.~\ref{fig:qmu}), so the numerator of Eq.~\eqref{eq:vcdm_factor} also remains positive. Consequently, over this interval,
\begin{equation}
0<\mathcal F_{\rm VCDM}(k,\tau)<1
\end{equation}
during the sector-level phantom-like phase, describing a scale-dependent suppression of the momentum-constraint source that disappears as $k/\mathcal H\to\infty$.

Because the departure of $\mathcal F_{\rm VCDM}$ from unity is concentrated on near- and super-horizon scales during the sector-level phantom-like phase, its direct perturbative imprint is expected to be largest at low multipoles, notably through the late-time ISW contribution to the CMB temperature anisotropies. A multi-probe analysis of a $\Ls$CDM-type history including the VCDM perturbation dynamics and its large-angle CMB effects is presented in Ref.~\cite{Akarsu:2024eoo}. Related sensitivity of the CMB temperature--lensing cross spectrum and the lensing--ISW bispectrum to the transition redshift has been studied for phenomenological sharp-transition $\Ls$CDM backgrounds in Ref.~\cite{Forconi:2025gwo}. That analysis treats the sharp phenomenological background rather than the VCDM-specific scalar constraint and is therefore complementary to the VCDM completion discussed here. A multi-probe ISW study of DE models with negative energy density---including a phenomenological $\Ls$CDM background, but without the VCDM-specific scalar constraint---combines galaxy cross-correlations, the lensing--ISW bispectrum, and the Planck ISW--lensing likelihood~\cite{Ghafari:2025eql}. Cross-correlations of the ISW signal with large scale structure could therefore provide a dedicated test of this VCDM completion. In the short-wavelength limit, the modified force law approaches its GR form, $G_{\rm eff}/G_{\rm N}\to1$, although the nonstandard background expansion continues to affect the growth history. These features offer one possible observational discriminator between a VCDM-type completion and fluid-type completions of the same background history, in which the DE sector carries its own perturbations. A dedicated ISW--large-scale structure cross-correlation analysis within the VCDM embedding considered here remains beyond the scope of the present work.

At linear scalar order, implementing the prescribed background $\rhode(z)$ together with the replacement~\eqref{eq:vcdm_momentum}---or, equivalently, Eqs.~\eqref{eq:vcdm_momentum_factor} and~\eqref{eq:vcdm_factor}---provides a controlled VCDM completion for a broad class of sign-switching effective density histories for which the background reconstruction remains regular. Since no additional propagating gravitational scalar mode is present, the no-ghost and gradient-stability conditions associated with such an extra mode are absent, while the standard matter and radiation sectors retain their usual conservation equations and microphysical properties. This is particularly useful for background histories motivated by constructions whose native perturbation theory would otherwise be technically demanding, such as the extra-dimensional model underlying the minimal phantom brane considered in the main text. A complete observational analysis of any particular embedded history nevertheless remains model dependent and lies beyond the scope of the present work.

\bibliographystyle{apsrev4-2_modified}
\bibliography{biblio_arxiv}

\begin{thebibliography}{128}%
\makeatletter
\providecommand \@ifxundefined [1]{%
 \@ifx{#1\undefined}
}%
\providecommand \@ifnum [1]{%
 \ifnum #1\expandafter \@firstoftwo
 \else \expandafter \@secondoftwo
 \fi
}%
\providecommand \@ifx [1]{%
 \ifx #1\expandafter \@firstoftwo
 \else \expandafter \@secondoftwo
 \fi
}%
\providecommand \natexlab [1]{#1}%
\providecommand \enquote  [1]{``#1''}%
\providecommand \bibnamefont  [1]{#1}%
\providecommand \bibfnamefont [1]{#1}%
\providecommand \citenamefont [1]{#1}%
\providecommand \href@noop [0]{\@secondoftwo}%
\providecommand \href [0]{\begingroup \@sanitize@url \@href}%
\providecommand \@href[1]{\@@startlink{#1}\@@href}%
\providecommand \@@href[1]{\endgroup#1\@@endlink}%
\providecommand \@sanitize@url [0]{\catcode `\\12\catcode `\$12\catcode `\&12\catcode `\#12\catcode `\^12\catcode `\_12\catcode `\%12\relax}%
\providecommand \@@startlink[1]{}%
\providecommand \@@endlink[0]{}%
\providecommand \url  [0]{\begingroup\@sanitize@url \@url }%
\providecommand \@url [1]{\endgroup\@href {#1}{\urlprefix }}%
\providecommand \urlprefix  [0]{URL }%
\providecommand \Eprint [0]{\href }%
\providecommand \doibase [0]{https://doi.org/}%
\providecommand \selectlanguage [0]{\@gobble}%
\providecommand \bibinfo  [0]{\@secondoftwo}%
\providecommand \bibfield  [0]{\@secondoftwo}%
\providecommand \translation [1]{[#1]}%
\providecommand \BibitemOpen [0]{}%
\providecommand \bibitemStop [0]{}%
\providecommand \bibitemNoStop [0]{.\EOS\space}%
\providecommand \EOS [0]{\spacefactor3000\relax}%
\providecommand \BibitemShut  [1]{\csname bibitem#1\endcsname}%
\let\auto@bib@innerbib\@empty
\bibitem [{\citenamefont {Riess}\ \emph {et~al.}(1998)\citenamefont {Riess} \emph {et~al.}}]{SupernovaSearchTeam:1998fmf}%
  \BibitemOpen
  \bibfield  {author} {\bibinfo {author} {\bibfnamefont {A.~G.}\ \bibnamefont {Riess}} \emph {et~al.} (\bibinfo {collaboration} {Supernova Search Team}),\ }\bibfield  {title} {\bibinfo {title} {{Observational evidence from supernovae for an accelerating universe and a cosmological constant}},\ }\href {https://doi.org/10.1086/300499} {\bibfield  {journal} {\bibinfo  {journal} {Astron. J.}\ }\textbf {\bibinfo {volume} {116}},\ \bibinfo {pages} {1009} (\bibinfo {year} {1998})},\ \Eprint {https://arxiv.org/abs/astro-ph/9805201} {astro-ph/9805201} \BibitemShut {NoStop}%
\bibitem [{\citenamefont {Perlmutter}\ \emph {et~al.}(1999)\citenamefont {Perlmutter} \emph {et~al.}}]{SupernovaCosmologyProject:1998vns}%
  \BibitemOpen
  \bibfield  {author} {\bibinfo {author} {\bibfnamefont {S.}~\bibnamefont {Perlmutter}} \emph {et~al.} (\bibinfo {collaboration} {Supernova Cosmology Project}),\ }\bibfield  {title} {\bibinfo {title} {{Measurements of $\Omega$ and $\Lambda$ from 42 high redshift supernovae}},\ }\href {https://doi.org/10.1086/307221} {\bibfield  {journal} {\bibinfo  {journal} {Astrophys. J.}\ }\textbf {\bibinfo {volume} {517}},\ \bibinfo {pages} {565} (\bibinfo {year} {1999})},\ \Eprint {https://arxiv.org/abs/astro-ph/9812133} {astro-ph/9812133} \BibitemShut {NoStop}%
\bibitem [{\citenamefont {Ratra}\ and\ \citenamefont {Peebles}(1988)}]{Ratra:1987rm}%
  \BibitemOpen
  \bibfield  {author} {\bibinfo {author} {\bibfnamefont {B.}~\bibnamefont {Ratra}}\ and\ \bibinfo {author} {\bibfnamefont {P.~J.~E.}\ \bibnamefont {Peebles}},\ }\bibfield  {title} {\bibinfo {title} {{Cosmological Consequences of a Rolling Homogeneous Scalar Field}},\ }\href {https://doi.org/10.1103/PhysRevD.37.3406} {\bibfield  {journal} {\bibinfo  {journal} {Phys. Rev. D}\ }\textbf {\bibinfo {volume} {37}},\ \bibinfo {pages} {3406} (\bibinfo {year} {1988})}\BibitemShut {NoStop}%
\bibitem [{\citenamefont {Caldwell}\ \emph {et~al.}(1998)\citenamefont {Caldwell}, \citenamefont {Dave},\ and\ \citenamefont {Steinhardt}}]{Caldwell:1997ii}%
  \BibitemOpen
  \bibfield  {author} {\bibinfo {author} {\bibfnamefont {R.~R.}\ \bibnamefont {Caldwell}}, \bibinfo {author} {\bibfnamefont {R.}~\bibnamefont {Dave}},\ and\ \bibinfo {author} {\bibfnamefont {P.~J.}\ \bibnamefont {Steinhardt}},\ }\bibfield  {title} {\bibinfo {title} {{Cosmological imprint of an energy component with general equation of state}},\ }\href {https://doi.org/10.1103/PhysRevLett.80.1582} {\bibfield  {journal} {\bibinfo  {journal} {Phys. Rev. Lett.}\ }\textbf {\bibinfo {volume} {80}},\ \bibinfo {pages} {1582} (\bibinfo {year} {1998})},\ \Eprint {https://arxiv.org/abs/astro-ph/9708069} {astro-ph/9708069} \BibitemShut {NoStop}%
\bibitem [{\citenamefont {Caldwell}(2002)}]{Caldwell:1999ew}%
  \BibitemOpen
  \bibfield  {author} {\bibinfo {author} {\bibfnamefont {R.~R.}\ \bibnamefont {Caldwell}},\ }\bibfield  {title} {\bibinfo {title} {{A Phantom menace? Cosmological consequences of a dark energy component with super-negative equation of state}},\ }\href {https://doi.org/10.1016/S0370-2693(02)02589-3} {\bibfield  {journal} {\bibinfo  {journal} {Phys. Lett. B}\ }\textbf {\bibinfo {volume} {545}},\ \bibinfo {pages} {23} (\bibinfo {year} {2002})},\ \Eprint {https://arxiv.org/abs/astro-ph/9908168} {astro-ph/9908168} \BibitemShut {NoStop}%
\bibitem [{\citenamefont {Weinberg}(1989)}]{Weinberg:1988cp}%
  \BibitemOpen
  \bibfield  {author} {\bibinfo {author} {\bibfnamefont {S.}~\bibnamefont {Weinberg}},\ }\bibfield  {title} {\bibinfo {title} {{The Cosmological Constant Problem}},\ }\href {https://doi.org/10.1103/RevModPhys.61.1} {\bibfield  {journal} {\bibinfo  {journal} {Rev. Mod. Phys.}\ }\textbf {\bibinfo {volume} {61}},\ \bibinfo {pages} {1} (\bibinfo {year} {1989})}\BibitemShut {NoStop}%
\bibitem [{\citenamefont {Peebles}\ and\ \citenamefont {Ratra}(2003)}]{Peebles:2002gy}%
  \BibitemOpen
  \bibfield  {author} {\bibinfo {author} {\bibfnamefont {P.~J.~E.}\ \bibnamefont {Peebles}}\ and\ \bibinfo {author} {\bibfnamefont {B.}~\bibnamefont {Ratra}},\ }\bibfield  {title} {\bibinfo {title} {{The Cosmological Constant and Dark Energy}},\ }\href {https://doi.org/10.1103/RevModPhys.75.559} {\bibfield  {journal} {\bibinfo  {journal} {Rev. Mod. Phys.}\ }\textbf {\bibinfo {volume} {75}},\ \bibinfo {pages} {559} (\bibinfo {year} {2003})},\ \Eprint {https://arxiv.org/abs/astro-ph/0207347} {astro-ph/0207347} \BibitemShut {NoStop}%
\bibitem [{\citenamefont {Copeland}\ \emph {et~al.}(2006)\citenamefont {Copeland}, \citenamefont {Sami},\ and\ \citenamefont {Tsujikawa}}]{Copeland:2006wr}%
  \BibitemOpen
  \bibfield  {author} {\bibinfo {author} {\bibfnamefont {E.~J.}\ \bibnamefont {Copeland}}, \bibinfo {author} {\bibfnamefont {M.}~\bibnamefont {Sami}},\ and\ \bibinfo {author} {\bibfnamefont {S.}~\bibnamefont {Tsujikawa}},\ }\bibfield  {title} {\bibinfo {title} {{Dynamics of dark energy}},\ }\href {https://doi.org/10.1142/S021827180600942X} {\bibfield  {journal} {\bibinfo  {journal} {Int. J. Mod. Phys. D}\ }\textbf {\bibinfo {volume} {15}},\ \bibinfo {pages} {1753} (\bibinfo {year} {2006})},\ \Eprint {https://arxiv.org/abs/hep-th/0603057} {hep-th/0603057} \BibitemShut {NoStop}%
\bibitem [{\citenamefont {Frieman}\ \emph {et~al.}(2008)\citenamefont {Frieman}, \citenamefont {Turner},\ and\ \citenamefont {Huterer}}]{Frieman:2008sn}%
  \BibitemOpen
  \bibfield  {author} {\bibinfo {author} {\bibfnamefont {J.}~\bibnamefont {Frieman}}, \bibinfo {author} {\bibfnamefont {M.}~\bibnamefont {Turner}},\ and\ \bibinfo {author} {\bibfnamefont {D.}~\bibnamefont {Huterer}},\ }\bibfield  {title} {\bibinfo {title} {{Dark Energy and the Accelerating Universe}},\ }\href {https://doi.org/10.1146/annurev.astro.46.060407.145243} {\bibfield  {journal} {\bibinfo  {journal} {Ann. Rev. Astron. Astrophys.}\ }\textbf {\bibinfo {volume} {46}},\ \bibinfo {pages} {385} (\bibinfo {year} {2008})},\ \Eprint {https://arxiv.org/abs/0803.0982} {0803.0982} \BibitemShut {NoStop}%
\bibitem [{\citenamefont {Chevallier}\ and\ \citenamefont {Polarski}(2001)}]{Chevallier:2000qy}%
  \BibitemOpen
  \bibfield  {author} {\bibinfo {author} {\bibfnamefont {M.}~\bibnamefont {Chevallier}}\ and\ \bibinfo {author} {\bibfnamefont {D.}~\bibnamefont {Polarski}},\ }\bibfield  {title} {\bibinfo {title} {{Accelerating universes with scaling dark matter}},\ }\href {https://doi.org/10.1142/S0218271801000822} {\bibfield  {journal} {\bibinfo  {journal} {Int. J. Mod. Phys. D}\ }\textbf {\bibinfo {volume} {10}},\ \bibinfo {pages} {213} (\bibinfo {year} {2001})},\ \Eprint {https://arxiv.org/abs/gr-qc/0009008} {gr-qc/0009008} \BibitemShut {NoStop}%
\bibitem [{\citenamefont {Linder}(2003)}]{Linder:2002et}%
  \BibitemOpen
  \bibfield  {author} {\bibinfo {author} {\bibfnamefont {E.~V.}\ \bibnamefont {Linder}},\ }\bibfield  {title} {\bibinfo {title} {{Exploring the expansion history of the universe}},\ }\href {https://doi.org/10.1103/PhysRevLett.90.091301} {\bibfield  {journal} {\bibinfo  {journal} {Phys. Rev. Lett.}\ }\textbf {\bibinfo {volume} {90}},\ \bibinfo {pages} {091301} (\bibinfo {year} {2003})},\ \Eprint {https://arxiv.org/abs/astro-ph/0208512} {astro-ph/0208512} \BibitemShut {NoStop}%
\bibitem [{\citenamefont {Linder}(2008)}]{Linder:2008pp}%
  \BibitemOpen
  \bibfield  {author} {\bibinfo {author} {\bibfnamefont {E.~V.}\ \bibnamefont {Linder}},\ }\bibfield  {title} {\bibinfo {title} {{Mapping the Cosmological Expansion}},\ }\href {https://doi.org/10.1088/0034-4885/71/5/056901} {\bibfield  {journal} {\bibinfo  {journal} {Rept. Prog. Phys.}\ }\textbf {\bibinfo {volume} {71}},\ \bibinfo {pages} {056901} (\bibinfo {year} {2008})},\ \Eprint {https://arxiv.org/abs/0801.2968} {0801.2968} \BibitemShut {NoStop}%
\bibitem [{\citenamefont {Visinelli}\ \emph {et~al.}(2019)\citenamefont {Visinelli}, \citenamefont {Vagnozzi},\ and\ \citenamefont {Danielsson}}]{Visinelli:2019qqu}%
  \BibitemOpen
  \bibfield  {author} {\bibinfo {author} {\bibfnamefont {L.}~\bibnamefont {Visinelli}}, \bibinfo {author} {\bibfnamefont {S.}~\bibnamefont {Vagnozzi}},\ and\ \bibinfo {author} {\bibfnamefont {U.}~\bibnamefont {Danielsson}},\ }\bibfield  {title} {\bibinfo {title} {{Revisiting a negative cosmological constant from low-redshift data}},\ }\href {https://doi.org/10.3390/sym11081035} {\bibfield  {journal} {\bibinfo  {journal} {Symmetry}\ }\textbf {\bibinfo {volume} {11}},\ \bibinfo {pages} {1035} (\bibinfo {year} {2019})},\ \Eprint {https://arxiv.org/abs/1907.07953} {1907.07953} \BibitemShut {NoStop}%
\bibitem [{\citenamefont {Akarsu}\ \emph {et~al.}(2020)\citenamefont {Akarsu}, \citenamefont {Barrow}, \citenamefont {Escamilla},\ and\ \citenamefont {Vazquez}}]{Akarsu:2019hmw}%
  \BibitemOpen
  \bibfield  {author} {\bibinfo {author} {\bibfnamefont {{\"O}.}~\bibnamefont {Akarsu}}, \bibinfo {author} {\bibfnamefont {J.~D.}\ \bibnamefont {Barrow}}, \bibinfo {author} {\bibfnamefont {L.~A.}\ \bibnamefont {Escamilla}},\ and\ \bibinfo {author} {\bibfnamefont {J.~A.}\ \bibnamefont {Vazquez}},\ }\bibfield  {title} {\bibinfo {title} {{Graduated dark energy: Observational hints of a spontaneous sign switch in the cosmological constant}},\ }\href {https://doi.org/10.1103/PhysRevD.101.063528} {\bibfield  {journal} {\bibinfo  {journal} {Phys. Rev. D}\ }\textbf {\bibinfo {volume} {101}},\ \bibinfo {pages} {063528} (\bibinfo {year} {2020})},\ \Eprint {https://arxiv.org/abs/1912.08751} {1912.08751} \BibitemShut {NoStop}%
\bibitem [{\citenamefont {Calder\'on}\ \emph {et~al.}(2021)\citenamefont {Calder\'on}, \citenamefont {Gannouji}, \citenamefont {L'Huillier},\ and\ \citenamefont {Polarski}}]{Calderon:2020hoc}%
  \BibitemOpen
  \bibfield  {author} {\bibinfo {author} {\bibfnamefont {R.}~\bibnamefont {Calder\'on}}, \bibinfo {author} {\bibfnamefont {R.}~\bibnamefont {Gannouji}}, \bibinfo {author} {\bibfnamefont {B.}~\bibnamefont {L'Huillier}},\ and\ \bibinfo {author} {\bibfnamefont {D.}~\bibnamefont {Polarski}},\ }\bibfield  {title} {\bibinfo {title} {{Negative cosmological constant in the dark sector?}},\ }\href {https://doi.org/10.1103/PhysRevD.103.023526} {\bibfield  {journal} {\bibinfo  {journal} {Phys. Rev. D}\ }\textbf {\bibinfo {volume} {103}},\ \bibinfo {pages} {023526} (\bibinfo {year} {2021})},\ \Eprint {https://arxiv.org/abs/2008.10237} {2008.10237} \BibitemShut {NoStop}%
\bibitem [{\citenamefont {Sen}\ \emph {et~al.}(2022)\citenamefont {Sen}, \citenamefont {Adil},\ and\ \citenamefont {Sen}}]{Sen:2021wld}%
  \BibitemOpen
  \bibfield  {author} {\bibinfo {author} {\bibfnamefont {A.~A.}\ \bibnamefont {Sen}}, \bibinfo {author} {\bibfnamefont {S.~A.}\ \bibnamefont {Adil}},\ and\ \bibinfo {author} {\bibfnamefont {S.}~\bibnamefont {Sen}},\ }\bibfield  {title} {\bibinfo {title} {{Do cosmological observations allow a negative $\Lambda$?}},\ }\href {https://doi.org/10.1093/mnras/stac2796} {\bibfield  {journal} {\bibinfo  {journal} {Mon. Not. Roy. Astron. Soc.}\ }\textbf {\bibinfo {volume} {518}},\ \bibinfo {pages} {1098} (\bibinfo {year} {2022})},\ \Eprint {https://arxiv.org/abs/2112.10641} {2112.10641} \BibitemShut {NoStop}%
\bibitem [{\citenamefont {Malekjani}\ \emph {et~al.}(2024)\citenamefont {Malekjani}, \citenamefont {Conville}, \citenamefont {Colg{\'a}in}, \citenamefont {Pourojaghi},\ and\ \citenamefont {Sheikh-Jabbari}}]{Malekjani:2023ple}%
  \BibitemOpen
  \bibfield  {author} {\bibinfo {author} {\bibfnamefont {M.}~\bibnamefont {Malekjani}}, \bibinfo {author} {\bibfnamefont {R.~M.}\ \bibnamefont {Conville}}, \bibinfo {author} {\bibfnamefont {E.~{\'O}.}\ \bibnamefont {Colg{\'a}in}}, \bibinfo {author} {\bibfnamefont {S.}~\bibnamefont {Pourojaghi}},\ and\ \bibinfo {author} {\bibfnamefont {M.~M.}\ \bibnamefont {Sheikh-Jabbari}},\ }\bibfield  {title} {\bibinfo {title} {{On redshift evolution and negative dark energy density in Pantheon + Supernovae}},\ }\href {https://doi.org/10.1140/epjc/s10052-024-12667-z} {\bibfield  {journal} {\bibinfo  {journal} {Eur. Phys. J. C}\ }\textbf {\bibinfo {volume} {84}},\ \bibinfo {pages} {317} (\bibinfo {year} {2024})},\ \Eprint {https://arxiv.org/abs/2301.12725} {2301.12725} \BibitemShut {NoStop}%
\bibitem [{\citenamefont {Alam}\ and\ \citenamefont {Sahni}(2006)}]{Alam:2005pb}%
  \BibitemOpen
  \bibfield  {author} {\bibinfo {author} {\bibfnamefont {U.}~\bibnamefont {Alam}}\ and\ \bibinfo {author} {\bibfnamefont {V.}~\bibnamefont {Sahni}},\ }\bibfield  {title} {\bibinfo {title} {{Confronting braneworld cosmology with supernova data and baryon oscillations}},\ }\href {https://doi.org/10.1103/PhysRevD.73.084024} {\bibfield  {journal} {\bibinfo  {journal} {Phys. Rev. D}\ }\textbf {\bibinfo {volume} {73}},\ \bibinfo {pages} {084024} (\bibinfo {year} {2006})},\ \Eprint {https://arxiv.org/abs/astro-ph/0511473} {astro-ph/0511473} \BibitemShut {NoStop}%
\bibitem [{\citenamefont {Alam}\ \emph {et~al.}(2017)\citenamefont {Alam}, \citenamefont {Bag},\ and\ \citenamefont {Sahni}}]{Alam:2016wpf}%
  \BibitemOpen
  \bibfield  {author} {\bibinfo {author} {\bibfnamefont {U.}~\bibnamefont {Alam}}, \bibinfo {author} {\bibfnamefont {S.}~\bibnamefont {Bag}},\ and\ \bibinfo {author} {\bibfnamefont {V.}~\bibnamefont {Sahni}},\ }\bibfield  {title} {\bibinfo {title} {{Constraining the Cosmology of the Phantom Brane using Distance Measures}},\ }\href {https://doi.org/10.1103/PhysRevD.95.023524} {\bibfield  {journal} {\bibinfo  {journal} {Phys. Rev. D}\ }\textbf {\bibinfo {volume} {95}},\ \bibinfo {pages} {023524} (\bibinfo {year} {2017})},\ \Eprint {https://arxiv.org/abs/1605.04707} {1605.04707} \BibitemShut {NoStop}%
\bibitem [{\citenamefont {{\"O}z{\"u}lker}(2022)}]{Ozulker:2022slu}%
  \BibitemOpen
  \bibfield  {author} {\bibinfo {author} {\bibfnamefont {E.}~\bibnamefont {{\"O}z{\"u}lker}},\ }\bibfield  {title} {\bibinfo {title} {{Is the dark energy equation of state parameter singular?}},\ }\href {https://doi.org/10.1103/PhysRevD.106.063509} {\bibfield  {journal} {\bibinfo  {journal} {Phys. Rev. D}\ }\textbf {\bibinfo {volume} {106}},\ \bibinfo {pages} {063509} (\bibinfo {year} {2022})},\ \Eprint {https://arxiv.org/abs/2203.04167} {2203.04167} \BibitemShut {NoStop}%
\bibitem [{\citenamefont {Raychaudhuri}(1955)}]{Raychaudhuri:1953yv}%
  \BibitemOpen
  \bibfield  {author} {\bibinfo {author} {\bibfnamefont {A.}~\bibnamefont {Raychaudhuri}},\ }\bibfield  {title} {\bibinfo {title} {{Relativistic cosmology. I}},\ }\href {https://doi.org/10.1103/PhysRev.98.1123} {\bibfield  {journal} {\bibinfo  {journal} {Phys. Rev.}\ }\textbf {\bibinfo {volume} {98}},\ \bibinfo {pages} {1123} (\bibinfo {year} {1955})}\BibitemShut {NoStop}%
\bibitem [{\citenamefont {Wald}(1984)}]{Wald:1984rg}%
  \BibitemOpen
  \bibfield  {author} {\bibinfo {author} {\bibfnamefont {R.~M.}\ \bibnamefont {Wald}},\ }\href {https://doi.org/10.7208/chicago/9780226870373.001.0001} {\emph {\bibinfo {title} {{General Relativity}}}}\ (\bibinfo  {publisher} {Chicago Univ. Pr.},\ \bibinfo {address} {Chicago, USA},\ \bibinfo {year} {1984})\BibitemShut {NoStop}%
\bibitem [{\citenamefont {Hawking}\ and\ \citenamefont {Ellis}(1973)}]{Hawking:1973uf}%
  \BibitemOpen
  \bibfield  {author} {\bibinfo {author} {\bibfnamefont {S.~W.}\ \bibnamefont {Hawking}}\ and\ \bibinfo {author} {\bibfnamefont {G.~F.~R.}\ \bibnamefont {Ellis}},\ }\href {https://doi.org/10.1017/CBO9780511524646} {\emph {\bibinfo {title} {{The Large Scale Structure of Space-Time}}}},\ Cambridge Monographs on Mathematical Physics\ (\bibinfo  {publisher} {Cambridge University Press},\ \bibinfo {year} {1973})\BibitemShut {NoStop}%
\bibitem [{\citenamefont {Akarsu}\ \emph {et~al.}(2026{\natexlab{a}})\citenamefont {Akarsu}, \citenamefont {Caruana}, \citenamefont {Dialektopoulos}, \citenamefont {Escamilla}, \citenamefont {Kahya},\ and\ \citenamefont {Levi~Said}}]{Akarsu:2026pom}%
  \BibitemOpen
  \bibfield  {author} {\bibinfo {author} {\bibfnamefont {{\"O}.}~\bibnamefont {Akarsu}}, \bibinfo {author} {\bibfnamefont {M.}~\bibnamefont {Caruana}}, \bibinfo {author} {\bibfnamefont {K.~F.}\ \bibnamefont {Dialektopoulos}}, \bibinfo {author} {\bibfnamefont {L.~A.}\ \bibnamefont {Escamilla}}, \bibinfo {author} {\bibfnamefont {E.~O.}\ \bibnamefont {Kahya}},\ and\ \bibinfo {author} {\bibfnamefont {J.}~\bibnamefont {Levi~Said}},\ }\href@noop {} {\bibinfo {title} {{Do equation of state parametrizations of dark energy faithfully capture the dynamics of the late universe?}}} (\bibinfo {year} {2026}{\natexlab{a}}),\ \Eprint {https://arxiv.org/abs/2604.12987} {2604.12987} \BibitemShut {NoStop}%
\bibitem [{\citenamefont {Akarsu}\ \emph {et~al.}(2026{\natexlab{b}})\citenamefont {Akarsu}, \citenamefont {Caruana}, \citenamefont {Dialektopoulos}, \citenamefont {Escamilla}, \citenamefont {Kahya},\ and\ \citenamefont {Levi~Said}}]{Akarsu:2026anp}%
  \BibitemOpen
  \bibfield  {author} {\bibinfo {author} {\bibfnamefont {{\"O}.}~\bibnamefont {Akarsu}}, \bibinfo {author} {\bibfnamefont {M.}~\bibnamefont {Caruana}}, \bibinfo {author} {\bibfnamefont {K.~F.}\ \bibnamefont {Dialektopoulos}}, \bibinfo {author} {\bibfnamefont {L.~A.}\ \bibnamefont {Escamilla}}, \bibinfo {author} {\bibfnamefont {E.~O.}\ \bibnamefont {Kahya}},\ and\ \bibinfo {author} {\bibfnamefont {J.}~\bibnamefont {Levi~Said}},\ }\href@noop {} {\bibinfo {title} {{Hints of sign-changing scalar field energy density and a transient acceleration phase at $z\sim 2$ from model-agnostic reconstructions}}} (\bibinfo {year} {2026}{\natexlab{b}}),\ \Eprint {https://arxiv.org/abs/2602.08928} {2602.08928} \BibitemShut {NoStop}%
\bibitem [{\citenamefont {Gupta~Choudhury}\ \emph {et~al.}(2026)\citenamefont {Gupta~Choudhury}, \citenamefont {Mukherjee}, \citenamefont {Di~Valentino},\ and\ \citenamefont {Sen}}]{GuptaChoudhury:2026gsl}%
  \BibitemOpen
  \bibfield  {author} {\bibinfo {author} {\bibfnamefont {S.}~\bibnamefont {Gupta~Choudhury}}, \bibinfo {author} {\bibfnamefont {P.}~\bibnamefont {Mukherjee}}, \bibinfo {author} {\bibfnamefont {E.}~\bibnamefont {Di~Valentino}},\ and\ \bibinfo {author} {\bibfnamefont {A.~A.}\ \bibnamefont {Sen}},\ }\href@noop {} {\bibinfo {title} {{Model-Independent Indication for a Localized Anomaly in the Late-Time Expansion History}}} (\bibinfo {year} {2026}),\ \Eprint {https://arxiv.org/abs/2607.13009} {2607.13009} \BibitemShut {NoStop}%
\bibitem [{\citenamefont {Adil}\ \emph {et~al.}(2026)\citenamefont {Adil}, \citenamefont {Zapata}, \citenamefont {Akarsu},\ and\ \citenamefont {Vazquez}}]{Adil:2026kfn}%
  \BibitemOpen
  \bibfield  {author} {\bibinfo {author} {\bibfnamefont {S.~A.}\ \bibnamefont {Adil}}, \bibinfo {author} {\bibfnamefont {M.~A.}\ \bibnamefont {Zapata}}, \bibinfo {author} {\bibfnamefont {{\"O}.}~\bibnamefont {Akarsu}},\ and\ \bibinfo {author} {\bibfnamefont {J.~A.}\ \bibnamefont {Vazquez}},\ }\bibfield  {title} {\bibinfo {title} {{Background-level reconstruction of scalar-field potentials from dark-energy histories and comparison with analytic potential families}},\ }\href {https://doi.org/10.1016/j.dark.2026.102387} {\bibfield  {journal} {\bibinfo  {journal} {Phys. Dark Univ.}\ }\textbf {\bibinfo {volume} {53}},\ \bibinfo {pages} {102387} (\bibinfo {year} {2026})},\ \Eprint {https://arxiv.org/abs/2603.14693} {2603.14693} \BibitemShut {NoStop}%
\bibitem [{\citenamefont {Kunz}(2009)}]{Kunz:2007rk}%
  \BibitemOpen
  \bibfield  {author} {\bibinfo {author} {\bibfnamefont {M.}~\bibnamefont {Kunz}},\ }\bibfield  {title} {\bibinfo {title} {{The dark degeneracy: On the number and nature of dark components}},\ }\href {https://doi.org/10.1103/PhysRevD.80.123001} {\bibfield  {journal} {\bibinfo  {journal} {Phys. Rev. D}\ }\textbf {\bibinfo {volume} {80}},\ \bibinfo {pages} {123001} (\bibinfo {year} {2009})},\ \Eprint {https://arxiv.org/abs/astro-ph/0702615} {astro-ph/0702615} \BibitemShut {NoStop}%
\bibitem [{\citenamefont {Caldwell}\ and\ \citenamefont {Linder}(2026)}]{Caldwell:2025inn}%
  \BibitemOpen
  \bibfield  {author} {\bibinfo {author} {\bibfnamefont {R.~R.}\ \bibnamefont {Caldwell}}\ and\ \bibinfo {author} {\bibfnamefont {E.~V.}\ \bibnamefont {Linder}},\ }\bibfield  {title} {\bibinfo {title} {{Null impact of the null energy condition in current cosmology}},\ }\href {https://doi.org/10.1088/1475-7516/2026/05/008} {\bibfield  {journal} {\bibinfo  {journal} {J. Cosmol. Astropart. Phys.}\ }\textbf {\bibinfo {volume} {05}},\ \bibinfo {pages} {008} (\bibinfo {year} {2026})},\ \Eprint {https://arxiv.org/abs/2511.07526} {2511.07526} \BibitemShut {NoStop}%
\bibitem [{\citenamefont {Mishra}(2026)}]{Mishra:2026tzn}%
  \BibitemOpen
  \bibfield  {author} {\bibinfo {author} {\bibfnamefont {S.~S.}\ \bibnamefont {Mishra}},\ }\href@noop {} {\bibinfo {title} {{Effective Phantom Dark Energy: What Cosmological Reconstruction Does and Does Not Imply}}} (\bibinfo {year} {2026}),\ \Eprint {https://arxiv.org/abs/2605.27301} {2605.27301} \BibitemShut {NoStop}%
\bibitem [{\citenamefont {Adame}\ \emph {et~al.}(2025)\citenamefont {Adame} \emph {et~al.}}]{DESI:2024mwx}%
  \BibitemOpen
  \bibfield  {author} {\bibinfo {author} {\bibfnamefont {A.~G.}\ \bibnamefont {Adame}} \emph {et~al.} (\bibinfo {collaboration} {DESI}),\ }\bibfield  {title} {\bibinfo {title} {{DESI 2024 VI: cosmological constraints from the measurements of baryon acoustic oscillations}},\ }\href {https://doi.org/10.1088/1475-7516/2025/02/021} {\bibfield  {journal} {\bibinfo  {journal} {J. Cosmol. Astropart. Phys.}\ }\textbf {\bibinfo {volume} {2025}}\bibfield  {number} {\bibinfo  {number} { (02)},\ \bibinfo {pages} {021}} (\bibinfo {year} {2025})},\ \Eprint {https://arxiv.org/abs/2404.03002} {2404.03002} \BibitemShut {NoStop}%
\bibitem [{\citenamefont {Abdul~Karim}\ \emph {et~al.}(2025)\citenamefont {Abdul~Karim} \emph {et~al.}}]{DESI:2025zgx}%
  \BibitemOpen
  \bibfield  {author} {\bibinfo {author} {\bibfnamefont {M.}~\bibnamefont {Abdul~Karim}} \emph {et~al.} (\bibinfo {collaboration} {DESI}),\ }\bibfield  {title} {\bibinfo {title} {{DESI DR2 results. II. Measurements of baryon acoustic oscillations and cosmological constraints}},\ }\href {https://doi.org/10.1103/tr6y-kpc6} {\bibfield  {journal} {\bibinfo  {journal} {Phys. Rev. D}\ }\textbf {\bibinfo {volume} {112}},\ \bibinfo {pages} {083515} (\bibinfo {year} {2025})},\ \Eprint {https://arxiv.org/abs/2503.14738} {2503.14738} \BibitemShut {NoStop}%
\bibitem [{\citenamefont {Xu}\ \emph {et~al.}(2026)\citenamefont {Xu}, \citenamefont {Kumar}, \citenamefont {Chen}, \citenamefont {Capistrano},\ and\ \citenamefont {Akarsu}}]{Xu:2026sbw}%
  \BibitemOpen
  \bibfield  {author} {\bibinfo {author} {\bibfnamefont {T.}~\bibnamefont {Xu}}, \bibinfo {author} {\bibfnamefont {S.}~\bibnamefont {Kumar}}, \bibinfo {author} {\bibfnamefont {Y.}~\bibnamefont {Chen}}, \bibinfo {author} {\bibfnamefont {A.~J.~S.}\ \bibnamefont {Capistrano}},\ and\ \bibinfo {author} {\bibfnamefont {{\"O}.}~\bibnamefont {Akarsu}},\ }\href@noop {} {\bibinfo {title} {{Probing Dynamical Dark Energy with Late-Time Data: Evidence, Tensions, and the Limits of the $w_0w_a$CDM Framework}}} (\bibinfo {year} {2026}),\ \Eprint {https://arxiv.org/abs/2602.11936} {2602.11936} \BibitemShut {NoStop}%
\bibitem [{\citenamefont {{\"O}z{\"u}lker}\ \emph {et~al.}(2025)\citenamefont {{\"O}z{\"u}lker}, \citenamefont {Di~Valentino},\ and\ \citenamefont {Giar{\`e}}}]{Ozulker:2025ehg}%
  \BibitemOpen
  \bibfield  {author} {\bibinfo {author} {\bibfnamefont {E.}~\bibnamefont {{\"O}z{\"u}lker}}, \bibinfo {author} {\bibfnamefont {E.}~\bibnamefont {Di~Valentino}},\ and\ \bibinfo {author} {\bibfnamefont {W.}~\bibnamefont {Giar{\`e}}},\ }\href@noop {} {\bibinfo {title} {{Dark Energy Crosses the Line: Quantifying and Testing the Evidence for Phantom Crossing}}} (\bibinfo {year} {2025}),\ \Eprint {https://arxiv.org/abs/2506.19053} {2506.19053} \BibitemShut {NoStop}%
\bibitem [{\citenamefont {G{\"o}k{\c{c}}en}\ \emph {et~al.}(2026)\citenamefont {G{\"o}k{\c{c}}en}, \citenamefont {Akarsu},\ and\ \citenamefont {Di~Valentino}}]{Gokcen:2026pkq}%
  \BibitemOpen
  \bibfield  {author} {\bibinfo {author} {\bibfnamefont {M.}~\bibnamefont {G{\"o}k{\c{c}}en}}, \bibinfo {author} {\bibfnamefont {{\"O}.}~\bibnamefont {Akarsu}},\ and\ \bibinfo {author} {\bibfnamefont {E.}~\bibnamefont {Di~Valentino}},\ }\bibfield  {title} {\bibinfo {title} {{Revisiting CPL with sign-switching density: To cross or not to cross the NECB}},\ }\href {https://doi.org/10.1016/j.dark.2026.102273} {\bibfield  {journal} {\bibinfo  {journal} {Phys. Dark Univ.}\ }\textbf {\bibinfo {volume} {52}},\ \bibinfo {pages} {102273} (\bibinfo {year} {2026})},\ \Eprint {https://arxiv.org/abs/2602.21169} {2602.21169} \BibitemShut {NoStop}%
\bibitem [{\citenamefont {Jassal}\ \emph {et~al.}(2005{\natexlab{a}})\citenamefont {Jassal}, \citenamefont {Bagla},\ and\ \citenamefont {Padmanabhan}}]{Jassal:2004ej}%
  \BibitemOpen
  \bibfield  {author} {\bibinfo {author} {\bibfnamefont {H.~K.}\ \bibnamefont {Jassal}}, \bibinfo {author} {\bibfnamefont {J.~S.}\ \bibnamefont {Bagla}},\ and\ \bibinfo {author} {\bibfnamefont {T.}~\bibnamefont {Padmanabhan}},\ }\bibfield  {title} {\bibinfo {title} {{WMAP constraints on low redshift evolution of dark energy}},\ }\href {https://doi.org/10.1111/j.1745-3933.2005.08577.x} {\bibfield  {journal} {\bibinfo  {journal} {Mon. Not. Roy. Astron. Soc.}\ }\textbf {\bibinfo {volume} {356}},\ \bibinfo {pages} {L11} (\bibinfo {year} {2005}{\natexlab{a}})},\ \Eprint {https://arxiv.org/abs/astro-ph/0404378} {astro-ph/0404378} \BibitemShut {NoStop}%
\bibitem [{\citenamefont {Jassal}\ \emph {et~al.}(2005{\natexlab{b}})\citenamefont {Jassal}, \citenamefont {Bagla},\ and\ \citenamefont {Padmanabhan}}]{Jassal:2005qc}%
  \BibitemOpen
  \bibfield  {author} {\bibinfo {author} {\bibfnamefont {H.~K.}\ \bibnamefont {Jassal}}, \bibinfo {author} {\bibfnamefont {J.~S.}\ \bibnamefont {Bagla}},\ and\ \bibinfo {author} {\bibfnamefont {T.}~\bibnamefont {Padmanabhan}},\ }\bibfield  {title} {\bibinfo {title} {{Observational constraints on low redshift evolution of dark energy: How consistent are different observations?}},\ }\href {https://doi.org/10.1103/PhysRevD.72.103503} {\bibfield  {journal} {\bibinfo  {journal} {Phys. Rev. D}\ }\textbf {\bibinfo {volume} {72}},\ \bibinfo {pages} {103503} (\bibinfo {year} {2005}{\natexlab{b}})},\ \Eprint {https://arxiv.org/abs/astro-ph/0506748} {astro-ph/0506748} \BibitemShut {NoStop}%
\bibitem [{\citenamefont {Barboza}\ and\ \citenamefont {Alcaniz}(2008)}]{Barboza:2008rh}%
  \BibitemOpen
  \bibfield  {author} {\bibinfo {author} {\bibfnamefont {E.~M.}\ \bibnamefont {Barboza}, \bibfnamefont {Jr.}}\ and\ \bibinfo {author} {\bibfnamefont {J.~S.}\ \bibnamefont {Alcaniz}},\ }\bibfield  {title} {\bibinfo {title} {{A parametric model for dark energy}},\ }\href {https://doi.org/10.1016/j.physletb.2008.08.012} {\bibfield  {journal} {\bibinfo  {journal} {Phys. Lett. B}\ }\textbf {\bibinfo {volume} {666}},\ \bibinfo {pages} {415} (\bibinfo {year} {2008})},\ \Eprint {https://arxiv.org/abs/0805.1713} {0805.1713} \BibitemShut {NoStop}%
\bibitem [{\citenamefont {Pan}\ \emph {et~al.}(2020)\citenamefont {Pan}, \citenamefont {Yang},\ and\ \citenamefont {Paliathanasis}}]{Pan:2019brc}%
  \BibitemOpen
  \bibfield  {author} {\bibinfo {author} {\bibfnamefont {S.}~\bibnamefont {Pan}}, \bibinfo {author} {\bibfnamefont {W.}~\bibnamefont {Yang}},\ and\ \bibinfo {author} {\bibfnamefont {A.}~\bibnamefont {Paliathanasis}},\ }\bibfield  {title} {\bibinfo {title} {{Imprints of an extended Chevallier{\textendash}Polarski{\textendash}Linder parametrization on the large scale of our universe}},\ }\href {https://doi.org/10.1140/epjc/s10052-020-7832-y} {\bibfield  {journal} {\bibinfo  {journal} {Eur. Phys. J. C}\ }\textbf {\bibinfo {volume} {80}},\ \bibinfo {pages} {274} (\bibinfo {year} {2020})},\ \Eprint {https://arxiv.org/abs/1902.07108} {1902.07108} \BibitemShut {NoStop}%
\bibitem [{\citenamefont {Najafi}\ \emph {et~al.}(2024)\citenamefont {Najafi}, \citenamefont {Pan}, \citenamefont {Di~Valentino},\ and\ \citenamefont {Firouzjaee}}]{Najafi:2024qzm}%
  \BibitemOpen
  \bibfield  {author} {\bibinfo {author} {\bibfnamefont {M.}~\bibnamefont {Najafi}}, \bibinfo {author} {\bibfnamefont {S.}~\bibnamefont {Pan}}, \bibinfo {author} {\bibfnamefont {E.}~\bibnamefont {Di~Valentino}},\ and\ \bibinfo {author} {\bibfnamefont {J.~T.}\ \bibnamefont {Firouzjaee}},\ }\bibfield  {title} {\bibinfo {title} {{Dynamical dark energy confronted with multiple CMB missions}},\ }\href {https://doi.org/10.1016/j.dark.2024.101539} {\bibfield  {journal} {\bibinfo  {journal} {Phys. Dark Univ.}\ }\textbf {\bibinfo {volume} {45}},\ \bibinfo {pages} {101539} (\bibinfo {year} {2024})},\ \Eprint {https://arxiv.org/abs/2407.14939} {2407.14939} \BibitemShut {NoStop}%
\bibitem [{\citenamefont {Tripathi}\ \emph {et~al.}(2017)\citenamefont {Tripathi}, \citenamefont {Sangwan},\ and\ \citenamefont {Jassal}}]{Tripathi:2016slv}%
  \BibitemOpen
  \bibfield  {author} {\bibinfo {author} {\bibfnamefont {A.}~\bibnamefont {Tripathi}}, \bibinfo {author} {\bibfnamefont {A.}~\bibnamefont {Sangwan}},\ and\ \bibinfo {author} {\bibfnamefont {H.~K.}\ \bibnamefont {Jassal}},\ }\bibfield  {title} {\bibinfo {title} {{Dark energy equation of state parameter and its evolution at low redshift}},\ }\href {https://doi.org/10.1088/1475-7516/2017/06/012} {\bibfield  {journal} {\bibinfo  {journal} {J. Cosmol. Astropart. Phys.}\ }\textbf {\bibinfo {volume} {06}},\ \bibinfo {pages} {012} (\bibinfo {year} {2017})},\ \Eprint {https://arxiv.org/abs/1611.01899} {1611.01899} \BibitemShut {NoStop}%
\bibitem [{\citenamefont {Di~Valentino}\ \emph {et~al.}(2021{\natexlab{a}})\citenamefont {Di~Valentino}, \citenamefont {Mukherjee},\ and\ \citenamefont {Sen}}]{DiValentino:2020naf}%
  \BibitemOpen
  \bibfield  {author} {\bibinfo {author} {\bibfnamefont {E.}~\bibnamefont {Di~Valentino}}, \bibinfo {author} {\bibfnamefont {A.}~\bibnamefont {Mukherjee}},\ and\ \bibinfo {author} {\bibfnamefont {A.~A.}\ \bibnamefont {Sen}},\ }\bibfield  {title} {\bibinfo {title} {{Dark Energy with Phantom Crossing and the $H_0$ Tension}},\ }\href {https://doi.org/10.3390/e23040404} {\bibfield  {journal} {\bibinfo  {journal} {Entropy}\ }\textbf {\bibinfo {volume} {23}},\ \bibinfo {pages} {404} (\bibinfo {year} {2021}{\natexlab{a}})},\ \Eprint {https://arxiv.org/abs/2005.12587} {2005.12587} \BibitemShut {NoStop}%
\bibitem [{\citenamefont {Adil}\ \emph {et~al.}(2024)\citenamefont {Adil}, \citenamefont {Akarsu}, \citenamefont {Di~Valentino}, \citenamefont {Nunes}, \citenamefont {{\"O}z{\"u}lker}, \citenamefont {Sen},\ and\ \citenamefont {Specogna}}]{Adil:2023exv}%
  \BibitemOpen
  \bibfield  {author} {\bibinfo {author} {\bibfnamefont {S.~A.}\ \bibnamefont {Adil}}, \bibinfo {author} {\bibfnamefont {{\"O}.}~\bibnamefont {Akarsu}}, \bibinfo {author} {\bibfnamefont {E.}~\bibnamefont {Di~Valentino}}, \bibinfo {author} {\bibfnamefont {R.~C.}\ \bibnamefont {Nunes}}, \bibinfo {author} {\bibfnamefont {E.}~\bibnamefont {{\"O}z{\"u}lker}}, \bibinfo {author} {\bibfnamefont {A.~A.}\ \bibnamefont {Sen}},\ and\ \bibinfo {author} {\bibfnamefont {E.}~\bibnamefont {Specogna}},\ }\bibfield  {title} {\bibinfo {title} {{Omnipotent dark energy: A phenomenological answer to the Hubble tension}},\ }\href {https://doi.org/10.1103/PhysRevD.109.023527} {\bibfield  {journal} {\bibinfo  {journal} {Phys. Rev. D}\ }\textbf {\bibinfo {volume} {109}},\ \bibinfo {pages} {023527} (\bibinfo {year} {2024})},\ \Eprint {https://arxiv.org/abs/2306.08046} {2306.08046} \BibitemShut {NoStop}%
\bibitem [{\citenamefont {Montefalcone}\ and\ \citenamefont {Stiskalek}(2026)}]{Montefalcone:2026iga}%
  \BibitemOpen
  \bibfield  {author} {\bibinfo {author} {\bibfnamefont {G.}~\bibnamefont {Montefalcone}}\ and\ \bibinfo {author} {\bibfnamefont {R.}~\bibnamefont {Stiskalek}},\ }\href@noop {} {\bibinfo {title} {{Parameterizing Dark Energy at the density level: A two-parameter alternative to CPL}}} (\bibinfo {year} {2026}),\ \Eprint {https://arxiv.org/abs/2603.25735} {2603.25735} \BibitemShut {NoStop}%
\bibitem [{\citenamefont {Sen}(2008)}]{Sen:2007gk}%
  \BibitemOpen
  \bibfield  {author} {\bibinfo {author} {\bibfnamefont {A.~A.}\ \bibnamefont {Sen}},\ }\bibfield  {title} {\bibinfo {title} {{Deviation From LambdaCDM: Pressure Parametrization}},\ }\href {https://doi.org/10.1103/PhysRevD.77.043508} {\bibfield  {journal} {\bibinfo  {journal} {Phys. Rev. D}\ }\textbf {\bibinfo {volume} {77}},\ \bibinfo {pages} {043508} (\bibinfo {year} {2008})},\ \Eprint {https://arxiv.org/abs/0708.1072} {0708.1072} \BibitemShut {NoStop}%
\bibitem [{\citenamefont {Cheng}\ \emph {et~al.}(2025)\citenamefont {Cheng}, \citenamefont {Di~Valentino}, \citenamefont {Escamilla}, \citenamefont {Sen},\ and\ \citenamefont {Visinelli}}]{Cheng:2025lod}%
  \BibitemOpen
  \bibfield  {author} {\bibinfo {author} {\bibfnamefont {H.}~\bibnamefont {Cheng}}, \bibinfo {author} {\bibfnamefont {E.}~\bibnamefont {Di~Valentino}}, \bibinfo {author} {\bibfnamefont {L.~A.}\ \bibnamefont {Escamilla}}, \bibinfo {author} {\bibfnamefont {A.~A.}\ \bibnamefont {Sen}},\ and\ \bibinfo {author} {\bibfnamefont {L.}~\bibnamefont {Visinelli}},\ }\bibfield  {title} {\bibinfo {title} {{Pressure parametrization of dark energy: first and second-order constraints with latest cosmological data}},\ }\href {https://doi.org/10.1088/1475-7516/2025/09/031} {\bibfield  {journal} {\bibinfo  {journal} {J. Cosmol. Astropart. Phys.}\ }\textbf {\bibinfo {volume} {09}},\ \bibinfo {pages} {031} (\bibinfo {year} {2025})},\ \Eprint {https://arxiv.org/abs/2505.02932} {2505.02932} \BibitemShut {NoStop}%
\bibitem [{\citenamefont {Acquaviva}\ \emph {et~al.}(2021)\citenamefont {Acquaviva}, \citenamefont {Akarsu}, \citenamefont {Katirci},\ and\ \citenamefont {Vazquez}}]{Acquaviva:2021jov}%
  \BibitemOpen
  \bibfield  {author} {\bibinfo {author} {\bibfnamefont {G.}~\bibnamefont {Acquaviva}}, \bibinfo {author} {\bibfnamefont {{\"O}.}~\bibnamefont {Akarsu}}, \bibinfo {author} {\bibfnamefont {N.}~\bibnamefont {Katirci}},\ and\ \bibinfo {author} {\bibfnamefont {J.~A.}\ \bibnamefont {Vazquez}},\ }\bibfield  {title} {\bibinfo {title} {{Simple-graduated dark energy and spatial curvature}},\ }\href {https://doi.org/10.1103/PhysRevD.104.023505} {\bibfield  {journal} {\bibinfo  {journal} {Phys. Rev. D}\ }\textbf {\bibinfo {volume} {104}},\ \bibinfo {pages} {023505} (\bibinfo {year} {2021})},\ \Eprint {https://arxiv.org/abs/2104.02623} {2104.02623} \BibitemShut {NoStop}%
\bibitem [{\citenamefont {Escamilla}\ \emph {et~al.}(2026)\citenamefont {Escamilla}, \citenamefont {Karadavut},\ and\ \citenamefont {Kat{\i}rc{\i}}}]{Escamilla:2026eks}%
  \BibitemOpen
  \bibfield  {author} {\bibinfo {author} {\bibfnamefont {L.~A.}\ \bibnamefont {Escamilla}}, \bibinfo {author} {\bibfnamefont {B.}~\bibnamefont {Karadavut}},\ and\ \bibinfo {author} {\bibfnamefont {N.}~\bibnamefont {Kat{\i}rc{\i}}},\ }\href@noop {} {\bibinfo {title} {{Dark Energy with Constant Inertial Mass Density: Updated Constraints and Curvature-Induced Sign Transitions in $\rho_{\rm DE}$ and $\rho_{\rm DE}+p_{\rm DE}$}}} (\bibinfo {year} {2026}),\ \Eprint {https://arxiv.org/abs/2603.15868} {2603.15868} \BibitemShut {NoStop}%
\bibitem [{\citenamefont {Escamilla}\ \emph {et~al.}(2023)\citenamefont {Escamilla}, \citenamefont {Akarsu}, \citenamefont {Di~Valentino},\ and\ \citenamefont {Vazquez}}]{Escamilla:2023shf}%
  \BibitemOpen
  \bibfield  {author} {\bibinfo {author} {\bibfnamefont {L.~A.}\ \bibnamefont {Escamilla}}, \bibinfo {author} {\bibfnamefont {O.}~\bibnamefont {Akarsu}}, \bibinfo {author} {\bibfnamefont {E.}~\bibnamefont {Di~Valentino}},\ and\ \bibinfo {author} {\bibfnamefont {J.~A.}\ \bibnamefont {Vazquez}},\ }\bibfield  {title} {\bibinfo {title} {{Model-independent reconstruction of the interacting dark energy kernel: Binned and Gaussian process}},\ }\href {https://doi.org/10.1088/1475-7516/2023/11/051} {\bibfield  {journal} {\bibinfo  {journal} {JCAP}\ }\textbf {\bibinfo {volume} {11}},\ \bibinfo {pages} {051} (\bibinfo {year} {2023})},\ \Eprint {https://arxiv.org/abs/2305.16290} {2305.16290} \BibitemShut {NoStop}%
\bibitem [{\citenamefont {Sabogal}\ \emph {et~al.}(2024)\citenamefont {Sabogal}, \citenamefont {Akarsu}, \citenamefont {Bonilla}, \citenamefont {Di~Valentino},\ and\ \citenamefont {Nunes}}]{Sabogal:2024qxs}%
  \BibitemOpen
  \bibfield  {author} {\bibinfo {author} {\bibfnamefont {M.~A.}\ \bibnamefont {Sabogal}}, \bibinfo {author} {\bibfnamefont {{\"O}.}~\bibnamefont {Akarsu}}, \bibinfo {author} {\bibfnamefont {A.}~\bibnamefont {Bonilla}}, \bibinfo {author} {\bibfnamefont {E.}~\bibnamefont {Di~Valentino}},\ and\ \bibinfo {author} {\bibfnamefont {R.~C.}\ \bibnamefont {Nunes}},\ }\bibfield  {title} {\bibinfo {title} {{Exploring new physics in the late Universe{\textquoteright}s expansion through non-parametric inference}},\ }\href {https://doi.org/10.1140/epjc/s10052-024-13081-1} {\bibfield  {journal} {\bibinfo  {journal} {Eur. Phys. J. C}\ }\textbf {\bibinfo {volume} {84}},\ \bibinfo {pages} {703} (\bibinfo {year} {2024})},\ \Eprint {https://arxiv.org/abs/2407.04223} {2407.04223} \BibitemShut {NoStop}%
\bibitem [{\citenamefont {Verde}\ \emph {et~al.}(2019)\citenamefont {Verde}, \citenamefont {Treu},\ and\ \citenamefont {Riess}}]{Verde:2019ivm}%
  \BibitemOpen
  \bibfield  {author} {\bibinfo {author} {\bibfnamefont {L.}~\bibnamefont {Verde}}, \bibinfo {author} {\bibfnamefont {T.}~\bibnamefont {Treu}},\ and\ \bibinfo {author} {\bibfnamefont {A.~G.}\ \bibnamefont {Riess}},\ }\bibfield  {title} {\bibinfo {title} {{Tensions between the Early and the Late Universe}},\ }\href {https://doi.org/10.1038/s41550-019-0902-0} {\bibfield  {journal} {\bibinfo  {journal} {Nature Astron.}\ }\textbf {\bibinfo {volume} {3}},\ \bibinfo {pages} {891} (\bibinfo {year} {2019})},\ \Eprint {https://arxiv.org/abs/1907.10625} {1907.10625} \BibitemShut {NoStop}%
\bibitem [{\citenamefont {Di~Valentino}\ \emph {et~al.}(2021{\natexlab{b}})\citenamefont {Di~Valentino} \emph {et~al.}}]{DiValentino:2020zio}%
  \BibitemOpen
  \bibfield  {author} {\bibinfo {author} {\bibfnamefont {E.}~\bibnamefont {Di~Valentino}} \emph {et~al.},\ }\bibfield  {title} {\bibinfo {title} {{Snowmass2021 - Letter of interest cosmology intertwined II: The hubble constant tension}},\ }\href {https://doi.org/10.1016/j.astropartphys.2021.102605} {\bibfield  {journal} {\bibinfo  {journal} {Astropart. Phys.}\ }\textbf {\bibinfo {volume} {131}},\ \bibinfo {pages} {102605} (\bibinfo {year} {2021}{\natexlab{b}})},\ \Eprint {https://arxiv.org/abs/2008.11284} {2008.11284} \BibitemShut {NoStop}%
\bibitem [{\citenamefont {Riess}\ \emph {et~al.}(2022)\citenamefont {Riess} \emph {et~al.}}]{Riess:2021jrx}%
  \BibitemOpen
  \bibfield  {author} {\bibinfo {author} {\bibfnamefont {A.~G.}\ \bibnamefont {Riess}} \emph {et~al.},\ }\bibfield  {title} {\bibinfo {title} {{A Comprehensive Measurement of the Local Value of the Hubble Constant with 1 km s$^{-1}$ Mpc$^{-1}$ Uncertainty from the Hubble Space Telescope and the SH0ES Team}},\ }\href {https://doi.org/10.3847/2041-8213/ac5c5b} {\bibfield  {journal} {\bibinfo  {journal} {Astrophys. J. Lett.}\ }\textbf {\bibinfo {volume} {934}},\ \bibinfo {pages} {L7} (\bibinfo {year} {2022})},\ \Eprint {https://arxiv.org/abs/2112.04510} {2112.04510} \BibitemShut {NoStop}%
\bibitem [{\citenamefont {Alam}\ \emph {et~al.}(2021)\citenamefont {Alam} \emph {et~al.}}]{eBOSS:2020yzd}%
  \BibitemOpen
  \bibfield  {author} {\bibinfo {author} {\bibfnamefont {S.}~\bibnamefont {Alam}} \emph {et~al.} (\bibinfo {collaboration} {eBOSS Collaboration}),\ }\bibfield  {title} {\bibinfo {title} {{Completed SDSS-IV extended Baryon Oscillation Spectroscopic Survey: Cosmological implications from two decades of spectroscopic surveys at the Apache Point Observatory}},\ }\href {https://doi.org/10.1103/PhysRevD.103.083533} {\bibfield  {journal} {\bibinfo  {journal} {Phys. Rev. D}\ }\textbf {\bibinfo {volume} {103}},\ \bibinfo {pages} {083533} (\bibinfo {year} {2021})},\ \Eprint {https://arxiv.org/abs/2007.08991} {2007.08991} \BibitemShut {NoStop}%
\bibitem [{\citenamefont {Di~Valentino}\ \emph {et~al.}(2021{\natexlab{c}})\citenamefont {Di~Valentino}, \citenamefont {Mena}, \citenamefont {Pan}, \citenamefont {Visinelli}, \citenamefont {Yang}, \citenamefont {Melchiorri}, \citenamefont {Mota}, \citenamefont {Riess},\ and\ \citenamefont {Silk}}]{DiValentino:2021izs}%
  \BibitemOpen
  \bibfield  {author} {\bibinfo {author} {\bibfnamefont {E.}~\bibnamefont {Di~Valentino}}, \bibinfo {author} {\bibfnamefont {O.}~\bibnamefont {Mena}}, \bibinfo {author} {\bibfnamefont {S.}~\bibnamefont {Pan}}, \bibinfo {author} {\bibfnamefont {L.}~\bibnamefont {Visinelli}}, \bibinfo {author} {\bibfnamefont {W.}~\bibnamefont {Yang}}, \bibinfo {author} {\bibfnamefont {A.}~\bibnamefont {Melchiorri}}, \bibinfo {author} {\bibfnamefont {D.~F.}\ \bibnamefont {Mota}}, \bibinfo {author} {\bibfnamefont {A.~G.}\ \bibnamefont {Riess}},\ and\ \bibinfo {author} {\bibfnamefont {J.}~\bibnamefont {Silk}},\ }\bibfield  {title} {\bibinfo {title} {{In the realm of the Hubble tension: a review of solutions}},\ }\href {https://doi.org/10.1088/1361-6382/ac086d} {\bibfield  {journal} {\bibinfo  {journal} {Class. Quant. Grav.}\ }\textbf {\bibinfo {volume} {38}},\ \bibinfo {pages} {153001} (\bibinfo {year} {2021}{\natexlab{c}})},\ \Eprint {https://arxiv.org/abs/2103.01183} {2103.01183} \BibitemShut {NoStop}%
\bibitem [{\citenamefont {Perivolaropoulos}\ and\ \citenamefont {Skara}(2022)}]{Perivolaropoulos:2021jda}%
  \BibitemOpen
  \bibfield  {author} {\bibinfo {author} {\bibfnamefont {L.}~\bibnamefont {Perivolaropoulos}}\ and\ \bibinfo {author} {\bibfnamefont {F.}~\bibnamefont {Skara}},\ }\bibfield  {title} {\bibinfo {title} {{Challenges for \ensuremath{\Lambda}CDM: An update}},\ }\href {https://doi.org/10.1016/j.newar.2022.101659} {\bibfield  {journal} {\bibinfo  {journal} {New Astron. Rev.}\ }\textbf {\bibinfo {volume} {95}},\ \bibinfo {pages} {101659} (\bibinfo {year} {2022})},\ \Eprint {https://arxiv.org/abs/2105.05208} {2105.05208} \BibitemShut {NoStop}%
\bibitem [{\citenamefont {Abdalla}\ \emph {et~al.}(2022)\citenamefont {Abdalla} \emph {et~al.}}]{Abdalla:2022yfr}%
  \BibitemOpen
  \bibfield  {author} {\bibinfo {author} {\bibfnamefont {E.}~\bibnamefont {Abdalla}} \emph {et~al.},\ }\bibfield  {title} {\bibinfo {title} {{Cosmology intertwined: A review of the particle physics, astrophysics, and cosmology associated with the cosmological tensions and anomalies}},\ }\href {https://doi.org/10.1016/j.jheap.2022.04.002} {\bibfield  {journal} {\bibinfo  {journal} {J. High Energy Astrophys.}\ }\textbf {\bibinfo {volume} {34}},\ \bibinfo {pages} {49} (\bibinfo {year} {2022})},\ \Eprint {https://arxiv.org/abs/2203.06142} {2203.06142} \BibitemShut {NoStop}%
\bibitem [{\citenamefont {Akarsu}\ \emph {et~al.}(2024{\natexlab{a}})\citenamefont {Akarsu}, \citenamefont {Colg{\'a}in}, \citenamefont {Sen},\ and\ \citenamefont {Sheikh-Jabbari}}]{Akarsu:2024qiq}%
  \BibitemOpen
  \bibfield  {author} {\bibinfo {author} {\bibfnamefont {{\"O}.}~\bibnamefont {Akarsu}}, \bibinfo {author} {\bibfnamefont {E.~{\'O}.}\ \bibnamefont {Colg{\'a}in}}, \bibinfo {author} {\bibfnamefont {A.~A.}\ \bibnamefont {Sen}},\ and\ \bibinfo {author} {\bibfnamefont {M.~M.}\ \bibnamefont {Sheikh-Jabbari}},\ }\bibfield  {title} {\bibinfo {title} {{{\ensuremath{\Lambda}}CDM Tensions: Localising Missing Physics through Consistency Checks}},\ }\href {https://doi.org/10.3390/universe10080305} {\bibfield  {journal} {\bibinfo  {journal} {Universe}\ }\textbf {\bibinfo {volume} {10}},\ \bibinfo {pages} {305} (\bibinfo {year} {2024}{\natexlab{a}})},\ \Eprint {https://arxiv.org/abs/2402.04767} {2402.04767} \BibitemShut {NoStop}%
\bibitem [{\citenamefont {Di~Valentino}\ \emph {et~al.}(2025)\citenamefont {Di~Valentino} \emph {et~al.}}]{CosmoVerse:2025wtd}%
  \BibitemOpen
  \bibfield  {author} {\bibinfo {author} {\bibfnamefont {E.}~\bibnamefont {Di~Valentino}} \emph {et~al.} (\bibinfo {collaboration} {CosmoVerse Network}),\ }\bibfield  {title} {\bibinfo {title} {{The CosmoVerse White Paper: Addressing observational tensions in cosmology with systematics and fundamental physics}},\ }\href {https://doi.org/10.1016/j.dark.2025.101965} {\bibfield  {journal} {\bibinfo  {journal} {Phys. Dark Univ.}\ }\textbf {\bibinfo {volume} {49}},\ \bibinfo {pages} {101965} (\bibinfo {year} {2025})},\ \Eprint {https://arxiv.org/abs/2504.01669} {2504.01669} \BibitemShut {NoStop}%
\bibitem [{\citenamefont {Aghanim}\ \emph {et~al.}(2020)\citenamefont {Aghanim} \emph {et~al.}}]{Planck:2018vyg}%
  \BibitemOpen
  \bibfield  {author} {\bibinfo {author} {\bibfnamefont {N.}~\bibnamefont {Aghanim}} \emph {et~al.} (\bibinfo {collaboration} {Planck}),\ }\bibfield  {title} {\bibinfo {title} {{Planck 2018 results. VI. Cosmological parameters}},\ }\href {https://doi.org/10.1051/0004-6361/201833910} {\bibfield  {journal} {\bibinfo  {journal} {Astron. Astrophys.}\ }\textbf {\bibinfo {volume} {641}},\ \bibinfo {pages} {A6} (\bibinfo {year} {2020})},\ \bibinfo {note} {[Erratum: Astron. Astrophys. 652, C4 (2021)]},\ \Eprint {https://arxiv.org/abs/1807.06209} {1807.06209} \BibitemShut {NoStop}%
\bibitem [{\citenamefont {Akarsu}\ \emph {et~al.}(2025{\natexlab{a}})\citenamefont {Akarsu}, \citenamefont {{\c{C}}am}, \citenamefont {Paraskevas},\ and\ \citenamefont {Perivolaropoulos}}]{Akarsu:2025ijk}%
  \BibitemOpen
  \bibfield  {author} {\bibinfo {author} {\bibfnamefont {{\"O}.}~\bibnamefont {Akarsu}}, \bibinfo {author} {\bibfnamefont {A.}~\bibnamefont {{\c{C}}am}}, \bibinfo {author} {\bibfnamefont {E.~A.}\ \bibnamefont {Paraskevas}},\ and\ \bibinfo {author} {\bibfnamefont {L.}~\bibnamefont {Perivolaropoulos}},\ }\bibfield  {title} {\bibinfo {title} {{Linear matter density perturbations in the {\ensuremath{\Lambda}}$_{s}$CDM model: Examining growth dynamics and addressing the $S_{8}$ tension}},\ }\href {https://doi.org/10.1088/1475-7516/2025/08/089} {\bibfield  {journal} {\bibinfo  {journal} {J. Cosmol. Astropart. Phys.}\ }\textbf {\bibinfo {volume} {2025}}\bibfield  {number} {\bibinfo  {number} { (08)},\ \bibinfo {pages} {089}} (\bibinfo {year} {2025}{\natexlab{a}})},\ \Eprint {https://arxiv.org/abs/2502.20384} {2502.20384} \BibitemShut {NoStop}%
\bibitem [{\citenamefont {Akarsu}\ \emph {et~al.}(2021)\citenamefont {Akarsu}, \citenamefont {Kumar}, \citenamefont {{\"O}z{\"u}lker},\ and\ \citenamefont {Vazquez}}]{Akarsu:2021fol}%
  \BibitemOpen
  \bibfield  {author} {\bibinfo {author} {\bibfnamefont {{\"O}.}~\bibnamefont {Akarsu}}, \bibinfo {author} {\bibfnamefont {S.}~\bibnamefont {Kumar}}, \bibinfo {author} {\bibfnamefont {E.}~\bibnamefont {{\"O}z{\"u}lker}},\ and\ \bibinfo {author} {\bibfnamefont {J.~A.}\ \bibnamefont {Vazquez}},\ }\bibfield  {title} {\bibinfo {title} {{Relaxing cosmological tensions with a sign switching cosmological constant}},\ }\href {https://doi.org/10.1103/PhysRevD.104.123512} {\bibfield  {journal} {\bibinfo  {journal} {Phys. Rev. D}\ }\textbf {\bibinfo {volume} {104}},\ \bibinfo {pages} {123512} (\bibinfo {year} {2021})},\ \Eprint {https://arxiv.org/abs/2108.09239} {2108.09239} \BibitemShut {NoStop}%
\bibitem [{\citenamefont {Akarsu}\ \emph {et~al.}(2023{\natexlab{a}})\citenamefont {Akarsu}, \citenamefont {Kumar}, \citenamefont {{\"O}z{\"u}lker}, \citenamefont {Vazquez},\ and\ \citenamefont {Yadav}}]{Akarsu:2022typ}%
  \BibitemOpen
  \bibfield  {author} {\bibinfo {author} {\bibfnamefont {{\"O}.}~\bibnamefont {Akarsu}}, \bibinfo {author} {\bibfnamefont {S.}~\bibnamefont {Kumar}}, \bibinfo {author} {\bibfnamefont {E.}~\bibnamefont {{\"O}z{\"u}lker}}, \bibinfo {author} {\bibfnamefont {J.~A.}\ \bibnamefont {Vazquez}},\ and\ \bibinfo {author} {\bibfnamefont {A.}~\bibnamefont {Yadav}},\ }\bibfield  {title} {\bibinfo {title} {{Relaxing cosmological tensions with a sign switching cosmological constant: Improved results with Planck, BAO, and Pantheon data}},\ }\href {https://doi.org/10.1103/PhysRevD.108.023513} {\bibfield  {journal} {\bibinfo  {journal} {Phys. Rev. D}\ }\textbf {\bibinfo {volume} {108}},\ \bibinfo {pages} {023513} (\bibinfo {year} {2023}{\natexlab{a}})},\ \Eprint {https://arxiv.org/abs/2211.05742} {2211.05742} \BibitemShut {NoStop}%
\bibitem [{\citenamefont {Akarsu}\ \emph {et~al.}(2023{\natexlab{b}})\citenamefont {Akarsu}, \citenamefont {Di~Valentino}, \citenamefont {Kumar}, \citenamefont {Nunes}, \citenamefont {Vazquez},\ and\ \citenamefont {Yadav}}]{Akarsu:2023mfb}%
  \BibitemOpen
  \bibfield  {author} {\bibinfo {author} {\bibfnamefont {{\"O}.}~\bibnamefont {Akarsu}}, \bibinfo {author} {\bibfnamefont {E.}~\bibnamefont {Di~Valentino}}, \bibinfo {author} {\bibfnamefont {S.}~\bibnamefont {Kumar}}, \bibinfo {author} {\bibfnamefont {R.~C.}\ \bibnamefont {Nunes}}, \bibinfo {author} {\bibfnamefont {J.~A.}\ \bibnamefont {Vazquez}},\ and\ \bibinfo {author} {\bibfnamefont {A.}~\bibnamefont {Yadav}},\ }\href@noop {} {\bibinfo {title} {{$\Lambda_{\rm s}$CDM model: A promising scenario for alleviation of cosmological tensions}}} (\bibinfo {year} {2023}{\natexlab{b}}),\ \Eprint {https://arxiv.org/abs/2307.10899} {2307.10899} \BibitemShut {NoStop}%
\bibitem [{\citenamefont {Akarsu}\ \emph {et~al.}(2026{\natexlab{c}})\citenamefont {Akarsu}, \citenamefont {De~Felice}, \citenamefont {Di~Valentino}, \citenamefont {Kumar}, \citenamefont {Nunes}, \citenamefont {{\"O}z{\"u}lker}, \citenamefont {Vazquez},\ and\ \citenamefont {Yadav}}]{Akarsu:2024qsi}%
  \BibitemOpen
  \bibfield  {author} {\bibinfo {author} {\bibfnamefont {{\"O}.}~\bibnamefont {Akarsu}}, \bibinfo {author} {\bibfnamefont {A.}~\bibnamefont {De~Felice}}, \bibinfo {author} {\bibfnamefont {E.}~\bibnamefont {Di~Valentino}}, \bibinfo {author} {\bibfnamefont {S.}~\bibnamefont {Kumar}}, \bibinfo {author} {\bibfnamefont {R.~C.}\ \bibnamefont {Nunes}}, \bibinfo {author} {\bibfnamefont {E.}~\bibnamefont {{\"O}z{\"u}lker}}, \bibinfo {author} {\bibfnamefont {J.~A.}\ \bibnamefont {Vazquez}},\ and\ \bibinfo {author} {\bibfnamefont {A.}~\bibnamefont {Yadav}},\ }\bibfield  {title} {\bibinfo {title} {{$\Lambda_{\rm s}$CDM cosmology from a type-II minimally modified gravity}},\ }\href {https://doi.org/10.1093/mnras/staf2276} {\bibfield  {journal} {\bibinfo  {journal} {Mon. Not. Roy. Astron. Soc.}\ }\textbf {\bibinfo {volume} {546}},\ \bibinfo {pages} {staf2276} (\bibinfo {year} {2026}{\natexlab{c}})},\ \Eprint {https://arxiv.org/abs/2402.07716} {2402.07716} \BibitemShut {NoStop}%
\bibitem [{\citenamefont {Akarsu}\ \emph {et~al.}(2024{\natexlab{b}})\citenamefont {Akarsu}, \citenamefont {De~Felice}, \citenamefont {Di~Valentino}, \citenamefont {Kumar}, \citenamefont {Nunes}, \citenamefont {{\"O}z{\"u}lker}, \citenamefont {Vazquez},\ and\ \citenamefont {Yadav}}]{Akarsu:2024eoo}%
  \BibitemOpen
  \bibfield  {author} {\bibinfo {author} {\bibfnamefont {{\"O}.}~\bibnamefont {Akarsu}}, \bibinfo {author} {\bibfnamefont {A.}~\bibnamefont {De~Felice}}, \bibinfo {author} {\bibfnamefont {E.}~\bibnamefont {Di~Valentino}}, \bibinfo {author} {\bibfnamefont {S.}~\bibnamefont {Kumar}}, \bibinfo {author} {\bibfnamefont {R.~C.}\ \bibnamefont {Nunes}}, \bibinfo {author} {\bibfnamefont {E.}~\bibnamefont {{\"O}z{\"u}lker}}, \bibinfo {author} {\bibfnamefont {J.~A.}\ \bibnamefont {Vazquez}},\ and\ \bibinfo {author} {\bibfnamefont {A.}~\bibnamefont {Yadav}},\ }\bibfield  {title} {\bibinfo {title} {{Cosmological constraints on {\ensuremath{\Lambda}}sCDM scenario in a type II minimally modified gravity}},\ }\href {https://doi.org/10.1103/PhysRevD.110.103527} {\bibfield  {journal} {\bibinfo  {journal} {Phys. Rev. D}\ }\textbf {\bibinfo {volume} {110}},\ \bibinfo {pages} {103527} (\bibinfo {year} {2024}{\natexlab{b}})},\ \Eprint {https://arxiv.org/abs/2406.07526} {2406.07526} \BibitemShut {NoStop}%
\bibitem [{\citenamefont {Escamilla}\ \emph {et~al.}(2025)\citenamefont {Escamilla}, \citenamefont {Akarsu}, \citenamefont {Di~Valentino}, \citenamefont {{\"O}z{\"u}lker},\ and\ \citenamefont {Vazquez}}]{Escamilla:2025imi}%
  \BibitemOpen
  \bibfield  {author} {\bibinfo {author} {\bibfnamefont {L.~A.}\ \bibnamefont {Escamilla}}, \bibinfo {author} {\bibfnamefont {{\"O}.}~\bibnamefont {Akarsu}}, \bibinfo {author} {\bibfnamefont {E.}~\bibnamefont {Di~Valentino}}, \bibinfo {author} {\bibfnamefont {E.}~\bibnamefont {{\"O}z{\"u}lker}},\ and\ \bibinfo {author} {\bibfnamefont {J.~A.}\ \bibnamefont {Vazquez}},\ }\href@noop {} {\bibinfo {title} {{Exploring the Growth-Index ($\gamma$) Tension with $\Lambda_{\rm s}$CDM}}} (\bibinfo {year} {2025}),\ \Eprint {https://arxiv.org/abs/2503.12945} {2503.12945} \BibitemShut {NoStop}%
\bibitem [{\citenamefont {Paraskevas}\ \emph {et~al.}(2024)\citenamefont {Paraskevas}, \citenamefont {\c{C}am}, \citenamefont {Perivolaropoulos},\ and\ \citenamefont {Akarsu}}]{Paraskevas:2024ytz}%
  \BibitemOpen
  \bibfield  {author} {\bibinfo {author} {\bibfnamefont {E.~A.}\ \bibnamefont {Paraskevas}}, \bibinfo {author} {\bibfnamefont {A.}~\bibnamefont {\c{C}am}}, \bibinfo {author} {\bibfnamefont {L.}~\bibnamefont {Perivolaropoulos}},\ and\ \bibinfo {author} {\bibfnamefont {{\"O}.}~\bibnamefont {Akarsu}},\ }\bibfield  {title} {\bibinfo {title} {{Transition dynamics in the {\ensuremath{\Lambda}}sCDM model: Implications for bound cosmic structures}},\ }\href {https://doi.org/10.1103/PhysRevD.109.103522} {\bibfield  {journal} {\bibinfo  {journal} {Phys. Rev. D}\ }\textbf {\bibinfo {volume} {109}},\ \bibinfo {pages} {103522} (\bibinfo {year} {2024})},\ \Eprint {https://arxiv.org/abs/2402.05908} {2402.05908} \BibitemShut {NoStop}%
\bibitem [{\citenamefont {Akarsu}\ \emph {et~al.}(2026{\natexlab{d}})\citenamefont {Akarsu}, \citenamefont {Di~Valentino}, \citenamefont {Vysko{\v{c}}il}, \citenamefont {Y{\i}lmaz}, \citenamefont {Y{\"u}kselci},\ and\ \citenamefont {Zhuk}}]{Akarsu:2025nns}%
  \BibitemOpen
  \bibfield  {author} {\bibinfo {author} {\bibfnamefont {{\"O}.}~\bibnamefont {Akarsu}}, \bibinfo {author} {\bibfnamefont {E.}~\bibnamefont {Di~Valentino}}, \bibinfo {author} {\bibfnamefont {J.}~\bibnamefont {Vysko{\v{c}}il}}, \bibinfo {author} {\bibfnamefont {E.}~\bibnamefont {Y{\i}lmaz}}, \bibinfo {author} {\bibfnamefont {A.~E.}\ \bibnamefont {Y{\"u}kselci}},\ and\ \bibinfo {author} {\bibfnamefont {A.}~\bibnamefont {Zhuk}},\ }\bibfield  {title} {\bibinfo {title} {{Nonlinear matter power spectrum from relativistic N-body simulations: {\ensuremath{\Lambda}}sCDM versus {\ensuremath{\Lambda}}CDM}},\ }\href {https://doi.org/10.1103/lmt4-hshz} {\bibfield  {journal} {\bibinfo  {journal} {Phys. Rev. D}\ }\textbf {\bibinfo {volume} {113}},\ \bibinfo {pages} {083508} (\bibinfo {year} {2026}{\natexlab{d}})},\ \Eprint {https://arxiv.org/abs/2510.18741} {2510.18741} \BibitemShut {NoStop}%
\bibitem [{\citenamefont {Yadav}\ \emph {et~al.}(2025)\citenamefont {Yadav}, \citenamefont {Kumar}, \citenamefont {K{\i}br{\i}s},\ and\ \citenamefont {Akarsu}}]{Yadav:2024duq}%
  \BibitemOpen
  \bibfield  {author} {\bibinfo {author} {\bibfnamefont {A.}~\bibnamefont {Yadav}}, \bibinfo {author} {\bibfnamefont {S.}~\bibnamefont {Kumar}}, \bibinfo {author} {\bibfnamefont {C.}~\bibnamefont {K{\i}br{\i}s}},\ and\ \bibinfo {author} {\bibfnamefont {{\"O}.}~\bibnamefont {Akarsu}},\ }\bibfield  {title} {\bibinfo {title} {{{\ensuremath{\Lambda}}$_{s}$CDM cosmology: alleviating major cosmological tensions by predicting standard neutrino properties}},\ }\href {https://doi.org/10.1088/1475-7516/2025/01/042} {\bibfield  {journal} {\bibinfo  {journal} {J. Cosmol. Astropart. Phys.}\ }\textbf {\bibinfo {volume} {2025}}\bibfield  {number} {\bibinfo  {number} { (01)},\ \bibinfo {pages} {042}} (\bibinfo {year} {2025})},\ \Eprint {https://arxiv.org/abs/2406.18496} {2406.18496} \BibitemShut {NoStop}%
\bibitem [{\citenamefont {K{\i}br{\i}s}\ \emph {et~al.}(2026)\citenamefont {K{\i}br{\i}s}, \citenamefont {Elbers}, \citenamefont {Akarsu},\ and\ \citenamefont {Di~Valentino}}]{Kibris:2026cqq}%
  \BibitemOpen
  \bibfield  {author} {\bibinfo {author} {\bibfnamefont {C.}~\bibnamefont {K{\i}br{\i}s}}, \bibinfo {author} {\bibfnamefont {W.}~\bibnamefont {Elbers}}, \bibinfo {author} {\bibfnamefont {{\"O}.}~\bibnamefont {Akarsu}},\ and\ \bibinfo {author} {\bibfnamefont {E.}~\bibnamefont {Di~Valentino}},\ }\href@noop {} {\bibinfo {title} {{Negative neutrino mass or negative dark energy?}}} (\bibinfo {year} {2026}),\ \Eprint {https://arxiv.org/abs/2605.21456} {2605.21456} \BibitemShut {NoStop}%
\bibitem [{\citenamefont {Souza}\ \emph {et~al.}(2025)\citenamefont {Souza}, \citenamefont {Barcelos}, \citenamefont {Nunes}, \citenamefont {Akarsu},\ and\ \citenamefont {Kumar}}]{Souza:2024qwd}%
  \BibitemOpen
  \bibfield  {author} {\bibinfo {author} {\bibfnamefont {M.~S.}\ \bibnamefont {Souza}}, \bibinfo {author} {\bibfnamefont {A.~M.}\ \bibnamefont {Barcelos}}, \bibinfo {author} {\bibfnamefont {R.~C.}\ \bibnamefont {Nunes}}, \bibinfo {author} {\bibfnamefont {{\"O}.}~\bibnamefont {Akarsu}},\ and\ \bibinfo {author} {\bibfnamefont {S.}~\bibnamefont {Kumar}},\ }\bibfield  {title} {\bibinfo {title} {{Mapping the {\ensuremath{\Lambda}}$_{s}$CDM Scenario to f(T) Modified Gravity: Effects on Structure Growth Rate}},\ }\href {https://doi.org/10.3390/universe11010002} {\bibfield  {journal} {\bibinfo  {journal} {Universe}\ }\textbf {\bibinfo {volume} {11}},\ \bibinfo {pages} {2} (\bibinfo {year} {2025})},\ \Eprint {https://arxiv.org/abs/2501.18031} {2501.18031} \BibitemShut {NoStop}%
\bibitem [{\citenamefont {Awad}\ \emph {et~al.}(2018)\citenamefont {Awad}, \citenamefont {El~Hanafy}, \citenamefont {Nashed},\ and\ \citenamefont {Saridakis}}]{Awad:2017yod}%
  \BibitemOpen
  \bibfield  {author} {\bibinfo {author} {\bibfnamefont {A.}~\bibnamefont {Awad}}, \bibinfo {author} {\bibfnamefont {W.}~\bibnamefont {El~Hanafy}}, \bibinfo {author} {\bibfnamefont {G.~G.~L.}\ \bibnamefont {Nashed}},\ and\ \bibinfo {author} {\bibfnamefont {E.~N.}\ \bibnamefont {Saridakis}},\ }\bibfield  {title} {\bibinfo {title} {{Phase Portraits of general f(T) Cosmology}},\ }\href {https://doi.org/10.1088/1475-7516/2018/02/052} {\bibfield  {journal} {\bibinfo  {journal} {J. Cosmol. Astropart. Phys.}\ }\textbf {\bibinfo {volume} {02}},\ \bibinfo {pages} {052} (\bibinfo {year} {2018})},\ \Eprint {https://arxiv.org/abs/1710.10194} {1710.10194} \BibitemShut {NoStop}%
\bibitem [{\citenamefont {Hashim}\ \emph {et~al.}(2021{\natexlab{a}})\citenamefont {Hashim}, \citenamefont {El~Hanafy}, \citenamefont {Golovnev},\ and\ \citenamefont {El-Zant}}]{Hashim:2020sez}%
  \BibitemOpen
  \bibfield  {author} {\bibinfo {author} {\bibfnamefont {M.}~\bibnamefont {Hashim}}, \bibinfo {author} {\bibfnamefont {W.}~\bibnamefont {El~Hanafy}}, \bibinfo {author} {\bibfnamefont {A.}~\bibnamefont {Golovnev}},\ and\ \bibinfo {author} {\bibfnamefont {A.~A.}\ \bibnamefont {El-Zant}},\ }\bibfield  {title} {\bibinfo {title} {{Toward a concordance teleparallel cosmology. Part I. Background dynamics}},\ }\href {https://doi.org/10.1088/1475-7516/2021/07/052} {\bibfield  {journal} {\bibinfo  {journal} {J. Cosmol. Astropart. Phys.}\ }\textbf {\bibinfo {volume} {07}},\ \bibinfo {pages} {052} (\bibinfo {year} {2021}{\natexlab{a}})},\ \Eprint {https://arxiv.org/abs/2010.14964} {2010.14964} \BibitemShut {NoStop}%
\bibitem [{\citenamefont {Hashim}\ \emph {et~al.}(2021{\natexlab{b}})\citenamefont {Hashim}, \citenamefont {El-Zant}, \citenamefont {El~Hanafy},\ and\ \citenamefont {Golovnev}}]{Hashim:2021pkq}%
  \BibitemOpen
  \bibfield  {author} {\bibinfo {author} {\bibfnamefont {M.}~\bibnamefont {Hashim}}, \bibinfo {author} {\bibfnamefont {A.~A.}\ \bibnamefont {El-Zant}}, \bibinfo {author} {\bibfnamefont {W.}~\bibnamefont {El~Hanafy}},\ and\ \bibinfo {author} {\bibfnamefont {A.}~\bibnamefont {Golovnev}},\ }\bibfield  {title} {\bibinfo {title} {{Toward a concordance teleparallel cosmology. Part~II. Linear perturbation}},\ }\href {https://doi.org/10.1088/1475-7516/2021/07/053} {\bibfield  {journal} {\bibinfo  {journal} {J. Cosmol. Astropart. Phys.}\ }\textbf {\bibinfo {volume} {07}},\ \bibinfo {pages} {053} (\bibinfo {year} {2021}{\natexlab{b}})},\ \Eprint {https://arxiv.org/abs/2104.08311} {2104.08311} \BibitemShut {NoStop}%
\bibitem [{\citenamefont {Akarsu}\ \emph {et~al.}(2025{\natexlab{b}})\citenamefont {Akarsu}, \citenamefont {Bulduk}, \citenamefont {De~Felice}, \citenamefont {Kat{\i}rc{\i}},\ and\ \citenamefont {Uzun}}]{Akarsu:2024nas}%
  \BibitemOpen
  \bibfield  {author} {\bibinfo {author} {\bibfnamefont {{\"O}.}~\bibnamefont {Akarsu}}, \bibinfo {author} {\bibfnamefont {B.}~\bibnamefont {Bulduk}}, \bibinfo {author} {\bibfnamefont {A.}~\bibnamefont {De~Felice}}, \bibinfo {author} {\bibfnamefont {N.}~\bibnamefont {Kat{\i}rc{\i}}},\ and\ \bibinfo {author} {\bibfnamefont {N.~M.}\ \bibnamefont {Uzun}},\ }\bibfield  {title} {\bibinfo {title} {{Unexplored regions in teleparallel f(T) gravity: Sign-changing dark energy density}},\ }\href {https://doi.org/10.1103/1xd4-k91h} {\bibfield  {journal} {\bibinfo  {journal} {Phys. Rev. D}\ }\textbf {\bibinfo {volume} {112}},\ \bibinfo {pages} {083532} (\bibinfo {year} {2025}{\natexlab{b}})},\ \Eprint {https://arxiv.org/abs/2410.23068} {2410.23068} \BibitemShut {NoStop}%
\bibitem [{\citenamefont {Sahni}\ and\ \citenamefont {Shtanov}(2003)}]{Sahni:2002dx}%
  \BibitemOpen
  \bibfield  {author} {\bibinfo {author} {\bibfnamefont {V.}~\bibnamefont {Sahni}}\ and\ \bibinfo {author} {\bibfnamefont {Y.}~\bibnamefont {Shtanov}},\ }\bibfield  {title} {\bibinfo {title} {{Brane world models of dark energy}},\ }\href {https://doi.org/10.1088/1475-7516/2003/11/014} {\bibfield  {journal} {\bibinfo  {journal} {J. Cosmol. Astropart. Phys.}\ }\textbf {\bibinfo {volume} {2003}}\bibfield  {number} {\bibinfo  {number} { (11)},\ \bibinfo {pages} {014}} (\bibinfo {year} {2003})},\ \Eprint {https://arxiv.org/abs/astro-ph/0202346} {astro-ph/0202346} \BibitemShut {NoStop}%
\bibitem [{\citenamefont {Sahni}\ and\ \citenamefont {Shtanov}(2002)}]{Sahni:2000ubn}%
  \BibitemOpen
  \bibfield  {author} {\bibinfo {author} {\bibfnamefont {V.}~\bibnamefont {Sahni}}\ and\ \bibinfo {author} {\bibfnamefont {Y.}~\bibnamefont {Shtanov}},\ }\bibfield  {title} {\bibinfo {title} {{New vistas in brane world cosmology}},\ }\href {https://doi.org/10.1142/S0218271802002827} {\bibfield  {journal} {\bibinfo  {journal} {Int. J. Mod. Phys. D}\ }\textbf {\bibinfo {volume} {11}},\ \bibinfo {pages} {1515} (\bibinfo {year} {2002})},\ \Eprint {https://arxiv.org/abs/gr-qc/0205111} {gr-qc/0205111} \BibitemShut {NoStop}%
\bibitem [{\citenamefont {Bag}\ \emph {et~al.}(2021)\citenamefont {Bag}, \citenamefont {Sahni}, \citenamefont {Shafieloo},\ and\ \citenamefont {Shtanov}}]{Bag:2021cqm}%
  \BibitemOpen
  \bibfield  {author} {\bibinfo {author} {\bibfnamefont {S.}~\bibnamefont {Bag}}, \bibinfo {author} {\bibfnamefont {V.}~\bibnamefont {Sahni}}, \bibinfo {author} {\bibfnamefont {A.}~\bibnamefont {Shafieloo}},\ and\ \bibinfo {author} {\bibfnamefont {Y.}~\bibnamefont {Shtanov}},\ }\bibfield  {title} {\bibinfo {title} {{Phantom Braneworld and the Hubble Tension}},\ }\href {https://doi.org/10.3847/1538-4357/ac307e} {\bibfield  {journal} {\bibinfo  {journal} {Astrophys. J.}\ }\textbf {\bibinfo {volume} {923}},\ \bibinfo {pages} {212} (\bibinfo {year} {2021})},\ \Eprint {https://arxiv.org/abs/2107.03271} {2107.03271} \BibitemShut {NoStop}%
\bibitem [{\citenamefont {De~Felice}\ \emph {et~al.}(2020)\citenamefont {De~Felice}, \citenamefont {Doll},\ and\ \citenamefont {Mukohyama}}]{DeFelice:2020eju}%
  \BibitemOpen
  \bibfield  {author} {\bibinfo {author} {\bibfnamefont {A.}~\bibnamefont {De~Felice}}, \bibinfo {author} {\bibfnamefont {A.}~\bibnamefont {Doll}},\ and\ \bibinfo {author} {\bibfnamefont {S.}~\bibnamefont {Mukohyama}},\ }\bibfield  {title} {\bibinfo {title} {{A theory of type-II minimally modified gravity}},\ }\href {https://doi.org/10.1088/1475-7516/2020/09/034} {\bibfield  {journal} {\bibinfo  {journal} {J. Cosmol. Astropart. Phys.}\ }\textbf {\bibinfo {volume} {2020}}\bibfield  {number} {\bibinfo  {number} { (09)},\ \bibinfo {pages} {034}} (\bibinfo {year} {2020})},\ \Eprint {https://arxiv.org/abs/2004.12549} {2004.12549} \BibitemShut {NoStop}%
\bibitem [{\citenamefont {De~Felice}\ \emph {et~al.}(2022)\citenamefont {De~Felice}, \citenamefont {Maeda}, \citenamefont {Mukohyama},\ and\ \citenamefont {Pookkillath}}]{DeFelice:2022uxv}%
  \BibitemOpen
  \bibfield  {author} {\bibinfo {author} {\bibfnamefont {A.}~\bibnamefont {De~Felice}}, \bibinfo {author} {\bibfnamefont {K.-i.}\ \bibnamefont {Maeda}}, \bibinfo {author} {\bibfnamefont {S.}~\bibnamefont {Mukohyama}},\ and\ \bibinfo {author} {\bibfnamefont {M.~C.}\ \bibnamefont {Pookkillath}},\ }\bibfield  {title} {\bibinfo {title} {{Comparison of two theories of Type-IIa minimally modified gravity}},\ }\href {https://doi.org/10.1103/PhysRevD.106.024028} {\bibfield  {journal} {\bibinfo  {journal} {Phys. Rev. D}\ }\textbf {\bibinfo {volume} {106}},\ \bibinfo {pages} {024028} (\bibinfo {year} {2022})},\ \Eprint {https://arxiv.org/abs/2204.08294} {2204.08294} \BibitemShut {NoStop}%
\bibitem [{\citenamefont {Whittaker}(1935)}]{Whittaker:1935lr}%
  \BibitemOpen
  \bibfield  {author} {\bibinfo {author} {\bibfnamefont {E.~T.}\ \bibnamefont {Whittaker}},\ }\bibfield  {title} {\bibinfo {title} {{On Gauss' theorem and the concept of mass in general relativity}},\ }\href {https://doi.org/10.1098/rspa.1935.0069} {\bibfield  {journal} {\bibinfo  {journal} {Proc. Roy. Soc. Lond. A}\ }\textbf {\bibinfo {volume} {149}},\ \bibinfo {pages} {384} (\bibinfo {year} {1935})}\BibitemShut {NoStop}%
\bibitem [{\citenamefont {Tolman}(1934)}]{Tolman:1934bk}%
  \BibitemOpen
  \bibfield  {author} {\bibinfo {author} {\bibfnamefont {R.~C.}\ \bibnamefont {Tolman}},\ }\href@noop {} {\emph {\bibinfo {title} {{Relativity, Thermodynamics and Cosmology}}}}\ (\bibinfo  {publisher} {Clarendon Press},\ \bibinfo {address} {Oxford},\ \bibinfo {year} {1934})\BibitemShut {NoStop}%
\bibitem [{\citenamefont {Barcelo}\ and\ \citenamefont {Visser}(2002)}]{Barcelo:2002bv}%
  \BibitemOpen
  \bibfield  {author} {\bibinfo {author} {\bibfnamefont {C.}~\bibnamefont {Barcelo}}\ and\ \bibinfo {author} {\bibfnamefont {M.}~\bibnamefont {Visser}},\ }\bibfield  {title} {\bibinfo {title} {{Twilight for the energy conditions?}},\ }\href {https://doi.org/10.1142/S0218271802002888} {\bibfield  {journal} {\bibinfo  {journal} {Int. J. Mod. Phys. D}\ }\textbf {\bibinfo {volume} {11}},\ \bibinfo {pages} {1553} (\bibinfo {year} {2002})},\ \Eprint {https://arxiv.org/abs/gr-qc/0205066} {gr-qc/0205066} \BibitemShut {NoStop}%
\bibitem [{\citenamefont {Curiel}(2017)}]{Curiel:2014zba}%
  \BibitemOpen
  \bibfield  {author} {\bibinfo {author} {\bibfnamefont {E.}~\bibnamefont {Curiel}},\ }\bibfield  {title} {\bibinfo {title} {{A Primer on Energy Conditions}},\ }\href {https://doi.org/10.1007/978-1-4939-3210-8_3} {\bibfield  {journal} {\bibinfo  {journal} {Einstein Stud.}\ }\textbf {\bibinfo {volume} {13}},\ \bibinfo {pages} {43} (\bibinfo {year} {2017})},\ \Eprint {https://arxiv.org/abs/1405.0403} {1405.0403} \BibitemShut {NoStop}%
\bibitem [{\citenamefont {Kontou}\ and\ \citenamefont {Sanders}(2020)}]{Kontou:2020bta}%
  \BibitemOpen
  \bibfield  {author} {\bibinfo {author} {\bibfnamefont {E.-A.}\ \bibnamefont {Kontou}}\ and\ \bibinfo {author} {\bibfnamefont {K.}~\bibnamefont {Sanders}},\ }\bibfield  {title} {\bibinfo {title} {{Energy conditions in general relativity and quantum field theory}},\ }\href {https://doi.org/10.1088/1361-6382/ab8fcf} {\bibfield  {journal} {\bibinfo  {journal} {Class. Quant. Grav.}\ }\textbf {\bibinfo {volume} {37}},\ \bibinfo {pages} {193001} (\bibinfo {year} {2020})},\ \Eprint {https://arxiv.org/abs/2003.01815} {2003.01815} \BibitemShut {NoStop}%
\bibitem [{\citenamefont {Rubakov}(2014)}]{Rubakov:2014jja}%
  \BibitemOpen
  \bibfield  {author} {\bibinfo {author} {\bibfnamefont {V.~A.}\ \bibnamefont {Rubakov}},\ }\bibfield  {title} {\bibinfo {title} {{The Null Energy Condition and its violation}},\ }\href {https://doi.org/10.3367/UFNe.0184.201402b.0137} {\bibfield  {journal} {\bibinfo  {journal} {Phys. Usp.}\ }\textbf {\bibinfo {volume} {57}},\ \bibinfo {pages} {128} (\bibinfo {year} {2014})},\ \Eprint {https://arxiv.org/abs/1401.4024} {1401.4024} \BibitemShut {NoStop}%
\bibitem [{\citenamefont {Nojiri}\ \emph {et~al.}(2005)\citenamefont {Nojiri}, \citenamefont {Odintsov},\ and\ \citenamefont {Tsujikawa}}]{Nojiri:2005sx}%
  \BibitemOpen
  \bibfield  {author} {\bibinfo {author} {\bibfnamefont {S.}~\bibnamefont {Nojiri}}, \bibinfo {author} {\bibfnamefont {S.~D.}\ \bibnamefont {Odintsov}},\ and\ \bibinfo {author} {\bibfnamefont {S.}~\bibnamefont {Tsujikawa}},\ }\bibfield  {title} {\bibinfo {title} {{Properties of singularities in (phantom) dark energy universe}},\ }\href {https://doi.org/10.1103/PhysRevD.71.063004} {\bibfield  {journal} {\bibinfo  {journal} {Phys. Rev. D}\ }\textbf {\bibinfo {volume} {71}},\ \bibinfo {pages} {063004} (\bibinfo {year} {2005})},\ \Eprint {https://arxiv.org/abs/hep-th/0501025} {hep-th/0501025} \BibitemShut {NoStop}%
\bibitem [{\citenamefont {Ellis}\ and\ \citenamefont {van Elst}(1999)}]{Ellis:1998ct}%
  \BibitemOpen
  \bibfield  {author} {\bibinfo {author} {\bibfnamefont {G.~F.~R.}\ \bibnamefont {Ellis}}\ and\ \bibinfo {author} {\bibfnamefont {H.}~\bibnamefont {van Elst}},\ }\bibfield  {title} {\bibinfo {title} {{Cosmological models: Cargese lectures 1998}},\ }\href {https://doi.org/10.1007/978-94-011-4455-1_1} {\bibfield  {journal} {\bibinfo  {journal} {NATO Sci. Ser. C}\ }\textbf {\bibinfo {volume} {541}},\ \bibinfo {pages} {1} (\bibinfo {year} {1999})},\ \Eprint {https://arxiv.org/abs/gr-qc/9812046} {gr-qc/9812046} \BibitemShut {NoStop}%
\bibitem [{\citenamefont {Vikman}(2005)}]{Vikman:2004dc}%
  \BibitemOpen
  \bibfield  {author} {\bibinfo {author} {\bibfnamefont {A.}~\bibnamefont {Vikman}},\ }\bibfield  {title} {\bibinfo {title} {{Can dark energy evolve to the phantom?}},\ }\href {https://doi.org/10.1103/PhysRevD.71.023515} {\bibfield  {journal} {\bibinfo  {journal} {Phys. Rev. D}\ }\textbf {\bibinfo {volume} {71}},\ \bibinfo {pages} {023515} (\bibinfo {year} {2005})},\ \Eprint {https://arxiv.org/abs/astro-ph/0407107} {astro-ph/0407107} \BibitemShut {NoStop}%
\bibitem [{\citenamefont {Akarsu}\ \emph {et~al.}(2025{\natexlab{c}})\citenamefont {Akarsu}, \citenamefont {Perivolaropoulos}, \citenamefont {Tsikoundoura}, \citenamefont {Y{\"u}kselci},\ and\ \citenamefont {Zhuk}}]{Akarsu:2025gwi}%
  \BibitemOpen
  \bibfield  {author} {\bibinfo {author} {\bibfnamefont {{\"O}.}~\bibnamefont {Akarsu}}, \bibinfo {author} {\bibfnamefont {L.}~\bibnamefont {Perivolaropoulos}}, \bibinfo {author} {\bibfnamefont {A.}~\bibnamefont {Tsikoundoura}}, \bibinfo {author} {\bibfnamefont {A.~E.}\ \bibnamefont {Y{\"u}kselci}},\ and\ \bibinfo {author} {\bibfnamefont {A.}~\bibnamefont {Zhuk}},\ }\href@noop {} {\bibinfo {title} {{Dynamical dark energy with AdS-to-dS and dS-to-dS transitions: Implications for the $H_0$ tension}}} (\bibinfo {year} {2025}{\natexlab{c}}),\ \Eprint {https://arxiv.org/abs/2502.14667} {2502.14667} \BibitemShut {NoStop}%
\bibitem [{\citenamefont {Akarsu}\ \emph {et~al.}(2026{\natexlab{e}})\citenamefont {Akarsu}, \citenamefont {Perivolaropoulos}, \citenamefont {Y{\"u}kselci},\ and\ \citenamefont {Zhuk}}]{Akarsu:2026lva}%
  \BibitemOpen
  \bibfield  {author} {\bibinfo {author} {\bibfnamefont {{\"O}.}~\bibnamefont {Akarsu}}, \bibinfo {author} {\bibfnamefont {L.}~\bibnamefont {Perivolaropoulos}}, \bibinfo {author} {\bibfnamefont {A.~E.}\ \bibnamefont {Y{\"u}kselci}},\ and\ \bibinfo {author} {\bibfnamefont {A.}~\bibnamefont {Zhuk}},\ }\href@noop {} {\bibinfo {title} {{A Friendly Phantom: Late-time AdS-to-dS transition and cosmological tensions}}} (\bibinfo {year} {2026}{\natexlab{e}}),\ \Eprint {https://arxiv.org/abs/2606.11062} {2606.11062} \BibitemShut {NoStop}%
\bibitem [{\citenamefont {Fields}\ \emph {et~al.}(2020)\citenamefont {Fields}, \citenamefont {Olive}, \citenamefont {Yeh},\ and\ \citenamefont {Young}}]{Fields:2019pfx}%
  \BibitemOpen
  \bibfield  {author} {\bibinfo {author} {\bibfnamefont {B.~D.}\ \bibnamefont {Fields}}, \bibinfo {author} {\bibfnamefont {K.~A.}\ \bibnamefont {Olive}}, \bibinfo {author} {\bibfnamefont {T.-H.}\ \bibnamefont {Yeh}},\ and\ \bibinfo {author} {\bibfnamefont {C.}~\bibnamefont {Young}},\ }\bibfield  {title} {\bibinfo {title} {{Big-Bang Nucleosynthesis after Planck}},\ }\href {https://doi.org/10.1088/1475-7516/2020/03/010} {\bibfield  {journal} {\bibinfo  {journal} {J. Cosmol. Astropart. Phys.}\ }\textbf {\bibinfo {volume} {03}},\ \bibinfo {pages} {010} (\bibinfo {year} {2020})},\ \bibinfo {note} {[Erratum: J. Cosmol. Astropart. Phys. 11, E02 (2020)]},\ \Eprint {https://arxiv.org/abs/1912.01132} {1912.01132} \BibitemShut {NoStop}%
\bibitem [{\citenamefont {Bouhmadi-L{\'o}pez}\ and\ \citenamefont {Ibarra-Uriondo}(2025{\natexlab{a}})}]{Bouhmadi-Lopez:2025ggl}%
  \BibitemOpen
  \bibfield  {author} {\bibinfo {author} {\bibfnamefont {M.}~\bibnamefont {Bouhmadi-L{\'o}pez}}\ and\ \bibinfo {author} {\bibfnamefont {B.}~\bibnamefont {Ibarra-Uriondo}},\ }\bibfield  {title} {\bibinfo {title} {{Cosmographic analysis of sign-switching dark energy}},\ }\href {https://doi.org/10.1103/v1cl-pr54} {\bibfield  {journal} {\bibinfo  {journal} {Phys. Rev. D}\ }\textbf {\bibinfo {volume} {112}},\ \bibinfo {pages} {063559} (\bibinfo {year} {2025}{\natexlab{a}})},\ \Eprint {https://arxiv.org/abs/2506.12139} {2506.12139} \BibitemShut {NoStop}%
\bibitem [{\citenamefont {Bouhmadi-L{\'o}pez}\ and\ \citenamefont {Ibarra-Uriondo}(2025{\natexlab{b}})}]{Bouhmadi-Lopez:2025spo}%
  \BibitemOpen
  \bibfield  {author} {\bibinfo {author} {\bibfnamefont {M.}~\bibnamefont {Bouhmadi-L{\'o}pez}}\ and\ \bibinfo {author} {\bibfnamefont {B.}~\bibnamefont {Ibarra-Uriondo}},\ }\bibfield  {title} {\bibinfo {title} {{Cosmological perturbations for smooth sign-switching dark energy models}},\ }\href {https://doi.org/10.1016/j.dark.2025.102129} {\bibfield  {journal} {\bibinfo  {journal} {Phys. Dark Univ.}\ }\textbf {\bibinfo {volume} {50}},\ \bibinfo {pages} {102129} (\bibinfo {year} {2025}{\natexlab{b}})},\ \Eprint {https://arxiv.org/abs/2506.18992} {2506.18992} \BibitemShut {NoStop}%
\bibitem [{\citenamefont {Ibarra-Uriondo}\ and\ \citenamefont {Bouhmadi-L{\'o}pez}(2026)}]{Ibarra-Uriondo:2026zbp}%
  \BibitemOpen
  \bibfield  {author} {\bibinfo {author} {\bibfnamefont {B.}~\bibnamefont {Ibarra-Uriondo}}\ and\ \bibinfo {author} {\bibfnamefont {M.}~\bibnamefont {Bouhmadi-L{\'o}pez}},\ }\bibfield  {title} {\bibinfo {title} {{Sign-switching dark energy: Smooth transitions with recent DESI DR2 observations}},\ }\href {https://doi.org/10.1016/j.dark.2026.102351} {\bibfield  {journal} {\bibinfo  {journal} {Phys. Dark Univ.}\ }\textbf {\bibinfo {volume} {52}},\ \bibinfo {pages} {102351} (\bibinfo {year} {2026})},\ \Eprint {https://arxiv.org/abs/2602.12347} {2602.12347} \BibitemShut {NoStop}%
\bibitem [{\citenamefont {Bouhmadi-L{\'o}pez}\ \emph {et~al.}(2026)\citenamefont {Bouhmadi-L{\'o}pez}, \citenamefont {Chiang},\ and\ \citenamefont {Ibarra-Uriondo}}]{Bouhmadi-Lopez:2026vyc}%
  \BibitemOpen
  \bibfield  {author} {\bibinfo {author} {\bibfnamefont {M.}~\bibnamefont {Bouhmadi-L{\'o}pez}}, \bibinfo {author} {\bibfnamefont {H.-W.}\ \bibnamefont {Chiang}},\ and\ \bibinfo {author} {\bibfnamefont {B.}~\bibnamefont {Ibarra-Uriondo}},\ }\href@noop {} {\bibinfo {title} {{Alleviating the Hubble Tension with Smooth Sign-Switching Dark Energy: Full CMB Constraints with DESI and PantheonPlus}}} (\bibinfo {year} {2026}),\ \Eprint {https://arxiv.org/abs/2607.05044} {2607.05044} \BibitemShut {NoStop}%
\bibitem [{\citenamefont {Akarsu}\ \emph {et~al.}(2025{\natexlab{d}})\citenamefont {Akarsu}, \citenamefont {Eingorn}, \citenamefont {Perivolaropoulos}, \citenamefont {Y{\"u}kselci},\ and\ \citenamefont {Zhuk}}]{Akarsu:2025dmj}%
  \BibitemOpen
  \bibfield  {author} {\bibinfo {author} {\bibfnamefont {{\"O}.}~\bibnamefont {Akarsu}}, \bibinfo {author} {\bibfnamefont {M.}~\bibnamefont {Eingorn}}, \bibinfo {author} {\bibfnamefont {L.}~\bibnamefont {Perivolaropoulos}}, \bibinfo {author} {\bibfnamefont {A.~E.}\ \bibnamefont {Y{\"u}kselci}},\ and\ \bibinfo {author} {\bibfnamefont {A.}~\bibnamefont {Zhuk}},\ }\href@noop {} {\bibinfo {title} {{Dynamical dark energy with AdS-dS transitions vs. Baryon Acoustic Oscillations at $z =$ 2.3-2.4}}} (\bibinfo {year} {2025}{\natexlab{d}}),\ \Eprint {https://arxiv.org/abs/2504.07299} {2504.07299} \BibitemShut {NoStop}%
\bibitem [{\citenamefont {G{\'o}mez-Valent}\ and\ \citenamefont {Gonz{\'a}lez-Fuentes}(2026)}]{Gomez-Valent:2025mfl}%
  \BibitemOpen
  \bibfield  {author} {\bibinfo {author} {\bibfnamefont {A.}~\bibnamefont {G{\'o}mez-Valent}}\ and\ \bibinfo {author} {\bibfnamefont {A.}~\bibnamefont {Gonz{\'a}lez-Fuentes}},\ }\bibfield  {title} {\bibinfo {title} {{Effective phantom divide crossing with standard and negative quintessence}},\ }\href {https://doi.org/10.1016/j.physletb.2025.140096} {\bibfield  {journal} {\bibinfo  {journal} {Phys. Lett. B}\ }\textbf {\bibinfo {volume} {872}},\ \bibinfo {pages} {140096} (\bibinfo {year} {2026})},\ \Eprint {https://arxiv.org/abs/2508.00621} {2508.00621} \BibitemShut {NoStop}%
\bibitem [{\citenamefont {Hashim}\ \emph {et~al.}(2026)\citenamefont {Hashim}, \citenamefont {Di~Valentino}, \citenamefont {Levi~Said},\ and\ \citenamefont {El~Hanafy}}]{Hashim:2026yoy}%
  \BibitemOpen
  \bibfield  {author} {\bibinfo {author} {\bibfnamefont {M.}~\bibnamefont {Hashim}}, \bibinfo {author} {\bibfnamefont {E.}~\bibnamefont {Di~Valentino}}, \bibinfo {author} {\bibfnamefont {J.}~\bibnamefont {Levi~Said}},\ and\ \bibinfo {author} {\bibfnamefont {W.}~\bibnamefont {El~Hanafy}},\ }\href@noop {} {\bibinfo {title} {{Cosmological Viability of Exponential Infrared $f(T)$ Gravity}}} (\bibinfo {year} {2026}),\ \Eprint {https://arxiv.org/abs/2606.31324} {2606.31324} \BibitemShut {NoStop}%
\bibitem [{\citenamefont {Sahni}(2005)}]{Sahni:2005pf}%
  \BibitemOpen
  \bibfield  {author} {\bibinfo {author} {\bibfnamefont {V.}~\bibnamefont {Sahni}},\ }\bibinfo {title} {{Cosmological surprises from braneworld models of dark energy}},\ in\ \href@noop {} {\emph {\bibinfo {booktitle} {{14th Workshop on General Relativity and Gravitation}}}}\ (\bibinfo {year} {2005})\ pp.\ \bibinfo {pages} {95--115},\ \Eprint {https://arxiv.org/abs/astro-ph/0502032} {astro-ph/0502032} \BibitemShut {NoStop}%
\bibitem [{\citenamefont {Bag}\ \emph {et~al.}(2018)\citenamefont {Bag}, \citenamefont {Mishra},\ and\ \citenamefont {Sahni}}]{Bag:2018jle}%
  \BibitemOpen
  \bibfield  {author} {\bibinfo {author} {\bibfnamefont {S.}~\bibnamefont {Bag}}, \bibinfo {author} {\bibfnamefont {S.~S.}\ \bibnamefont {Mishra}},\ and\ \bibinfo {author} {\bibfnamefont {V.}~\bibnamefont {Sahni}},\ }\bibfield  {title} {\bibinfo {title} {{Emulating a {\ensuremath{\Lambda}}CDM-like expansion on the phantom brane}},\ }\href {https://doi.org/10.1103/PhysRevD.97.123537} {\bibfield  {journal} {\bibinfo  {journal} {Phys. Rev. D}\ }\textbf {\bibinfo {volume} {97}},\ \bibinfo {pages} {123537} (\bibinfo {year} {2018})},\ \Eprint {https://arxiv.org/abs/1807.00684} {1807.00684} \BibitemShut {NoStop}%
\bibitem [{\citenamefont {Hu}(2005)}]{Hu:2004kh}%
  \BibitemOpen
  \bibfield  {author} {\bibinfo {author} {\bibfnamefont {W.}~\bibnamefont {Hu}},\ }\bibfield  {title} {\bibinfo {title} {{Crossing the phantom divide: Dark energy internal degrees of freedom}},\ }\href {https://doi.org/10.1103/PhysRevD.71.047301} {\bibfield  {journal} {\bibinfo  {journal} {Phys. Rev. D}\ }\textbf {\bibinfo {volume} {71}},\ \bibinfo {pages} {047301} (\bibinfo {year} {2005})},\ \Eprint {https://arxiv.org/abs/astro-ph/0410680} {astro-ph/0410680} \BibitemShut {NoStop}%
\bibitem [{\citenamefont {Fang}\ \emph {et~al.}(2008)\citenamefont {Fang}, \citenamefont {Hu},\ and\ \citenamefont {Lewis}}]{Fang:2008sn}%
  \BibitemOpen
  \bibfield  {author} {\bibinfo {author} {\bibfnamefont {W.}~\bibnamefont {Fang}}, \bibinfo {author} {\bibfnamefont {W.}~\bibnamefont {Hu}},\ and\ \bibinfo {author} {\bibfnamefont {A.}~\bibnamefont {Lewis}},\ }\bibfield  {title} {\bibinfo {title} {{Crossing the Phantom Divide with Parameterized Post-Friedmann Dark Energy}},\ }\href {https://doi.org/10.1103/PhysRevD.78.087303} {\bibfield  {journal} {\bibinfo  {journal} {Phys. Rev. D}\ }\textbf {\bibinfo {volume} {78}},\ \bibinfo {pages} {087303} (\bibinfo {year} {2008})},\ \Eprint {https://arxiv.org/abs/0808.3125} {0808.3125} \BibitemShut {NoStop}%
\bibitem [{\citenamefont {Ma}\ and\ \citenamefont {Bertschinger}(1995)}]{Ma:1995ey}%
  \BibitemOpen
  \bibfield  {author} {\bibinfo {author} {\bibfnamefont {C.-P.}\ \bibnamefont {Ma}}\ and\ \bibinfo {author} {\bibfnamefont {E.}~\bibnamefont {Bertschinger}},\ }\bibfield  {title} {\bibinfo {title} {{Cosmological perturbation theory in the synchronous and conformal Newtonian gauges}},\ }\href {https://doi.org/10.1086/176550} {\bibfield  {journal} {\bibinfo  {journal} {Astrophys. J.}\ }\textbf {\bibinfo {volume} {455}},\ \bibinfo {pages} {7} (\bibinfo {year} {1995})},\ \Eprint {https://arxiv.org/abs/astro-ph/9506072} {astro-ph/9506072} \BibitemShut {NoStop}%
\bibitem [{\citenamefont {Mukhanov}(2005)}]{Mukhanov:2005sc}%
  \BibitemOpen
  \bibfield  {author} {\bibinfo {author} {\bibfnamefont {V.}~\bibnamefont {Mukhanov}},\ }\href {https://doi.org/10.1017/CBO9780511790553} {\emph {\bibinfo {title} {{Physical Foundations of Cosmology}}}}\ (\bibinfo  {publisher} {Cambridge University Press},\ \bibinfo {address} {Cambridge, England},\ \bibinfo {year} {2005})\BibitemShut {NoStop}%
\bibitem [{\citenamefont {Malik}\ and\ \citenamefont {Wands}(2009)}]{Malik:2008im}%
  \BibitemOpen
  \bibfield  {author} {\bibinfo {author} {\bibfnamefont {K.~A.}\ \bibnamefont {Malik}}\ and\ \bibinfo {author} {\bibfnamefont {D.}~\bibnamefont {Wands}},\ }\bibfield  {title} {\bibinfo {title} {{Cosmological perturbations}},\ }\href {https://doi.org/10.1016/j.physrep.2009.03.001} {\bibfield  {journal} {\bibinfo  {journal} {Phys. Rept.}\ }\textbf {\bibinfo {volume} {475}},\ \bibinfo {pages} {1} (\bibinfo {year} {2009})},\ \Eprint {https://arxiv.org/abs/0809.4944} {0809.4944} \BibitemShut {NoStop}%
\bibitem [{\citenamefont {De~Felice}\ \emph {et~al.}(2010)\citenamefont {De~Felice}, \citenamefont {Gerard},\ and\ \citenamefont {Suyama}}]{DeFelice:2009bx}%
  \BibitemOpen
  \bibfield  {author} {\bibinfo {author} {\bibfnamefont {A.}~\bibnamefont {De~Felice}}, \bibinfo {author} {\bibfnamefont {J.-M.}\ \bibnamefont {Gerard}},\ and\ \bibinfo {author} {\bibfnamefont {T.}~\bibnamefont {Suyama}},\ }\bibfield  {title} {\bibinfo {title} {{Cosmological perturbations of a perfect fluid and noncommutative variables}},\ }\href {https://doi.org/10.1103/PhysRevD.81.063527} {\bibfield  {journal} {\bibinfo  {journal} {Phys. Rev. D}\ }\textbf {\bibinfo {volume} {81}},\ \bibinfo {pages} {063527} (\bibinfo {year} {2010})},\ \Eprint {https://arxiv.org/abs/0908.3439} {0908.3439} \BibitemShut {NoStop}%
\bibitem [{\citenamefont {Carroll}\ \emph {et~al.}(2003)\citenamefont {Carroll}, \citenamefont {Hoffman},\ and\ \citenamefont {Trodden}}]{Carroll:2003st}%
  \BibitemOpen
  \bibfield  {author} {\bibinfo {author} {\bibfnamefont {S.~M.}\ \bibnamefont {Carroll}}, \bibinfo {author} {\bibfnamefont {M.}~\bibnamefont {Hoffman}},\ and\ \bibinfo {author} {\bibfnamefont {M.}~\bibnamefont {Trodden}},\ }\bibfield  {title} {\bibinfo {title} {{Can the dark energy equation-of-state parameter w be less than -1?}},\ }\href {https://doi.org/10.1103/PhysRevD.68.023509} {\bibfield  {journal} {\bibinfo  {journal} {Phys. Rev. D}\ }\textbf {\bibinfo {volume} {68}},\ \bibinfo {pages} {023509} (\bibinfo {year} {2003})},\ \Eprint {https://arxiv.org/abs/astro-ph/0301273} {astro-ph/0301273} \BibitemShut {NoStop}%
\bibitem [{\citenamefont {Cline}\ \emph {et~al.}(2004)\citenamefont {Cline}, \citenamefont {Jeon},\ and\ \citenamefont {Moore}}]{Cline:2003gs}%
  \BibitemOpen
  \bibfield  {author} {\bibinfo {author} {\bibfnamefont {J.~M.}\ \bibnamefont {Cline}}, \bibinfo {author} {\bibfnamefont {S.}~\bibnamefont {Jeon}},\ and\ \bibinfo {author} {\bibfnamefont {G.~D.}\ \bibnamefont {Moore}},\ }\bibfield  {title} {\bibinfo {title} {{The Phantom menaced: Constraints on low-energy effective ghosts}},\ }\href {https://doi.org/10.1103/PhysRevD.70.043543} {\bibfield  {journal} {\bibinfo  {journal} {Phys. Rev. D}\ }\textbf {\bibinfo {volume} {70}},\ \bibinfo {pages} {043543} (\bibinfo {year} {2004})},\ \Eprint {https://arxiv.org/abs/hep-ph/0311312} {hep-ph/0311312} \BibitemShut {NoStop}%
\bibitem [{\citenamefont {Lewis}\ and\ \citenamefont {Chamberlain}(2025)}]{Lewis:2024cqj}%
  \BibitemOpen
  \bibfield  {author} {\bibinfo {author} {\bibfnamefont {A.}~\bibnamefont {Lewis}}\ and\ \bibinfo {author} {\bibfnamefont {E.}~\bibnamefont {Chamberlain}},\ }\bibfield  {title} {\bibinfo {title} {{Understanding acoustic scale observations: the one-sided fight against {\ensuremath{\Lambda}}}},\ }\href {https://doi.org/10.1088/1475-7516/2025/05/065} {\bibfield  {journal} {\bibinfo  {journal} {J. Cosmol. Astropart. Phys.}\ }\textbf {\bibinfo {volume} {2025}}\bibfield  {number} {\bibinfo  {number} { (05)},\ \bibinfo {pages} {065}} (\bibinfo {year} {2025})},\ \Eprint {https://arxiv.org/abs/2412.13894} {2412.13894} \BibitemShut {NoStop}%
\bibitem [{\citenamefont {Horndeski}(1974)}]{Horndeski:1974wa}%
  \BibitemOpen
  \bibfield  {author} {\bibinfo {author} {\bibfnamefont {G.~W.}\ \bibnamefont {Horndeski}},\ }\bibfield  {title} {\bibinfo {title} {{Second-order scalar-tensor field equations in a four-dimensional space}},\ }\href {https://doi.org/10.1007/BF01807638} {\bibfield  {journal} {\bibinfo  {journal} {Int. J. Theor. Phys.}\ }\textbf {\bibinfo {volume} {10}},\ \bibinfo {pages} {363} (\bibinfo {year} {1974})}\BibitemShut {NoStop}%
\bibitem [{\citenamefont {Kobayashi}(2019)}]{Kobayashi:2019hrl}%
  \BibitemOpen
  \bibfield  {author} {\bibinfo {author} {\bibfnamefont {T.}~\bibnamefont {Kobayashi}},\ }\bibfield  {title} {\bibinfo {title} {{Horndeski theory and beyond: a review}},\ }\href {https://doi.org/10.1088/1361-6633/ab2429} {\bibfield  {journal} {\bibinfo  {journal} {Rept. Prog. Phys.}\ }\textbf {\bibinfo {volume} {82}},\ \bibinfo {pages} {086901} (\bibinfo {year} {2019})},\ \Eprint {https://arxiv.org/abs/1901.07183} {1901.07183} \BibitemShut {NoStop}%
\bibitem [{\citenamefont {Feng}\ \emph {et~al.}(2005)\citenamefont {Feng}, \citenamefont {Wang},\ and\ \citenamefont {Zhang}}]{Feng:2004ad}%
  \BibitemOpen
  \bibfield  {author} {\bibinfo {author} {\bibfnamefont {B.}~\bibnamefont {Feng}}, \bibinfo {author} {\bibfnamefont {X.-L.}\ \bibnamefont {Wang}},\ and\ \bibinfo {author} {\bibfnamefont {X.-M.}\ \bibnamefont {Zhang}},\ }\bibfield  {title} {\bibinfo {title} {{Dark energy constraints from the cosmic age and supernova}},\ }\href {https://doi.org/10.1016/j.physletb.2004.12.071} {\bibfield  {journal} {\bibinfo  {journal} {Phys. Lett. B}\ }\textbf {\bibinfo {volume} {607}},\ \bibinfo {pages} {35} (\bibinfo {year} {2005})},\ \Eprint {https://arxiv.org/abs/astro-ph/0404224} {astro-ph/0404224} \BibitemShut {NoStop}%
\bibitem [{\citenamefont {Cai}\ \emph {et~al.}(2010)\citenamefont {Cai}, \citenamefont {Saridakis}, \citenamefont {Setare},\ and\ \citenamefont {Xia}}]{Cai:2009zp}%
  \BibitemOpen
  \bibfield  {author} {\bibinfo {author} {\bibfnamefont {Y.-F.}\ \bibnamefont {Cai}}, \bibinfo {author} {\bibfnamefont {E.~N.}\ \bibnamefont {Saridakis}}, \bibinfo {author} {\bibfnamefont {M.~R.}\ \bibnamefont {Setare}},\ and\ \bibinfo {author} {\bibfnamefont {J.-Q.}\ \bibnamefont {Xia}},\ }\bibfield  {title} {\bibinfo {title} {{Quintom Cosmology: Theoretical implications and observations}},\ }\href {https://doi.org/10.1016/j.physrep.2010.04.001} {\bibfield  {journal} {\bibinfo  {journal} {Phys. Rept.}\ }\textbf {\bibinfo {volume} {493}},\ \bibinfo {pages} {1} (\bibinfo {year} {2010})},\ \Eprint {https://arxiv.org/abs/0909.2776} {0909.2776} \BibitemShut {NoStop}%
\bibitem [{\citenamefont {Wang}\ \emph {et~al.}(2016)\citenamefont {Wang}, \citenamefont {Abdalla}, \citenamefont {Atrio-Barandela},\ and\ \citenamefont {Pavon}}]{Wang:2016lxa}%
  \BibitemOpen
  \bibfield  {author} {\bibinfo {author} {\bibfnamefont {B.}~\bibnamefont {Wang}}, \bibinfo {author} {\bibfnamefont {E.}~\bibnamefont {Abdalla}}, \bibinfo {author} {\bibfnamefont {F.}~\bibnamefont {Atrio-Barandela}},\ and\ \bibinfo {author} {\bibfnamefont {D.}~\bibnamefont {Pavon}},\ }\bibfield  {title} {\bibinfo {title} {{Dark Matter and Dark Energy Interactions: Theoretical Challenges, Cosmological Implications and Observational Signatures}},\ }\href {https://doi.org/10.1088/0034-4885/79/9/096901} {\bibfield  {journal} {\bibinfo  {journal} {Rept. Prog. Phys.}\ }\textbf {\bibinfo {volume} {79}},\ \bibinfo {pages} {096901} (\bibinfo {year} {2016})},\ \Eprint {https://arxiv.org/abs/1603.08299} {1603.08299} \BibitemShut {NoStop}%
\bibitem [{\citenamefont {Creminelli}\ \emph {et~al.}(2006)\citenamefont {Creminelli}, \citenamefont {Luty}, \citenamefont {Nicolis},\ and\ \citenamefont {Senatore}}]{Creminelli:2006xe}%
  \BibitemOpen
  \bibfield  {author} {\bibinfo {author} {\bibfnamefont {P.}~\bibnamefont {Creminelli}}, \bibinfo {author} {\bibfnamefont {M.~A.}\ \bibnamefont {Luty}}, \bibinfo {author} {\bibfnamefont {A.}~\bibnamefont {Nicolis}},\ and\ \bibinfo {author} {\bibfnamefont {L.}~\bibnamefont {Senatore}},\ }\bibfield  {title} {\bibinfo {title} {{Starting the Universe: Stable Violation of the Null Energy Condition and Non-standard Cosmologies}},\ }\href {https://doi.org/10.1088/1126-6708/2006/12/080} {\bibfield  {journal} {\bibinfo  {journal} {J. High Energy Phys.}\ }\textbf {\bibinfo {volume} {2006}}\bibfield  {number} {\bibinfo  {number} { (12)},\ \bibinfo {pages} {080}} (\bibinfo {year} {2006})},\ \Eprint {https://arxiv.org/abs/hep-th/0606090} {hep-th/0606090} \BibitemShut {NoStop}%
\bibitem [{\citenamefont {Deffayet}\ \emph {et~al.}(2010)\citenamefont {Deffayet}, \citenamefont {Pujolas}, \citenamefont {Sawicki},\ and\ \citenamefont {Vikman}}]{Deffayet:2010qz}%
  \BibitemOpen
  \bibfield  {author} {\bibinfo {author} {\bibfnamefont {C.}~\bibnamefont {Deffayet}}, \bibinfo {author} {\bibfnamefont {O.}~\bibnamefont {Pujolas}}, \bibinfo {author} {\bibfnamefont {I.}~\bibnamefont {Sawicki}},\ and\ \bibinfo {author} {\bibfnamefont {A.}~\bibnamefont {Vikman}},\ }\bibfield  {title} {\bibinfo {title} {{Imperfect Dark Energy from Kinetic Gravity Braiding}},\ }\href {https://doi.org/10.1088/1475-7516/2010/10/026} {\bibfield  {journal} {\bibinfo  {journal} {J. Cosmol. Astropart. Phys.}\ }\textbf {\bibinfo {volume} {2010}}\bibfield  {number} {\bibinfo  {number} { (10)},\ \bibinfo {pages} {026}} (\bibinfo {year} {2010})},\ \Eprint {https://arxiv.org/abs/1008.0048} {1008.0048} \BibitemShut {NoStop}%
\bibitem [{\citenamefont {Gubitosi}\ \emph {et~al.}(2013)\citenamefont {Gubitosi}, \citenamefont {Piazza},\ and\ \citenamefont {Vernizzi}}]{Gubitosi:2012hu}%
  \BibitemOpen
  \bibfield  {author} {\bibinfo {author} {\bibfnamefont {G.}~\bibnamefont {Gubitosi}}, \bibinfo {author} {\bibfnamefont {F.}~\bibnamefont {Piazza}},\ and\ \bibinfo {author} {\bibfnamefont {F.}~\bibnamefont {Vernizzi}},\ }\bibfield  {title} {\bibinfo {title} {{The Effective Field Theory of Dark Energy}},\ }\href {https://doi.org/10.1088/1475-7516/2013/02/032} {\bibfield  {journal} {\bibinfo  {journal} {J. Cosmol. Astropart. Phys.}\ }\textbf {\bibinfo {volume} {2013}}\bibfield  {number} {\bibinfo  {number} { (02)},\ \bibinfo {pages} {032}} (\bibinfo {year} {2013})},\ \Eprint {https://arxiv.org/abs/1210.0201} {1210.0201} \BibitemShut {NoStop}%
\bibitem [{\citenamefont {Gleyzes}\ \emph {et~al.}(2013)\citenamefont {Gleyzes}, \citenamefont {Langlois}, \citenamefont {Piazza},\ and\ \citenamefont {Vernizzi}}]{Gleyzes:2013ooa}%
  \BibitemOpen
  \bibfield  {author} {\bibinfo {author} {\bibfnamefont {J.}~\bibnamefont {Gleyzes}}, \bibinfo {author} {\bibfnamefont {D.}~\bibnamefont {Langlois}}, \bibinfo {author} {\bibfnamefont {F.}~\bibnamefont {Piazza}},\ and\ \bibinfo {author} {\bibfnamefont {F.}~\bibnamefont {Vernizzi}},\ }\bibfield  {title} {\bibinfo {title} {{Essential Building Blocks of Dark Energy}},\ }\href {https://doi.org/10.1088/1475-7516/2013/08/025} {\bibfield  {journal} {\bibinfo  {journal} {J. Cosmol. Astropart. Phys.}\ }\textbf {\bibinfo {volume} {2013}}\bibfield  {number} {\bibinfo  {number} { (08)},\ \bibinfo {pages} {025}} (\bibinfo {year} {2013})},\ \Eprint {https://arxiv.org/abs/1304.4840} {1304.4840} \BibitemShut {NoStop}%
\bibitem [{\citenamefont {Heisenberg}(2014)}]{Heisenberg:2014rta}%
  \BibitemOpen
  \bibfield  {author} {\bibinfo {author} {\bibfnamefont {L.}~\bibnamefont {Heisenberg}},\ }\bibfield  {title} {\bibinfo {title} {{Generalization of the Proca Action}},\ }\href {https://doi.org/10.1088/1475-7516/2014/05/015} {\bibfield  {journal} {\bibinfo  {journal} {J. Cosmol. Astropart. Phys.}\ }\textbf {\bibinfo {volume} {2014}}\bibfield  {number} {\bibinfo  {number} { (05)},\ \bibinfo {pages} {015}} (\bibinfo {year} {2014})},\ \Eprint {https://arxiv.org/abs/1402.7026} {1402.7026} \BibitemShut {NoStop}%
\bibitem [{\citenamefont {De~Felice}\ \emph {et~al.}(2016)\citenamefont {De~Felice}, \citenamefont {Heisenberg}, \citenamefont {Kase}, \citenamefont {Mukohyama}, \citenamefont {Tsujikawa},\ and\ \citenamefont {Zhang}}]{DeFelice:2016yws}%
  \BibitemOpen
  \bibfield  {author} {\bibinfo {author} {\bibfnamefont {A.}~\bibnamefont {De~Felice}}, \bibinfo {author} {\bibfnamefont {L.}~\bibnamefont {Heisenberg}}, \bibinfo {author} {\bibfnamefont {R.}~\bibnamefont {Kase}}, \bibinfo {author} {\bibfnamefont {S.}~\bibnamefont {Mukohyama}}, \bibinfo {author} {\bibfnamefont {S.}~\bibnamefont {Tsujikawa}},\ and\ \bibinfo {author} {\bibfnamefont {Y.-l.}\ \bibnamefont {Zhang}},\ }\bibfield  {title} {\bibinfo {title} {{Cosmology in generalized Proca theories}},\ }\href {https://doi.org/10.1088/1475-7516/2016/06/048} {\bibfield  {journal} {\bibinfo  {journal} {J. Cosmol. Astropart. Phys.}\ }\textbf {\bibinfo {volume} {2016}}\bibfield  {number} {\bibinfo  {number} { (06)},\ \bibinfo {pages} {048}} (\bibinfo {year} {2016})},\ \Eprint {https://arxiv.org/abs/1603.05806} {1603.05806} \BibitemShut {NoStop}%
\bibitem [{\citenamefont {Creminelli}\ and\ \citenamefont {Vernizzi}(2017)}]{Creminelli:2017sry}%
  \BibitemOpen
  \bibfield  {author} {\bibinfo {author} {\bibfnamefont {P.}~\bibnamefont {Creminelli}}\ and\ \bibinfo {author} {\bibfnamefont {F.}~\bibnamefont {Vernizzi}},\ }\bibfield  {title} {\bibinfo {title} {{Dark Energy after GW170817 and GRB170817A}},\ }\href {https://doi.org/10.1103/PhysRevLett.119.251302} {\bibfield  {journal} {\bibinfo  {journal} {Phys. Rev. Lett.}\ }\textbf {\bibinfo {volume} {119}},\ \bibinfo {pages} {251302} (\bibinfo {year} {2017})},\ \Eprint {https://arxiv.org/abs/1710.05877} {1710.05877} \BibitemShut {NoStop}%
\bibitem [{\citenamefont {Ezquiaga}\ and\ \citenamefont {Zumalacarregui}(2017)}]{Ezquiaga:2017ekz}%
  \BibitemOpen
  \bibfield  {author} {\bibinfo {author} {\bibfnamefont {J.~M.}\ \bibnamefont {Ezquiaga}}\ and\ \bibinfo {author} {\bibfnamefont {M.}~\bibnamefont {Zumalacarregui}},\ }\bibfield  {title} {\bibinfo {title} {{Dark Energy After GW170817: Dead Ends and the Road Ahead}},\ }\href {https://doi.org/10.1103/PhysRevLett.119.251304} {\bibfield  {journal} {\bibinfo  {journal} {Phys. Rev. Lett.}\ }\textbf {\bibinfo {volume} {119}},\ \bibinfo {pages} {251304} (\bibinfo {year} {2017})},\ \Eprint {https://arxiv.org/abs/1710.05901} {1710.05901} \BibitemShut {NoStop}%
\bibitem [{\citenamefont {Garc{\'\i}a-Garc{\'\i}a}\ \emph {et~al.}(2026)\citenamefont {Garc{\'\i}a-Garc{\'\i}a}, \citenamefont {Ferreira},\ and\ \citenamefont {Wolf}}]{Garcia-Garcia:2026nzy}%
  \BibitemOpen
  \bibfield  {author} {\bibinfo {author} {\bibfnamefont {C.}~\bibnamefont {Garc{\'\i}a-Garc{\'\i}a}}, \bibinfo {author} {\bibfnamefont {P.~G.}\ \bibnamefont {Ferreira}},\ and\ \bibinfo {author} {\bibfnamefont {W.~J.}\ \bibnamefont {Wolf}},\ }\href@noop {} {\bibinfo {title} {{The Status of Single Scalar Field Dark Energy}}} (\bibinfo {year} {2026}),\ \Eprint {https://arxiv.org/abs/2607.07777} {2607.07777} \BibitemShut {NoStop}%
\bibitem [{\citenamefont {Blas}\ \emph {et~al.}(2011)\citenamefont {Blas}, \citenamefont {Lesgourgues},\ and\ \citenamefont {Tram}}]{Blas:2011rf}%
  \BibitemOpen
  \bibfield  {author} {\bibinfo {author} {\bibfnamefont {D.}~\bibnamefont {Blas}}, \bibinfo {author} {\bibfnamefont {J.}~\bibnamefont {Lesgourgues}},\ and\ \bibinfo {author} {\bibfnamefont {T.}~\bibnamefont {Tram}},\ }\bibfield  {title} {\bibinfo {title} {{The Cosmic Linear Anisotropy Solving System (CLASS) II: Approximation schemes}},\ }\href {https://doi.org/10.1088/1475-7516/2011/07/034} {\bibfield  {journal} {\bibinfo  {journal} {J. Cosmol. Astropart. Phys.}\ }\textbf {\bibinfo {volume} {2011}}\bibfield  {number} {\bibinfo  {number} { (07)},\ \bibinfo {pages} {034}} (\bibinfo {year} {2011})},\ \Eprint {https://arxiv.org/abs/1104.2933} {1104.2933} \BibitemShut {NoStop}%
\bibitem [{\citenamefont {Forconi}\ and\ \citenamefont {Melchiorri}(2025)}]{Forconi:2025gwo}%
  \BibitemOpen
  \bibfield  {author} {\bibinfo {author} {\bibfnamefont {M.}~\bibnamefont {Forconi}}\ and\ \bibinfo {author} {\bibfnamefont {A.}~\bibnamefont {Melchiorri}},\ }\bibfield  {title} {\bibinfo {title} {{The impact on non-Gaussianities of the ISW-Lensing correlation in non-standard cosmologies}},\ }\href {https://doi.org/10.1016/j.dark.2025.102126} {\bibfield  {journal} {\bibinfo  {journal} {Phys. Dark Univ.}\ }\textbf {\bibinfo {volume} {50}},\ \bibinfo {pages} {102126} (\bibinfo {year} {2025})}\BibitemShut {NoStop}%
\bibitem [{\citenamefont {Ghafari}\ \emph {et~al.}(2025)\citenamefont {Ghafari}, \citenamefont {Najafi}, \citenamefont {Ghodsi~Yengejeh}, \citenamefont {{\"O}z{\"u}lker}, \citenamefont {Di~Valentino},\ and\ \citenamefont {Firouzjaee}}]{Ghafari:2025eql}%
  \BibitemOpen
  \bibfield  {author} {\bibinfo {author} {\bibfnamefont {P.}~\bibnamefont {Ghafari}}, \bibinfo {author} {\bibfnamefont {M.}~\bibnamefont {Najafi}}, \bibinfo {author} {\bibfnamefont {M.}~\bibnamefont {Ghodsi~Yengejeh}}, \bibinfo {author} {\bibfnamefont {E.}~\bibnamefont {{\"O}z{\"u}lker}}, \bibinfo {author} {\bibfnamefont {E.}~\bibnamefont {Di~Valentino}},\ and\ \bibinfo {author} {\bibfnamefont {J.~T.}\ \bibnamefont {Firouzjaee}},\ }\href@noop {} {\bibinfo {title} {{A Multi-Probe ISW Study of Dark Energy Models with Negative Energy Density: Galaxy Correlations, Lensing Bispectrum, and Planck ISW-Lensing Likelihood}}} (\bibinfo {year} {2025}),\ \Eprint {https://arxiv.org/abs/2512.07060} {2512.07060} \BibitemShut {NoStop}%
\end{thebibliography}%

\end{document}